\documentclass[aps,prd,twocolumn,tightenlines,superscriptaddress,showpacs,byrevtex]{revtex4}

\usepackage{dcolumn}  % Align table columns on decimal point
\usepackage{epsfig}
\usepackage{floatflt}
\usepackage{amssymb,amsmath,indentfirst,enumerate}

\newcommand{\ee}{e^+e^-}
\newcommand{\uu}{\mu^+\mu^-}
\newcommand{\pp}{\pi^+\pi^-}
\newcommand{\Uf}{\Upsilon(10860)}
\newcommand{\Uu}{\Upsilon(1S)}
\newcommand{\Ud}{\Upsilon(2S)}
\newcommand{\Ut}{\Upsilon(3S)}
\newcommand{\Un}{\Upsilon(nS)}

\newcommand{\mpp}{M(\pi^+\pi^-)}
\newcommand{\mmpp}{M_{\rm miss}(\pi^+\pi^-)}

\newcommand{\hm}{h_b(mP)}

\newcommand{\etal}{\em et al.}

\begin{document}

%\vspace*{10mm}

\title{\boldmath Amplitude analysis of $\ee\to\Un\pp$ at $\sqrt{s}=10.865$~GeV}

%%% Paper:
%%% Journal:  Physical Review
%%% Contacts:
%%% Non-responding authors or those who said NO are commented out.
%%% ====================================================================
%%% Click the RELOAD button on your web browser to see the updated file.
%%% ====================================================================
%%% Use \input{author} to insert this material into your latex file.
%%%%% Force institutions to appear in alphabetical order when typeset.
\noaffiliation
\affiliation{University of the Basque Country UPV/EHU, 48080 Bilbao}
\affiliation{Beihang University, Beijing 100191}
\affiliation{University of Bonn, 53115 Bonn}
\affiliation{Budker Institute of Nuclear Physics SB RAS and Novosibirsk State University, Novosibirsk 630090}
\affiliation{Faculty of Mathematics and Physics, Charles University, 121 16 Prague}
\affiliation{Chiba University, Chiba 263-8522}
\affiliation{University of Cincinnati, Cincinnati, Ohio 45221}
\affiliation{Deutsches Elektronen--Synchrotron, 22607 Hamburg}
%%%\affiliation{Department of Physics, Fu Jen Catholic University, Taipei 24205}
\affiliation{Justus-Liebig-Universit\"at Gie\ss{}en, 35392 Gie\ss{}en}
%%%\affiliation{Gifu University, Gifu 501-1193}
\affiliation{II. Physikalisches Institut, Georg-August-Universit\"at G\"ottingen, 37073 G\"ottingen}
%%%\affiliation{The Graduate University for Advanced Studies, Hayama 240-0193}
%%%\affiliation{Gyeongsang National University, Chinju 660-701}
\affiliation{Hanyang University, Seoul 133-791}
\affiliation{University of Hawaii, Honolulu, Hawaii 96822}
\affiliation{High Energy Accelerator Research Organization (KEK), Tsukuba 305-0801}
%%%\affiliation{Hiroshima Institute of Technology, Hiroshima 731-5193}
\affiliation{IKERBASQUE, Basque Foundation for Science, 48011 Bilbao}
%%%\affiliation{University of Illinois at Urbana-Champaign, Urbana, Illinois 61801}
\affiliation{Indian Institute of Technology Guwahati, Assam 781039}
\affiliation{Indian Institute of Technology Madras, Chennai 600036}
%%%\affiliation{Indiana University, Bloomington, Indiana 47408}
\affiliation{Institute of High Energy Physics, Chinese Academy of Sciences, Beijing 100049}
\affiliation{Institute of High Energy Physics, Vienna 1050}
\affiliation{Institute for High Energy Physics, Protvino 142281}
%%%\affiliation{Institute of Mathematical Sciences, Chennai 600113}
\affiliation{INFN - Sezione di Torino, 10125 Torino}
\affiliation{Institute for Theoretical and Experimental Physics, Moscow 117218}
\affiliation{J. Stefan Institute, 1000 Ljubljana}
\affiliation{Kanagawa University, Yokohama 221-8686}
\affiliation{Institut f\"ur Experimentelle Kernphysik, Karlsruher Institut f\"ur Technologie, 76131 Karlsruhe}
\affiliation{Kavli Institute for the Physics and Mathematics of the Universe (WPI), University of Tokyo, Kashiwa 277-8583}
%%%\affiliation{Department of Physics, Faculty of Sciences, King Abdulaziz University, Jeddah 21589}
\affiliation{Korea Institute of Science and Technology Information, Daejeon 305-806}
\affiliation{Korea University, Seoul 136-713}
%%%\affiliation{Kyoto University, Kyoto 606-8502}
\affiliation{Kyungpook National University, Daegu 702-701}
\affiliation{\'Ecole Polytechnique F\'ed\'erale de Lausanne (EPFL), Lausanne 1015}
\affiliation{Faculty of Mathematics and Physics, University of Ljubljana, 1000 Ljubljana}
\affiliation{Luther College, Decorah, Iowa 52101}
\affiliation{University of Maribor, 2000 Maribor}
\affiliation{Max-Planck-Institut f\"ur Physik, 80805 M\"unchen}
\affiliation{School of Physics, University of Melbourne, Victoria 3010}
\affiliation{Moscow Physical Engineering Institute, Moscow 115409}
%%%\affiliation{Moscow Institute of Physics and Technology, Moscow Region 141700}
\affiliation{Graduate School of Science, Nagoya University, Nagoya 464-8602}
\affiliation{Kobayashi-Maskawa Institute, Nagoya University, Nagoya 464-8602}
%%%\affiliation{Nara University of Education, Nara 630-8528}
\affiliation{Nara Women's University, Nara 630-8506}
\affiliation{National Central University, Chung-li 32054}
\affiliation{National United University, Miao Li 36003}
\affiliation{Department of Physics, National Taiwan University, Taipei 10617}
\affiliation{H. Niewodniczanski Institute of Nuclear Physics, Krakow 31-342}
\affiliation{Nippon Dental University, Niigata 951-8580}
\affiliation{Niigata University, Niigata 950-2181}
\affiliation{University of Nova Gorica, 5000 Nova Gorica}
\affiliation{Osaka City University, Osaka 558-8585}
%%%\affiliation{Osaka University, Osaka 565-0871}
\affiliation{Pacific Northwest National Laboratory, Richland, Washington 99352}
\affiliation{Panjab University, Chandigarh 160014}
\affiliation{Peking University, Beijing 100871}
%%%\affiliation{University of Pittsburgh, Pittsburgh, Pennsylvania 15260}
%%%\affiliation{Punjab Agricultural University, Ludhiana 141004}
%%%\affiliation{Research Center for Electron Photon Science, Tohoku University, Sendai 980-8578}
%%%\affiliation{Research Center for Nuclear Physics, Osaka University, Osaka 567-0047}
%%%\affiliation{RIKEN BNL Research Center, Upton, New York 11973}
%%%\affiliation{Saga University, Saga 840-8502}
\affiliation{University of Science and Technology of China, Hefei 230026}
\affiliation{Seoul National University, Seoul 151-742}
%%%\affiliation{Shinshu University, Nagano 390-8621}
\affiliation{Soongsil University, Seoul 156-743}
\affiliation{Sungkyunkwan University, Suwon 440-746}
\affiliation{School of Physics, University of Sydney, NSW 2006}
\affiliation{Department of Physics, Faculty of Science, University of Tabuk, Tabuk 71451}
\affiliation{Tata Institute of Fundamental Research, Mumbai 400005}
\affiliation{Excellence Cluster Universe, Technische Universit\"at M\"unchen, 85748 Garching}
%%%\affiliation{Toho University, Funabashi 274-8510}
\affiliation{Tohoku Gakuin University, Tagajo 985-8537}
\affiliation{Tohoku University, Sendai 980-8578}
\affiliation{Department of Physics, University of Tokyo, Tokyo 113-0033}
\affiliation{Tokyo Institute of Technology, Tokyo 152-8550}
\affiliation{Tokyo Metropolitan University, Tokyo 192-0397}
\affiliation{Tokyo University of Agriculture and Technology, Tokyo 184-8588}
\affiliation{University of Torino, 10124 Torino}
%%%\affiliation{Toyama National College of Maritime Technology, Toyama 933-0293}
\affiliation{CNP, Virginia Polytechnic Institute and State University, Blacksburg, Virginia 24061}
\affiliation{Wayne State University, Detroit, Michigan 48202}
\affiliation{Yamagata University, Yamagata 990-8560}
\affiliation{Yonsei University, Seoul 120-749}
  \author{A.~Garmash}\affiliation{Budker Institute of Nuclear Physics SB RAS and
 Novosibirsk State University, Novosibirsk 630090} % BINP
  \author{A.~Bondar}\affiliation{Budker Institute of Nuclear Physics SB RAS and Novosibirsk State University, Novosibirsk 630090} % BINP
  \author{A.~Kuzmin}\affiliation{Budker Institute of Nuclear Physics SB RAS and Novosibirsk State University, Novosibirsk 630090} % BINP
  \author{A.~Abdesselam}\affiliation{Department of Physics, Faculty of Science, University of Tabuk, Tabuk 71451} % Tabuk
  \author{I.~Adachi}\affiliation{High Energy Accelerator Research Organization (KEK), Tsukuba 305-0801} % KEK
% \author{K.~Adamczyk}\affiliation{H. Niewodniczanski Institute of Nuclear Physics, Krakow 31-342} % Krakow
  \author{H.~Aihara}\affiliation{Department of Physics, University of Tokyo, Tokyo 113-0033} % Tokyo
  \author{S.~Al~Said}\affiliation{Department of Physics, Faculty of Science, University of Tabuk, Tabuk 71451}\affiliation{Department of Physics, Faculty of Science, King Abdulaziz University, Jeddah 21589}
% University, Jeddah 21589) % Tabuk
% \author{K.~Arinstein}\affiliation{Budker Institute of Nuclear Physics SB RAS and Novosibirsk State University, Novosibirsk 630090} % BINP
% \author{Y.~Arita}\affiliation{Graduate School of Science, Nagoya University, Nagoya 464-8602} % Nagoya
  \author{D.~M.~Asner}\affiliation{Pacific Northwest National Laboratory, Richland, Washington 99352} % PNNL
% \author{T.~Aso}\affiliation{Toyama National College of Maritime Technology, Toyama 933-0293} % Toyama
  \author{V.~Aulchenko}\affiliation{Budker Institute of Nuclear Physics SB RAS and Novosibirsk State University, Novosibirsk 630090} % BINP
  \author{T.~Aushev}\affiliation{Institute for Theoretical and Experimental Physics, Moscow 117218} % ITEP
% ...
% Ljubljana
% \author{N.~Zwahlen}\affiliation{\'Ecole Polytechnique F\'ed\'erale de Lausanne (EPFL), Lausanne 1015} % Lausanne
% \author{O.~Zyukova}\affiliation{Budker Institute of Nuclear Physics SB RAS and Novosibirsk State University, Novosibirsk 630090} % BINP
\collaboration{The Belle Collaboration}

\begin{abstract}
\noindent

We report results on studies of the $\ee$ annihilation into three-body 
$\Un\pp$ ($n=1,2,3$) final states including measurements of cross sections
and the full amplitude analysis. The cross sections measured at 
$\sqrt{s}=10.865$~GeV and corrected for the initial state radiation are 
$\sigma(\ee\to\Uu\pp)=(2.27\pm0.12\pm0.14)$~pb, 
$\sigma(\ee\to\Ud\pp)=(4.07\pm0.16\pm0.45)$~pb, and
$\sigma(\ee\to\Ut\pp)=(1.46\pm0.09\pm0.16)$~pb.
Amplitude analysis of the three-body $\Un\pp$ final states strongly favors 
$I^{G}(J^{P})$=$1^{+}(1^{+})$ quantum-number assignments for the two 
bottomonium-like $Z^\pm_b$ states, recently observed in the $\Un\pi^\pm$ 
and $\hm\pi^\pm$ ($m=1,2$) decay channels. The results are obtained with a 
$121.4\,{\rm fb}^{-1}$ data sample collected with the Belle detector at 
the KEKB asymmetric-energy $\ee$ collider.
\end{abstract}

\pacs{14.40.Pq, 13.25.Gv, 12.39.Pn}

\maketitle

{\renewcommand{\thefootnote}{\fnsymbol{footnote}}}
\setcounter{footnote}{0}

%============================================================================
%============================================================================
%============================================================================

\section{Introduction}

Analysis of the $\Uf$ decays to non-$B\bar{B}$ final states has led to
several surprises. Recently, the Belle Collaboration reported observation
of anomalously high rates for the $\ee\to\Un\pp$ ($n=1,2,3$)~\cite{Belle_ypp}
and $\ee\to\hm\pp$ ($m=1,2$)~\cite{Belle_hb} transitions measured in the
vicinity of the $\Uf$ peak. If the $\Un$ signals are attributed entirely to 
the $\Uf$ decays, the measured partial decay widths 
$\Gamma[\Uf\to\Un\pp]\sim0.5$~MeV are about two orders of magnitude larger
than the typical widths for the dipion transitions amongst $\Un$ states 
with $n\leq4$. In addition, the rates of the $\ee\to\hm\pp$ processes are
found to be comparable with those for $\ee\to\Un\pp$, and hence the 
process with a spin flip of the heavy quark (that is, $\hm$ production) 
is not suppressed. These unexpected observations indicate that an exotic 
mechanism might contribute to the $\Uf$ decays. A detailed analysis of 
the three-body $\ee\to\Un\pp$ and $\ee\to\hm\pp$ processes reported by 
Belle~\cite{Belle_Zbc} revealed the presence of two charged 
bottomonium-like states, denoted as $Z_b(10610)^\pm$ and $Z_b(10650)^\pm$.
These two resonances are observed in the decay chains 
$\ee\to Z^\pm_b\pi^\mp\to\Un\pp$ and $\ee\to Z^\pm_b\pi^\mp\to\hm\pp$. 
The non-resonant contribution is found to be sizable in the $\Un\pp$ 
channels and consistent with zero in the $\hm\pp$ ones. Masses and 
widths of the $Z_b^\pm$ states have been measured in a two-dimensional 
amplitude analysis of the three-body $\ee\to\Un\pp$ 
transitions~\cite{Belle_Zbc}. Also, observation of the neutral 
$Z_b(10610)^0$ partner has been reported recently by Belle~\cite{Belle_Zbn}. 
Although the simplified angular analysis in Ref.~\cite{Belle_ang}
favors the $J^P=1^+$ assignment for the two charged $Z_b$ states, the 
discrimination power against other possible combinations is not high 
enough to claim this assignment unequivocally.
 
Results of the analysis of three-body $\ee\to\Un\pp$ processes presented 
in this paper are obtained utilizing Dalitz techniques that not only allow
us to determine the relative fractions of intermediate components but 
also provide high sensitivity to the spin and parity of the $Z_b$ states. 
Results on the $\ee$ annihilation to the three-body $\Un\pp$ final states 
reported here supersede those published in Ref.~\cite{Belle_ypp}.

We use a data sample with an integrated luminosity of $121.4$~fb$^{-1}$ 
collected at the peak of the $\Uf$ resonance ($\sqrt{s}=10.865$~GeV/$c^2$) 
with the Belle detector at the KEKB asymmetric-energy $\ee$ 
collider~\cite{KEKB}. 

%============================================================================
%============================================================================
%============================================================================

\section{Belle Detector}

The Belle detector~\cite{Belle} is a large-solid-angle magnetic spectrometer
based on a 1.5~T superconducting solenoid magnet. Charged particle tracking
is provided by a four-layer silicon vertex detector and a 50-layer central
drift chamber (CDC) that surround the interaction point. The charged particle
acceptance covers laboratory polar angles between $\theta=17^{\circ}$ and
$150^{\circ}$, corresponding to about 92\% of the total solid angle in the
center-of-mass (c.m.) frame. 

Charged hadron identification is provided by $dE/dx$ measurements in the CDC,
an array of 1188 aerogel Cherenkov counters (ACC), and a barrel-like array
of 128 time-of-flight scintillation counters (TOF); information from the three
sub-detectors is combined to form likelihood ratios, which is then 
used for pion, kaon and proton discrimination. Electromagnetic showering
particles are detected in an array of 8736 CsI(Tl) crystals (ECL) that covers
the same solid angle as the charged particle tracking system. 
Electron identification in Belle is based on a combination of $dE/dx$
measurements in the CDC, the response of the ACC, and the position, shape 
and total energy deposition (i.e., $E/p$) of the shower detected in the ECL.
The electron identification efficiency is greater than 92\% for tracks with
$p_{\rm lab}>1.0$~GeV/$c$ and the hadron misidentification probability is 
below 0.3\%. The magnetic field is returned via an iron yoke that is 
instrumented to detect muons and $K^0_L$ mesons. Muons are identified based
on their penetration range and transverse scattering in the KLM detector. 
In the momentum region relevant to this analysis, the identification 
efficiency is about 90\% while the probability to misidentify a pion as a 
muon is below 2\%. 

We use the EvtGen event generator~\cite{EvtGen} with PHOTOS~\cite{PHOTOS} 
for radiative corrections and a GEANT-based Monte 
Carlo (MC) simulation~\cite{GEANT} to model the response of the detector
and determine the acceptance. The MC simulation includes run-dependent 
detector performance variations and background conditions.

\begin{figure}[!t]
  \centering
  \includegraphics[width=0.45\textwidth]{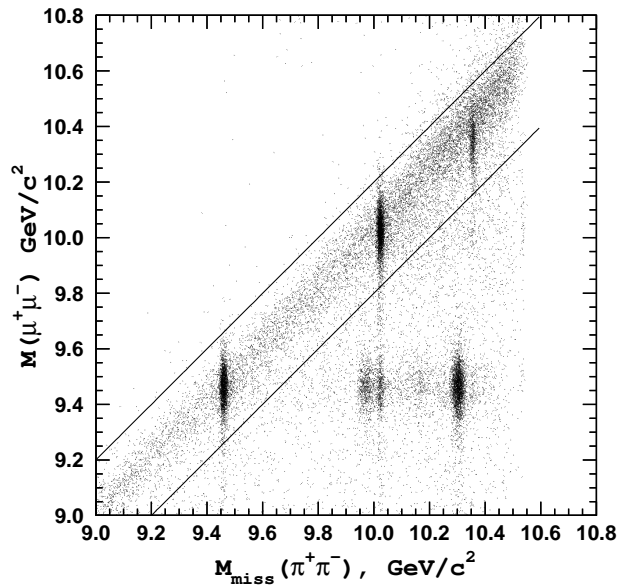}
  \caption{Scatter plot of all the $\ee\to\Un\pp$
           candidate events passed through initial selection criteria.
           The region between the two diagonal lines is defined as the
           signal region.}
\label{fig:ynspp-s-y5}
\end{figure}

%==============================================================================
%==============================================================================
%==============================================================================

\begin{figure*}[!ht]
  \centering
  \includegraphics[width=0.32\textwidth]{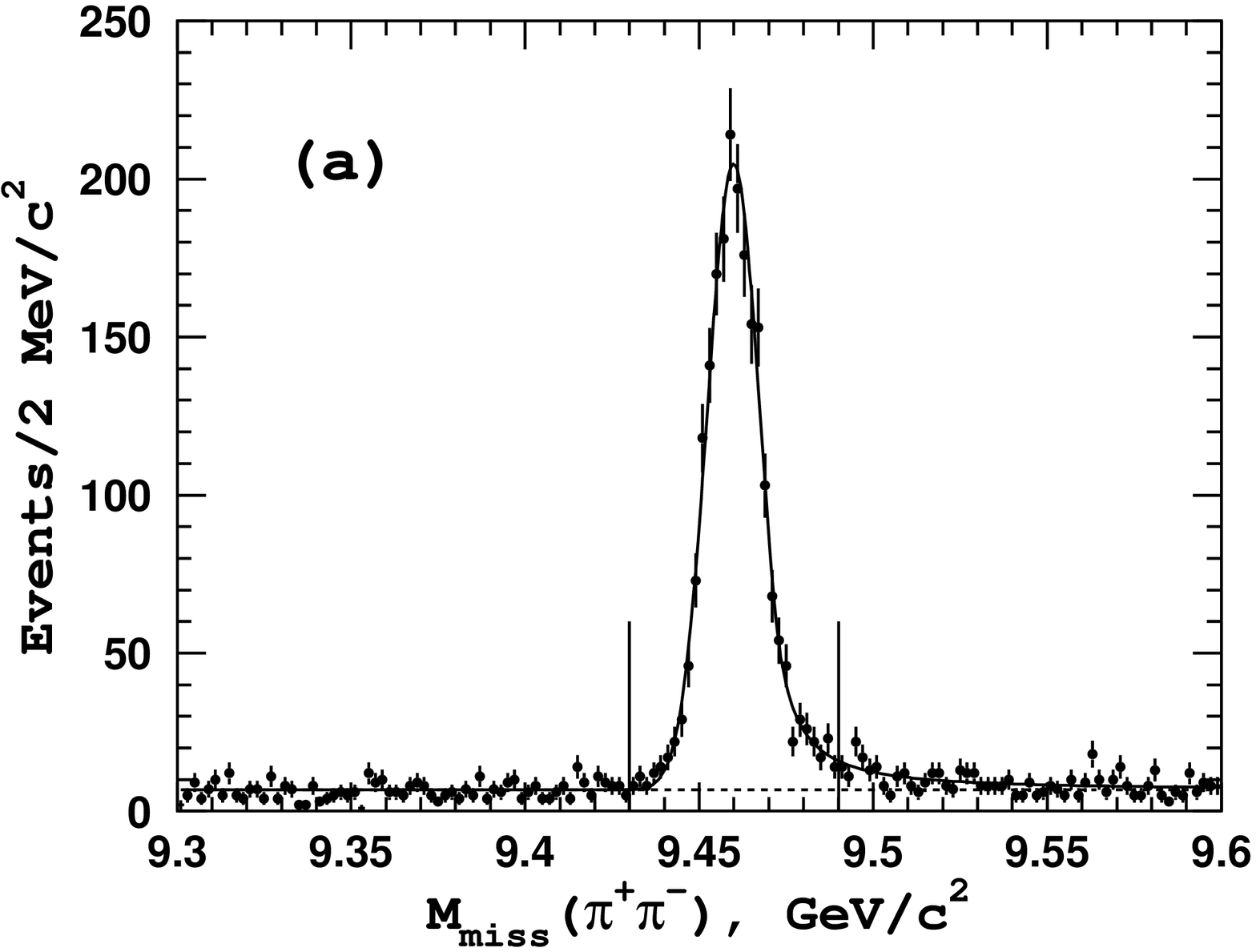} \hfill
  \includegraphics[width=0.32\textwidth]{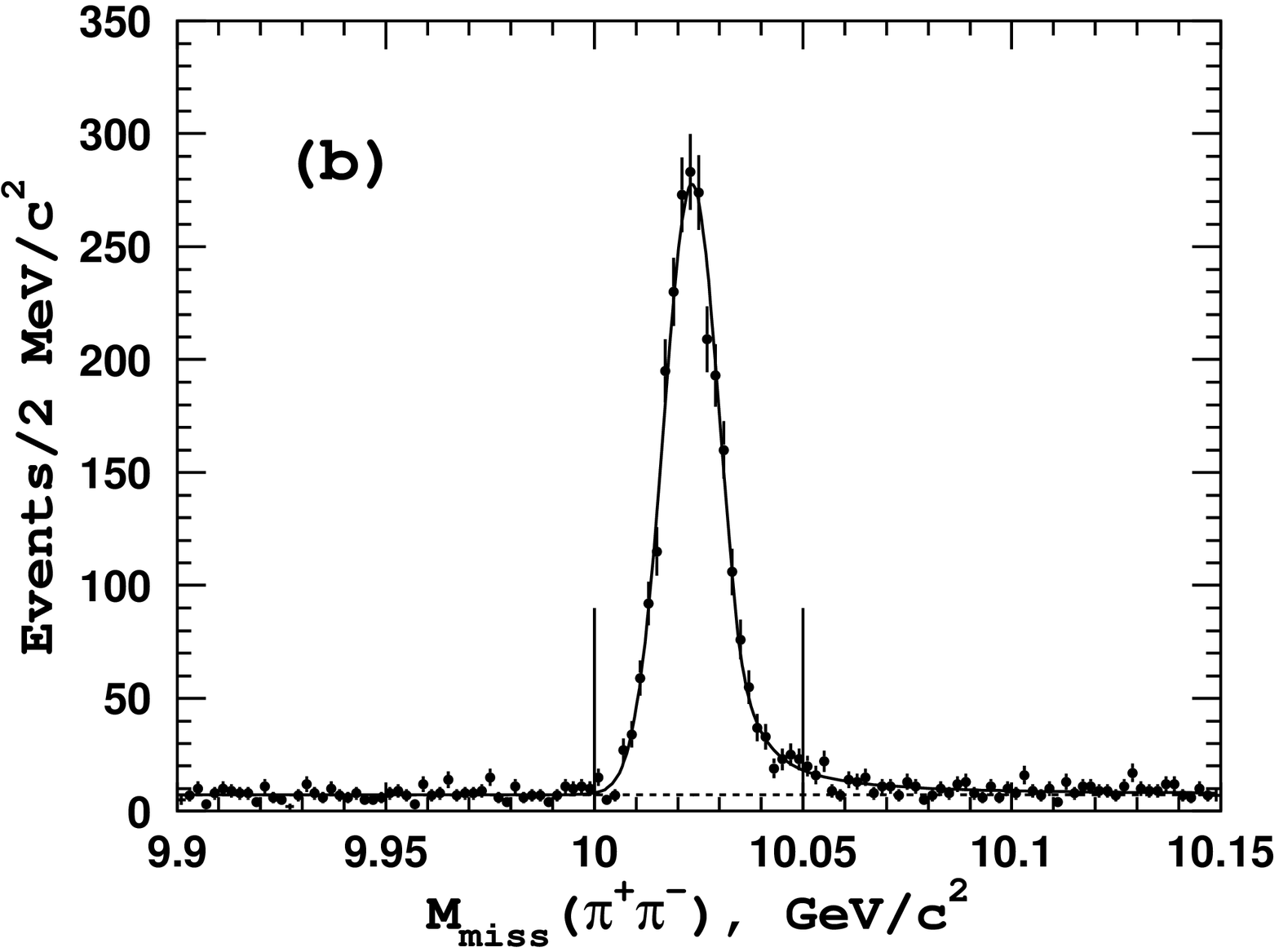} \hfill
  \includegraphics[width=0.32\textwidth]{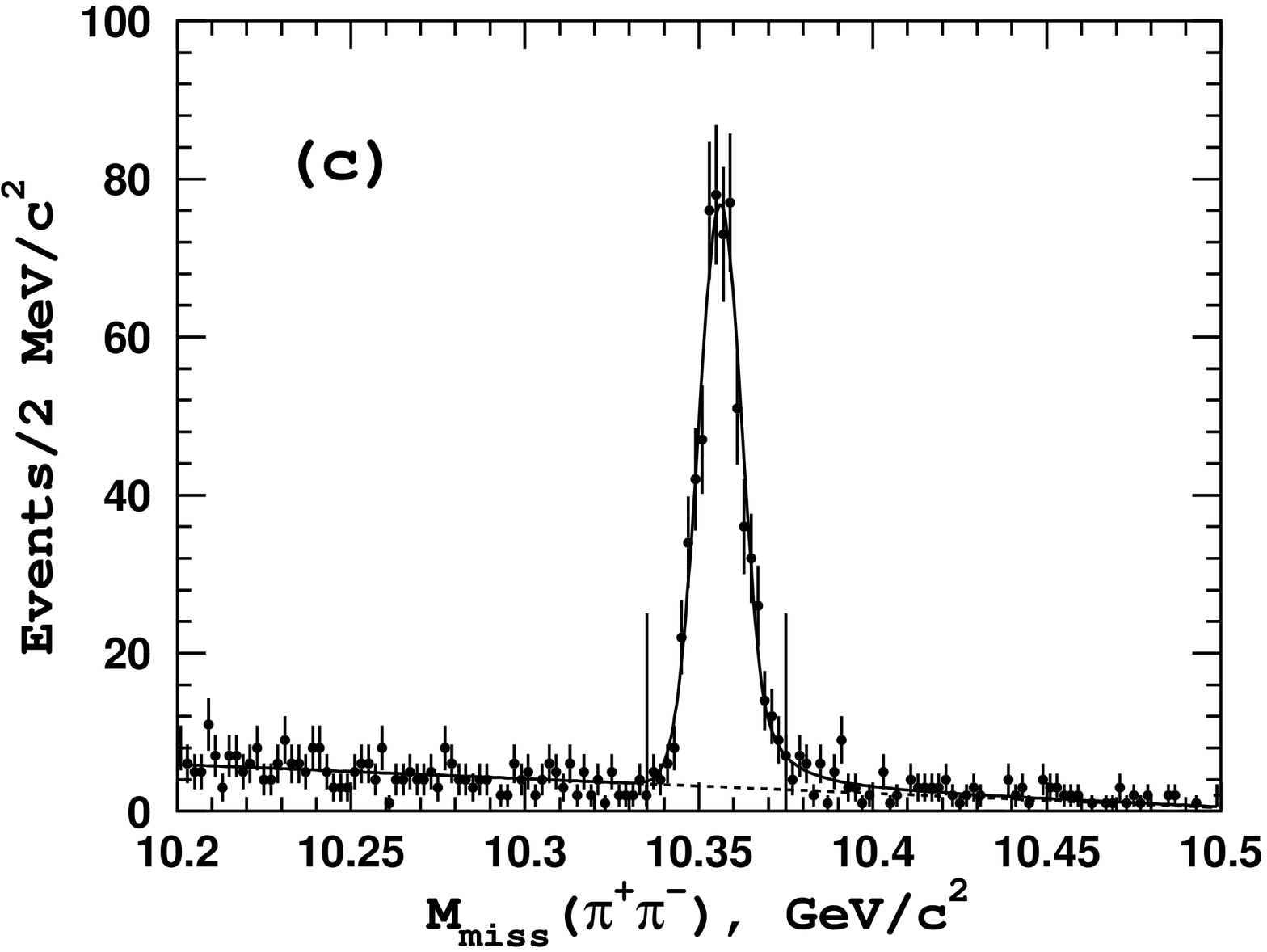}
  \caption{Distribution of missing mass associated with the $\pp$ 
           combination for $\ee\to\Un\pp$ candidate events in the
           (a) $\Uu$; (b) $\Ud$; (c) $\Ut$ mass region. 
           Points with error bars are the data, the solid line is 
           the fit, and the dashed line shows the background component. 
           Vertical lines define the corresponding signal region.}
\label{fig:ynspp-s-mm}
\end{figure*}

\begin{table*}[t]
  \caption{Summary of results from the analysis of the $\mmpp$ distribution.
           Quoted uncertainty is statistical only.}
  \medskip
  \label{tab:ynspp_sfrac}
\centering
  \begin{tabular}{lccc} \hline \hline
Final state &  $\Uu\pp$   &
               $\Ud\pp$   &
               $\Ut\pp$
\\ \hline
          $\mpp$  Signal, GeV/$c^2$          &
          $>0.45$                            &
          $>0.37$                            &
          $>0.32$    
\\
            $N_{\rm signal}$                   &
           $2090\pm115$                      &
           $2476\pm 97$                      &
           $ 628\pm 41$
\\
           $\Upsilon$ Peak,  MeV/$c^2$        &
           $ 9459.9\pm0.8$                   &
           $10023.4\pm0.4$                   &
           $10356.2\pm0.7$    
\\
           $\sigma$,    MeV/$c^2$            &
           $8.34$                            &
           $7.48$                            &
           $6.85$    
\\ \hline
         $\mmpp$ Signal, GeV/$c^2$           &
         $ (9.430,  9.490)$                  &
         $(10.000, 10.050)$                  &
         $(10.335, 10.375)$    
\\
            $N_{\rm events}$                   &
           $1905$                            &
           $2312$                            &
           $ 635$
\\
            $f_{\rm sig}$                     &
           $0.937\pm0.071$                   &
           $0.940\pm0.060$                   &
           $0.918\pm0.076$    
\\ \hline
          $\mmpp$ Sidebands, GeV/$c^2$        &
         ~~~$ (9.38,  9.43)$~~~               &
         ~~~$ (9.94,  9.99)$~~~               &
         ~~~$(10.30, 10.33)$~~~    
\\
                                              &
         ~~~$ (9.49, 9.53)$~~~                &
         ~~~$(10.06, 10.11)$~~~               &
         ~~~$(10.38, 10.41)$~~~    
\\
            $N_{\rm events}$                   &
           $ 272 $                            &
           $ 291 $                            &
           $  91 $
\\
\hline \hline
\end{tabular}
\end{table*}

%============================================================================
%============================================================================
%============================================================================

\section{Event Selection}

Charged tracks are selected with a set of track quality requirements based
on the average hit residual and on the distances of closest approach to 
the interaction point. We require four well reconstructed tracks with a 
net zero charge in the event with two of them, oppositely charged, 
identified as muons and the other two consistent with pions. We also require 
that none of the four tracks be identified as an electron (electron veto).

Candidate $\ee\to\Un\pp\to\uu\pp$ events are identified via the measured 
invariant mass of the $\uu$ combination and the recoil mass, $\mmpp$, 
associated with the $\pp$ system, defined by
\begin{equation}
\mmpp = \sqrt{(E_{\rm c.m.}-E_{\pi\pi}^*)^2-p_{\pi\pi}^{*2}},
\end{equation}
where $E_{\rm c.m.}$ is the c.m.\ energy and $E^*_{\pi\pi}$ and $p_{\pi\pi}^*$ 
are the energy and momentum of the $\pp$ system measured in the c.m.\ 
frame. The two-dimensional distribution of $M(\uu)$ versus $\mmpp$ for 
all selected candidates is shown in Fig.~\ref{fig:ynspp-s-y5}. Events 
originating from the $\ee\to\uu\pp$ process fall within a narrow diagonal 
band (signal region) that is defined as $|\mmpp-M(\uu)|<0.2$~GeV/$c^2$ 
(see Fig.~\ref{fig:ynspp-s-y5}). Concentrations of events within the  
signal region near the $\Un$ nominal masses are apparent on the plot. 
Clusters of events below the diagonal band are mainly due to initial 
state radiation $\ee\to\Upsilon(2S,3S)\pp\gamma$ processes and direct 
$\ee\to\Upsilon(2S,3S)\pp$ production with a subsequent dipion transition
of the $\Upsilon(2S,3S)$ state to the ground $\Uu$ state. 
The one-dimensional $\mmpp$ projections for events in the signal region 
are shown in Fig.~\ref{fig:ynspp-s-mm}, where an additional requirement 
on the invariant mass of the $\pp$ system, $M(\pp)$, is imposed 
(see Table~\ref{tab:ynspp_sfrac}) to 
suppress the background from photon conversion in the inner parts of the 
Belle detector. We perform a binned maximum likelihood fit to the $\mmpp$
distributions with a sum of a Crystal Ball function~\cite{CBFunction} for 
the $\Un$ signal and a linear function for the combinatorial background 
component. The Crystal Ball function is used to account for the 
asymmetric shape of the $\Un$ signal due to initial state radiation of 
soft photons. All parameters (seven in total) are free parameters 
of the fit. Results of the fits are shown in Fig.~\ref{fig:ynspp-s-mm} 
and summarized in Table~\ref{tab:ynspp_sfrac}.

For the subsequent analysis, we select events around the respective $\Un$ 
mass peak as specified in Table~\ref{tab:ynspp_sfrac}. After all the 
selections are applied, we are left with $1905$, $2312$, and $635$ candidate
events for the $\Uu\pp$, $\Ud\pp$, and $\Ut\pp$ final state, respectively. 
The fractions of signal events in the selected samples are determined using
results of the fit to the corresponding $\mmpp$ spectrum (see 
Table~\ref{tab:ynspp_sfrac}). For selected events, we perform a 
mass-constrained fit of the $\uu$ pair to the nominal mass of the 
corresponding $\Un$ state to improve the $\Un\pi$ invariant mass resolution.

%==============================================================================
%==============================================================================
%==============================================================================

\begin{figure*}[!t]
  \centering
  \includegraphics[width=0.32\textwidth]{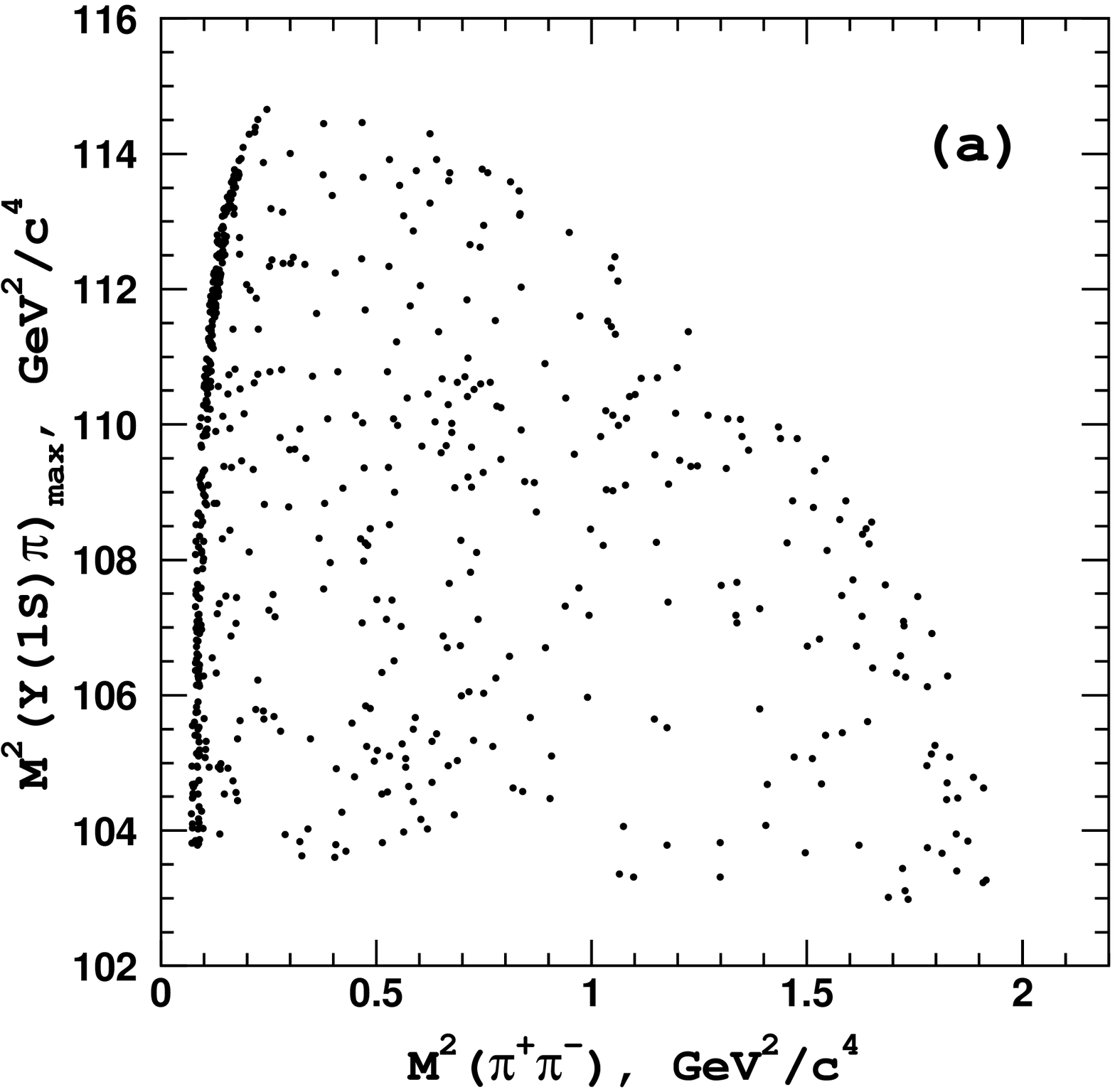} \hfill
  \includegraphics[width=0.32\textwidth]{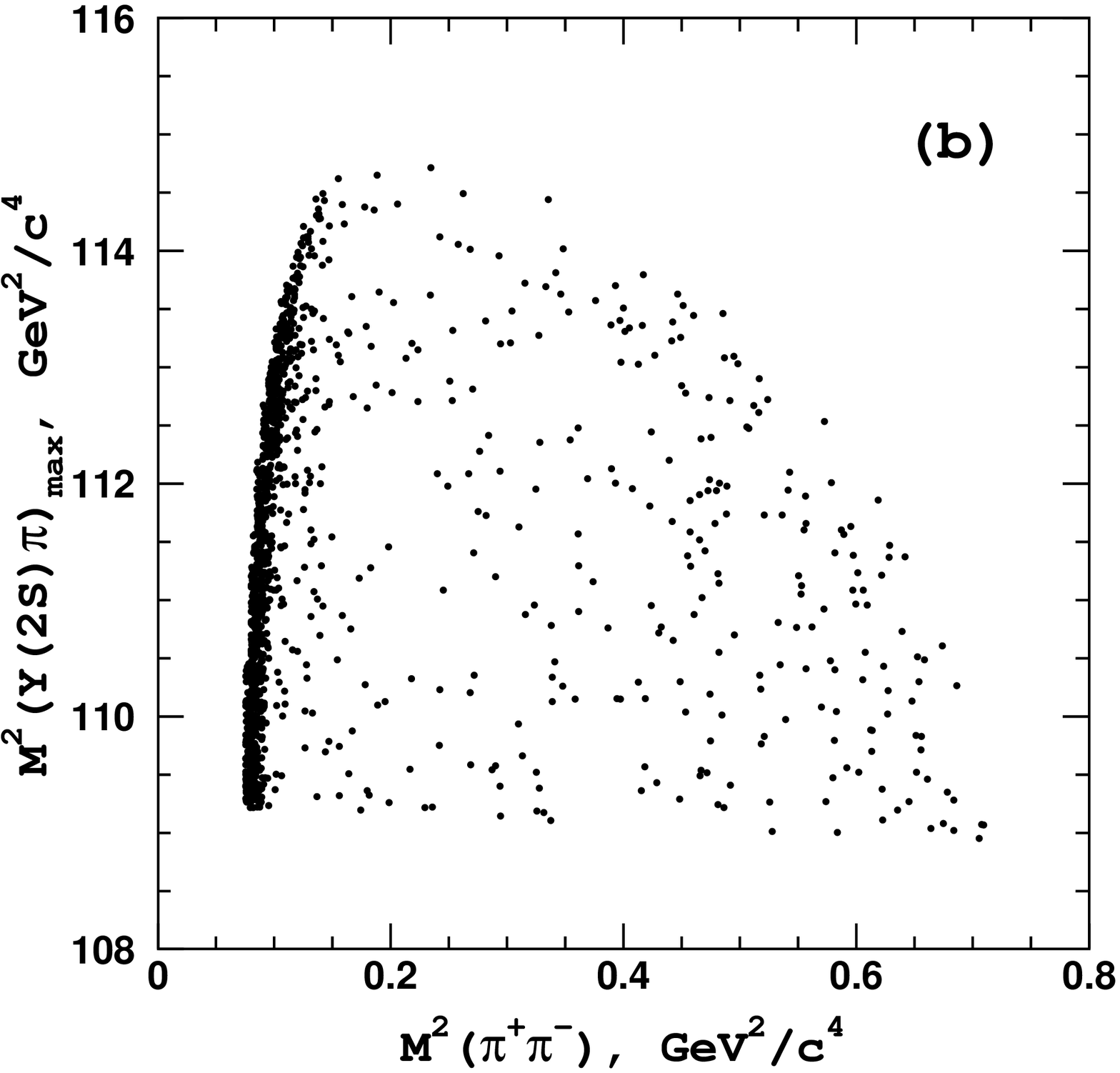} \hfill
  \includegraphics[width=0.32\textwidth]{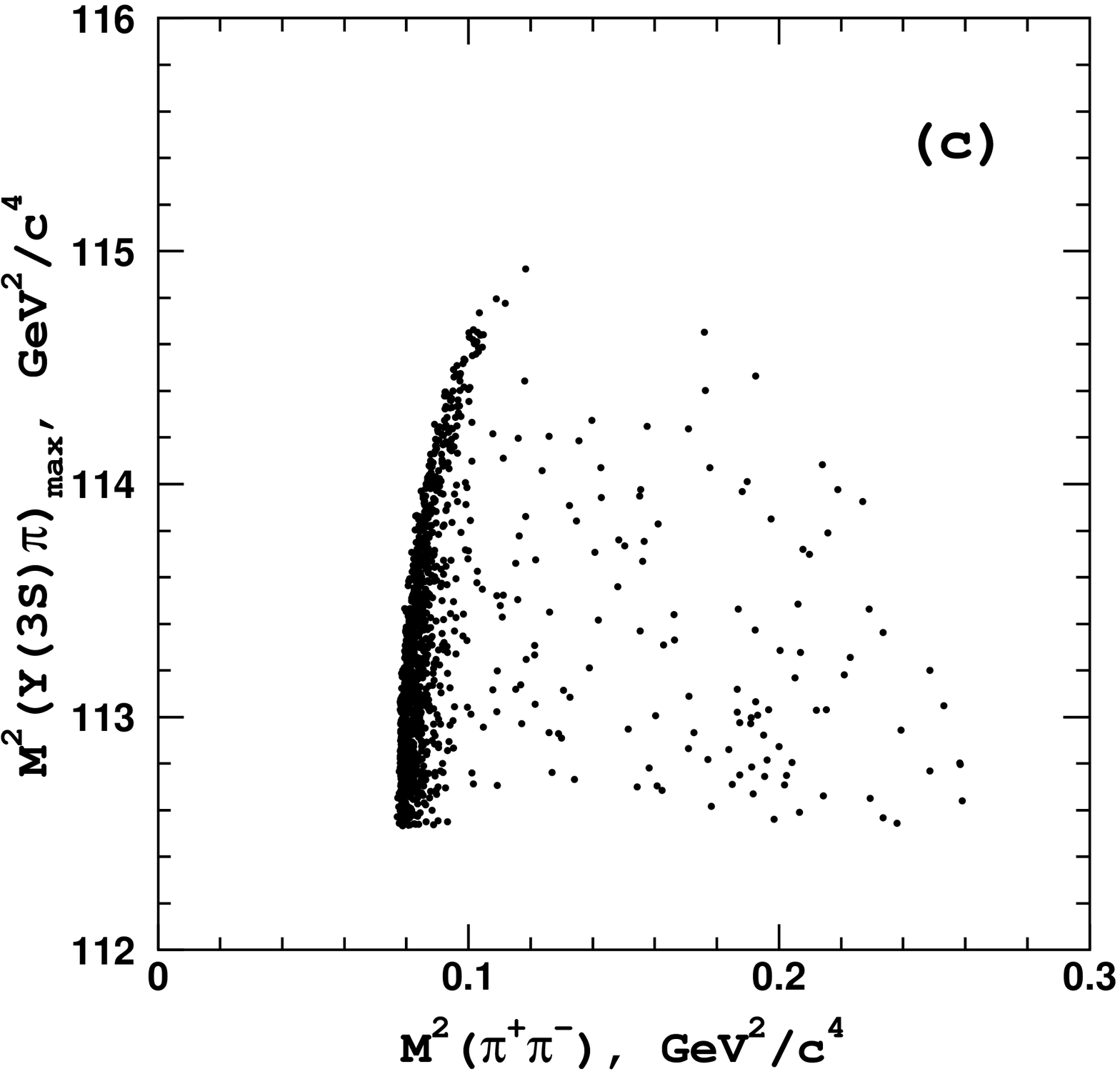} \\
  \includegraphics[width=0.32\textwidth]{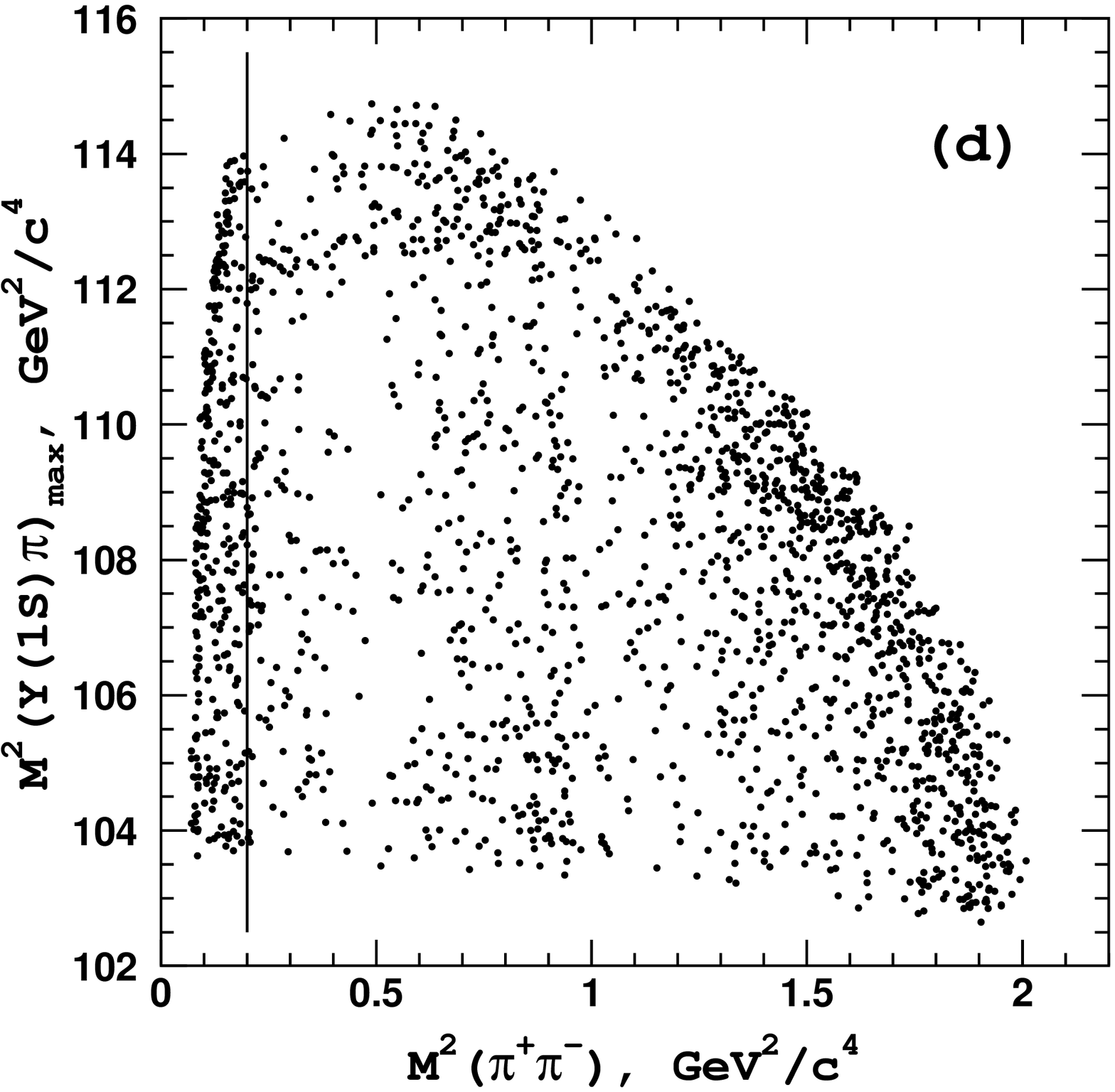} \hfill
  \includegraphics[width=0.32\textwidth]{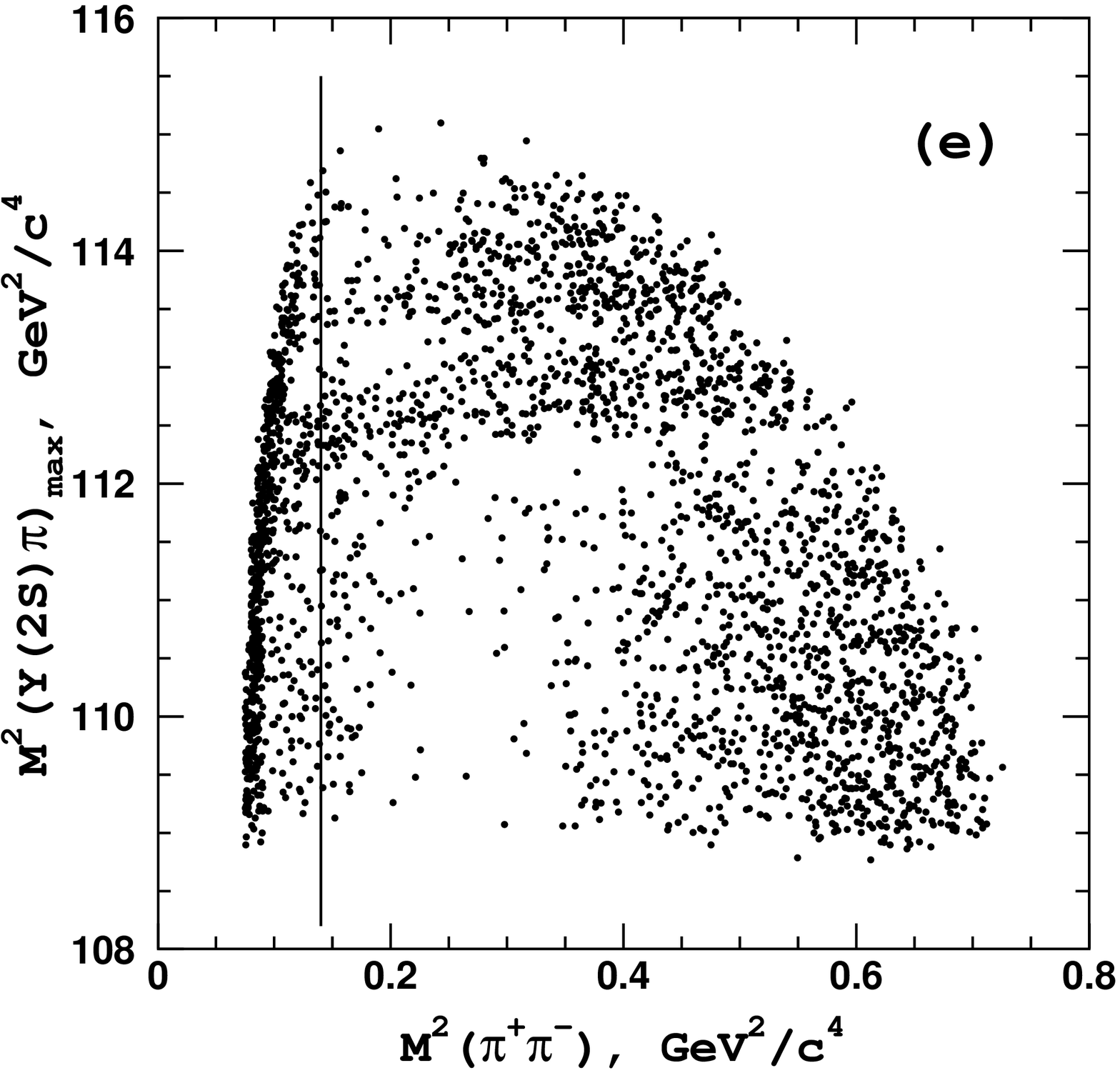} \hfill
  \includegraphics[width=0.32\textwidth]{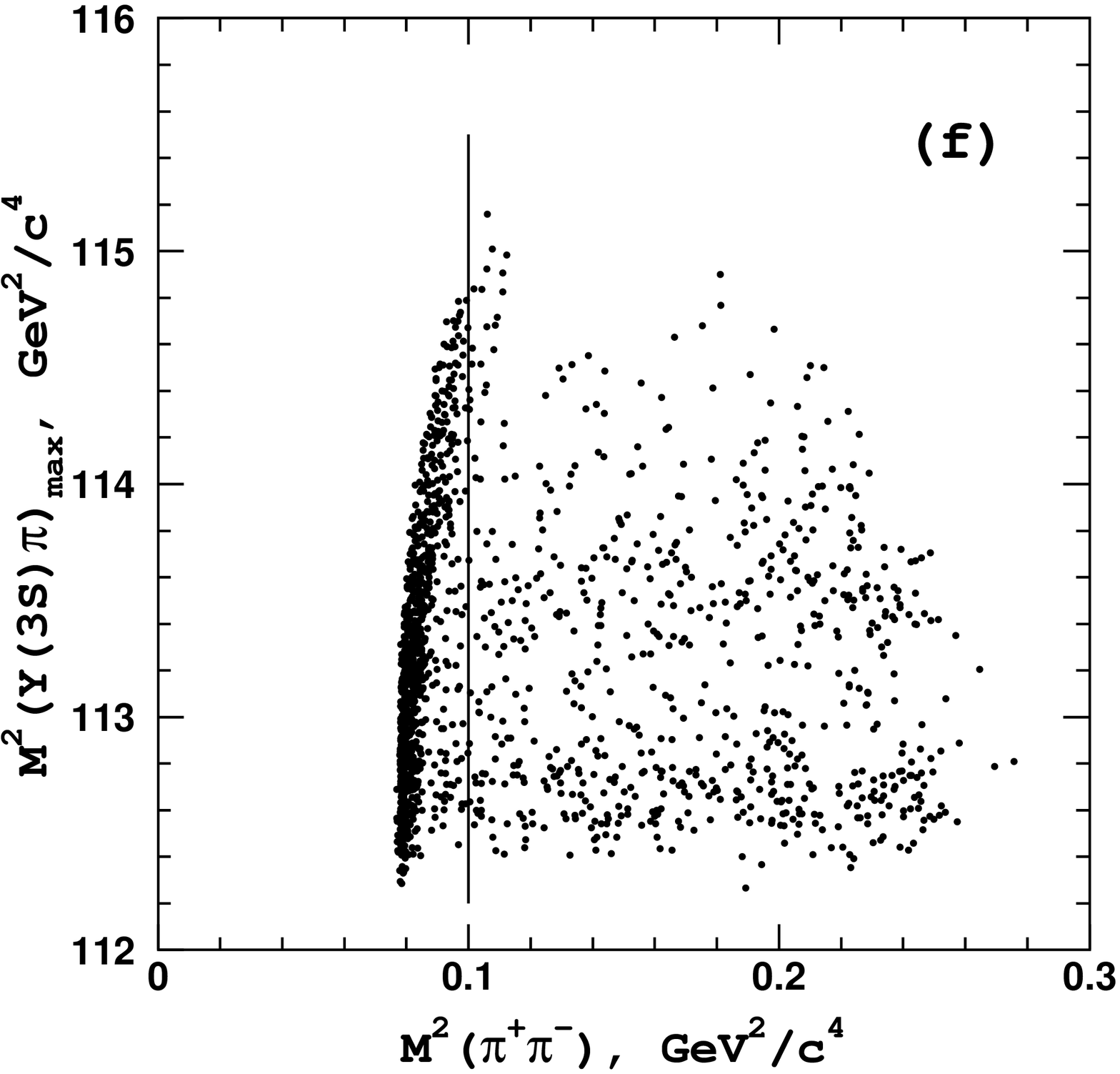}
  \caption{Dalitz plots for $\Un\pp$ events in sidebands of
           the (a) $\Uu$; (b) $\Ud$; (c) $\Ut$.
           Dalitz plots for $\Un\pp$ events in the signal region
           of the (d) $\Uu$; (e) $\Ud$; (f) $\Ut$.
           Regions of the Dalitz plots to the left of the respective 
           vertical line are excluded from the amplitude analyses.}
\label{fig:ynspp-dp}
\end{figure*}

\section{Amplitude analysis}

The $\ee$ annihilation to three-body $\Un\pp$ with a subsequent decay of
$\Un\to\uu$ results in four particles that are observed in the detector;
thus, there are six degrees of freedom in the system. In this analysis,
we use a Lorentz-invariant form of the transition amplitude (see the 
Appendix) and, for visualization of fit results, we make one-dimensional 
projections as described below. The amplitude analysis of the 
$\ee\to\Un\pp$ transitions reported here is performed by means of an 
unbinned maximum likelihood fit. 

Before analyzing events in the signal region, one needs to determine the 
distribution of background events over the phase space. Samples of 
background events are selected in $\Un$ mass sidebands and then fit 
to the nominal mass of the corresponding $\Un$ state to match the phase 
space boundaries for the signal. Definition of the mass sidebands and
the event yields are given in Table~\ref{tab:ynspp_sfrac}. Dalitz plots 
for the sideband events are shown in Figs.~\ref{fig:ynspp-dp}~(a, b, c),
where $M(\Un\pi)_{\rm max}$ is the maximum invariant mass of the two 
$\Un\pi$ combinations; here the requirement on $M(\pp)$ is relaxed. 
For visualization purposes, we plot the Dalitz distributions in terms 
of $M(\Un\pi)_{\rm max}$ in order to combine $\Un\pi^+$ and $\Un\pi^-$ 
events. As is apparent from these distributions, there is a strong 
enhancement in the level of the background just above the $\pp$ invariant
mass threshold. This enhancement is due to conversion of photons into 
an $\ee$ pair in the innermost parts of the Belle detector. Due to their
low momenta, conversion electrons and positrons are poorly identified 
by the CDC and so pass the electron veto requirement. We exclude this high
background region by applying a requirement on $\mpp$ as given in 
Table~\ref{tab:ynspp_sfrac}. The distribution of background events in the
remainder of the phase space is parametrized with the sum of a constant 
(that is uniform over phase space) and a term exponential in 
$M^2(\pp)$ to account for an excess of background events in the lower 
$M^2(\pp)$ region. In addition, in the $\Uu\pp$ sample, we include a 
contribution from $\rho(770)^0\to\pp$ decays. 

\begin{figure*}[!t]
  \centering
  \includegraphics[width=0.32\textwidth]{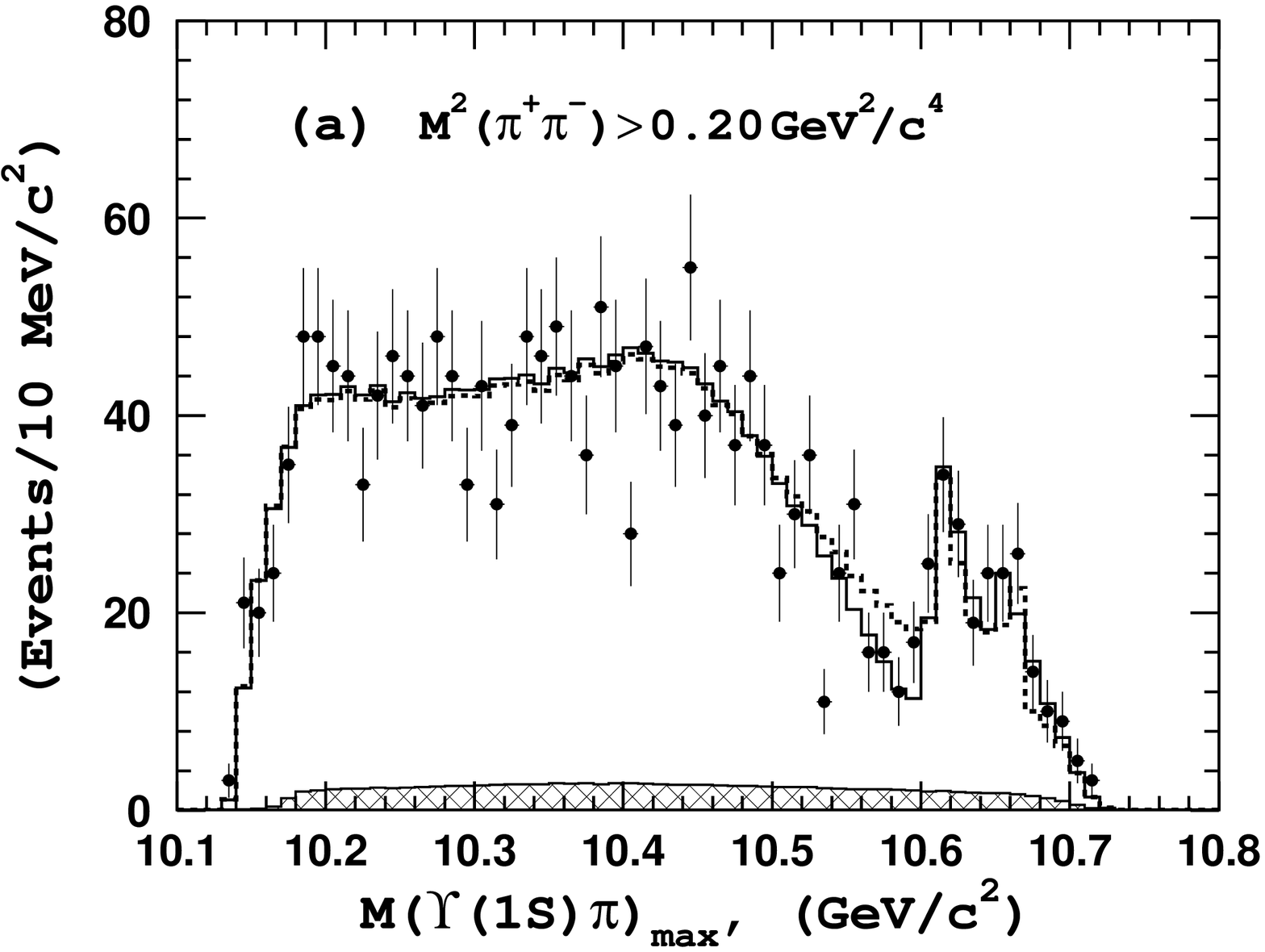} \hfill
  \includegraphics[width=0.32\textwidth]{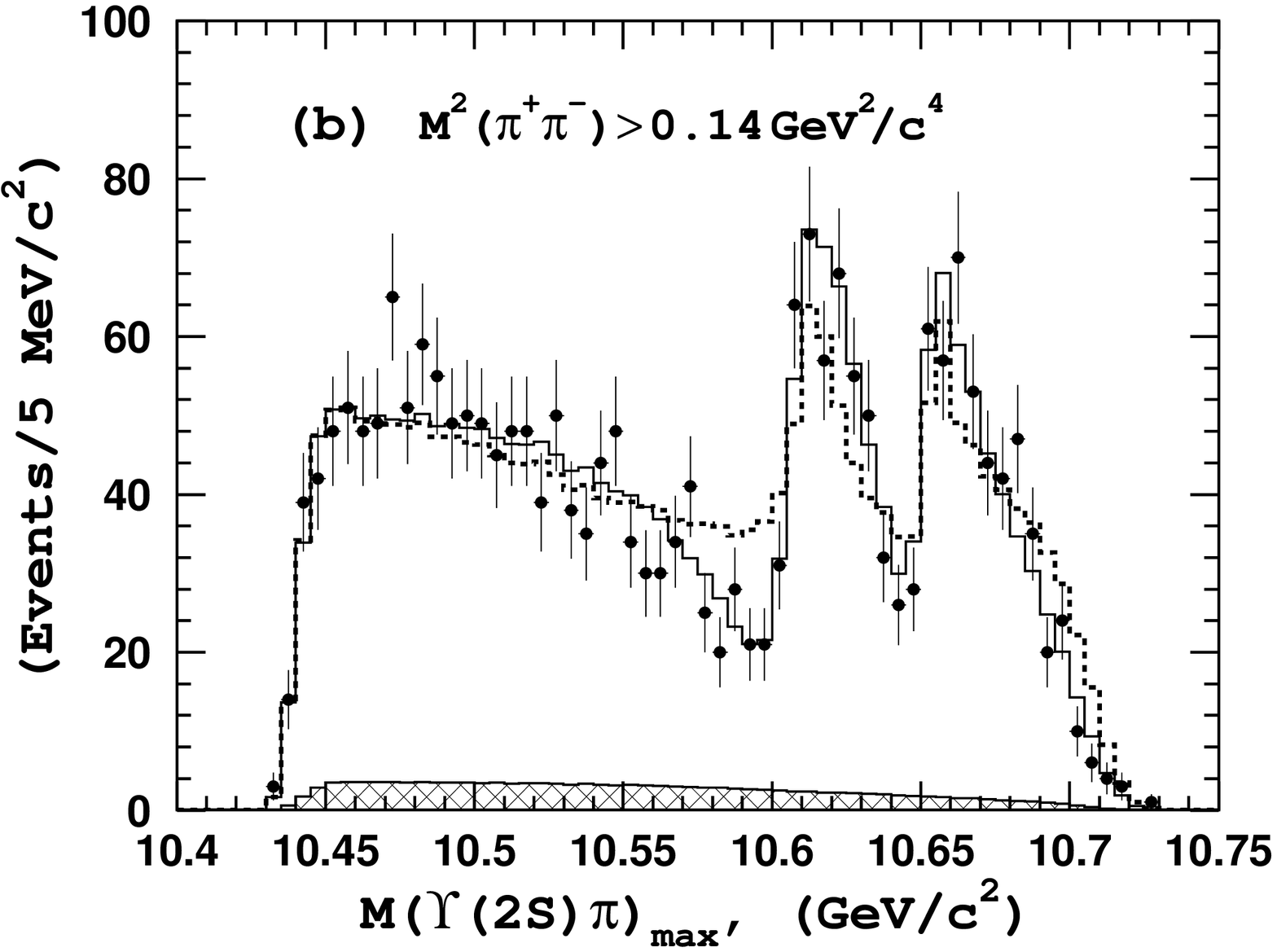} \hfill
  \includegraphics[width=0.32\textwidth]{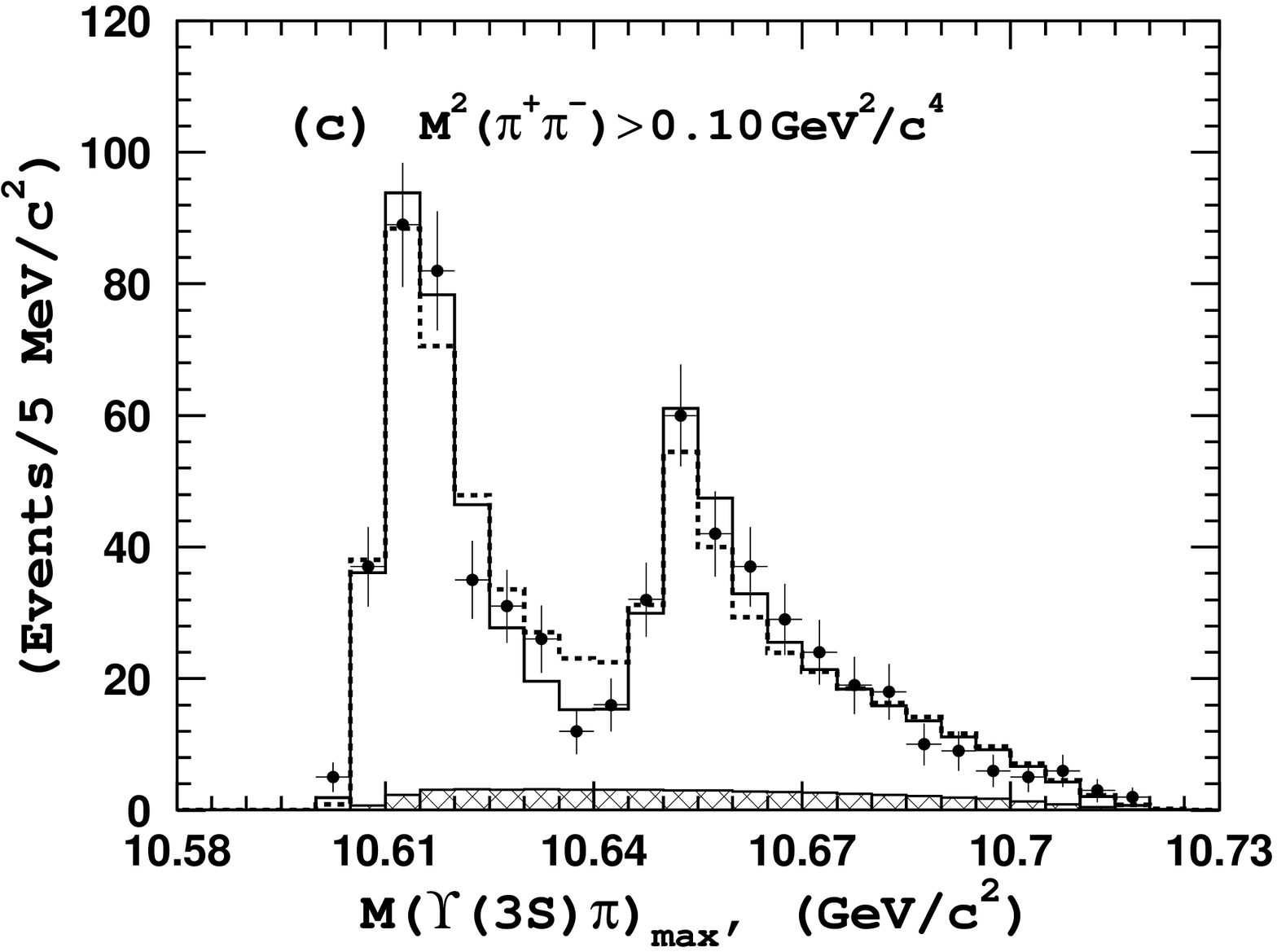} \\
  \includegraphics[width=0.32\textwidth]{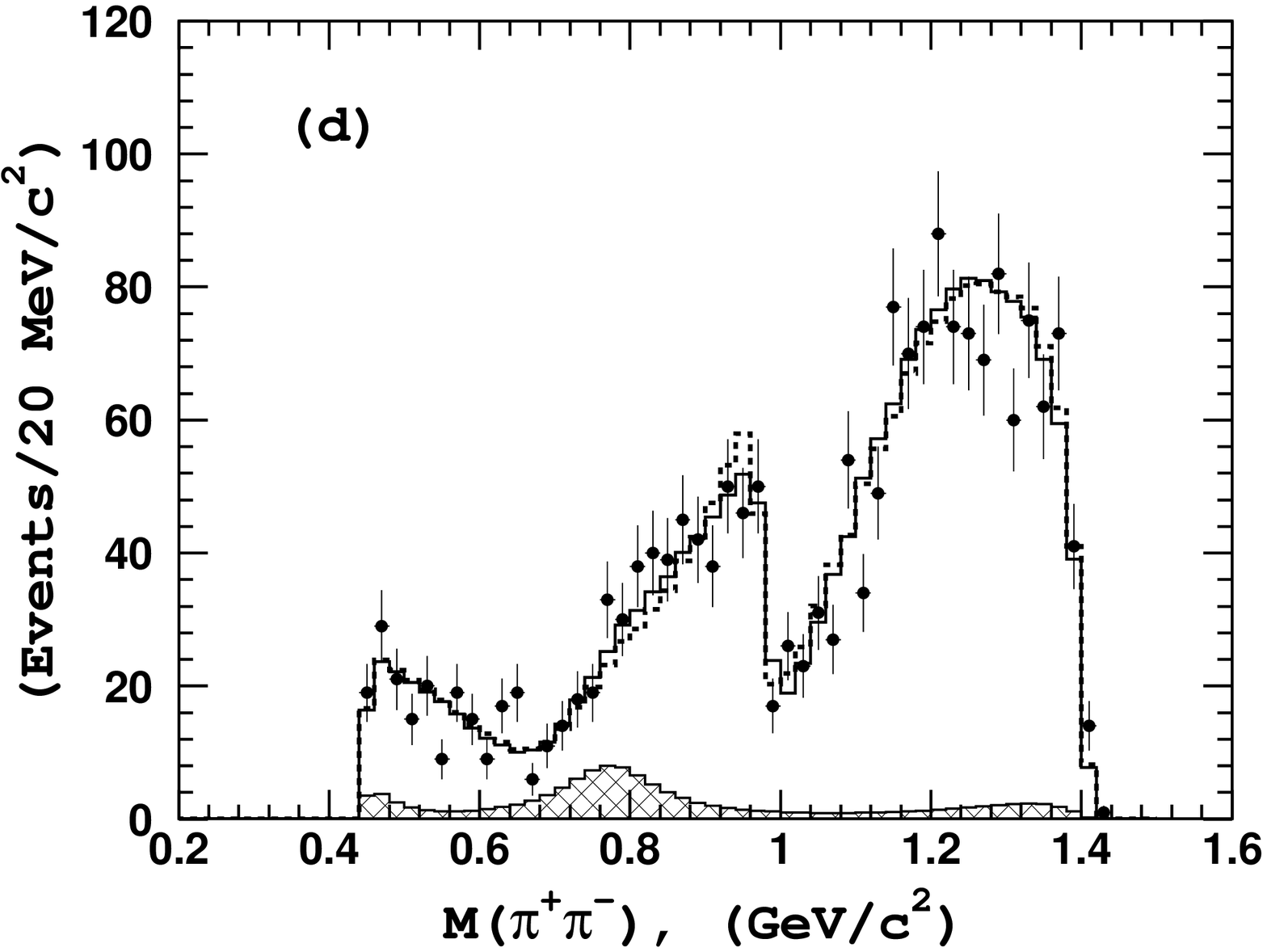} \hfill
  \includegraphics[width=0.32\textwidth]{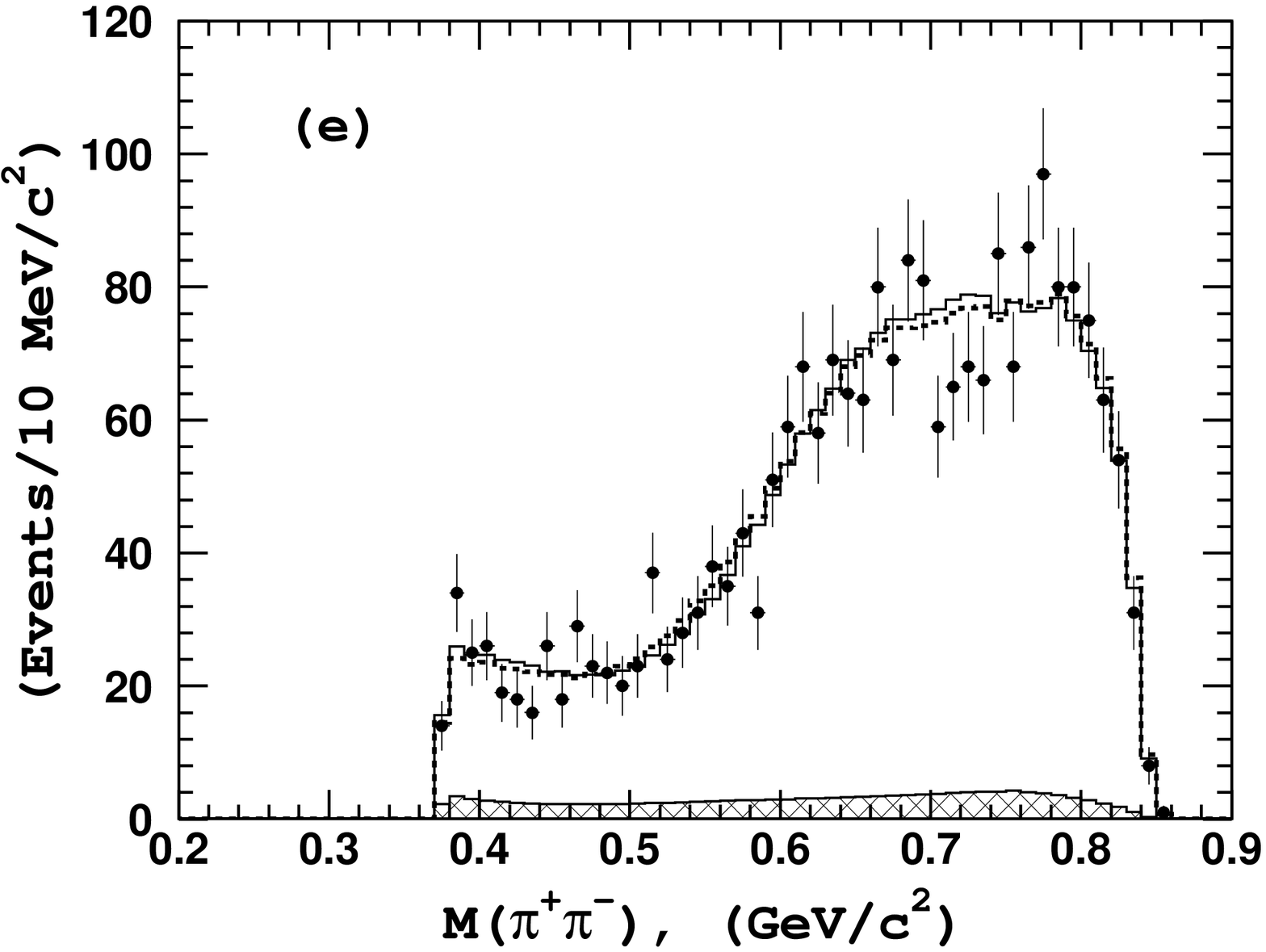} \hfill
  \includegraphics[width=0.32\textwidth]{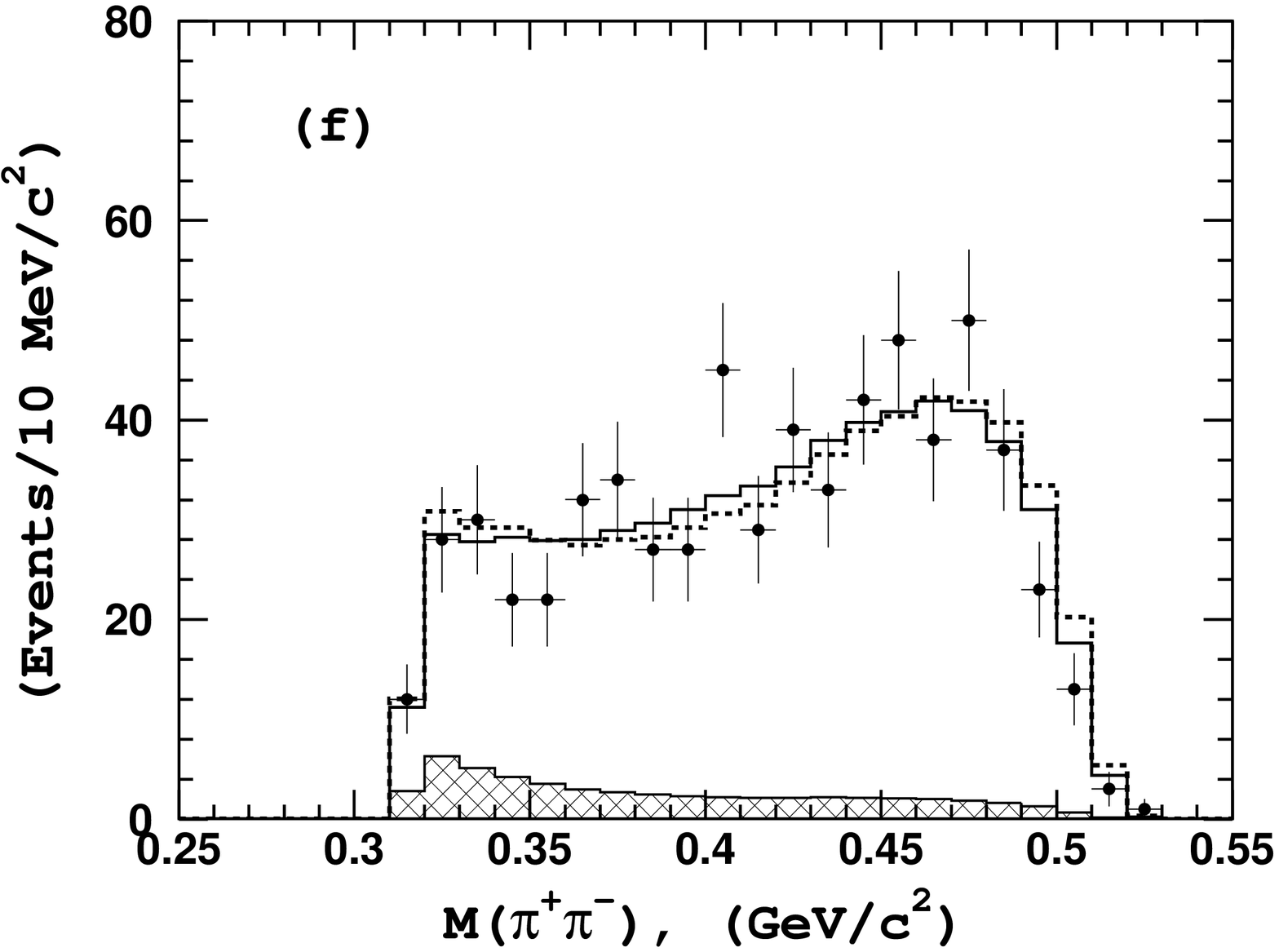}
  \caption{Comparison of fit results with the nominal model with $J^P=1^+$ 
           assigned to both $Z_b$ states (solid open histogram) 
           and the data (points with error bars) for events in the 
           (a, d) $\Uu\pp$, (b, e) $\Ud\pp$, (c, f) $\Ut\pp$ signal region. 
           Dashed histogram shows results of the fit with $J^P=2^+$ 
           assignment for the $Z_b$ states.
           Hatched histograms show the estimated background components.}
\label{fig:ynspp-f-hh}
\end{figure*}

Figures~\ref{fig:ynspp-dp}(d, e, f) show Dalitz plots for events in the 
signal regions for the three final states being considered here. In 
the fit to the $\ee\to\Un\pp$ data, we consider possible contributions 
from the following set of quasi-two-body modes: $Z_b(10610)^\pm\pi^\mp$, 
$Z_b(10650)^\pm\pi^\mp$, $\Un \sigma(500)$, $\Un f_0(980)$, $\Un f_2(1270)$,
and a non-resonant component. A detailed description of the transition 
amplitude is given in the Appendix.

For modes with higher $\Un$ states, the available phase space is 
very limited, making it impossible to distinguish unambiguously between 
multiple scalar components in the amplitude. Thus, in the nominal model 
used to fit the $\ee\to\Ud\pp$ data, we fix the $f_0(980)$ amplitude
at zero. In addition, in the nominal model used to fit the 
$\ee\to\Ut\pp$ data, we fix the $\sigma(500)$ and $f_2(1270)$
components at zero. As a result, the total numbers of fit parameters
are 16, 14, and 10 for the final states with $\Uu$, $\Ud$, and $\Ut$,
respectively. The effect of this reduction of the amplitude is 
considered in the evaluation of the systematic uncertainties.

In the fit to the data, we test the following assumptions on the spin 
and parity of the observed $Z_b$ states: $J^P=1^+$, $1^-$, $2^+$ and 
$2^-$. Note that $J^P=0^+$ and $0^-$ combinations are forbidden because 
of the observed $Z_b\to\Un\pi$ and $Z_b\to\hm\pi$ decay modes, 
respectively. (Since the masses and the widths of two resonances 
measured in the $\hm\pi$ and in the $\Un\pi$~\cite{Belle_Zbc} systems 
are consistent, we assume the same pair of $Z_b$ states is observed in 
these decay modes.) The simplified angular analysis reported in 
Ref.~\cite{Belle_ang} favors the $J^P=1^+$ hypothesis; thus, our nominal
model here adopts $J^P=1^+$.

%
%======================================================================
%
\begin{figure*}[!t]
  \centering
  \includegraphics[width=0.32\textwidth]{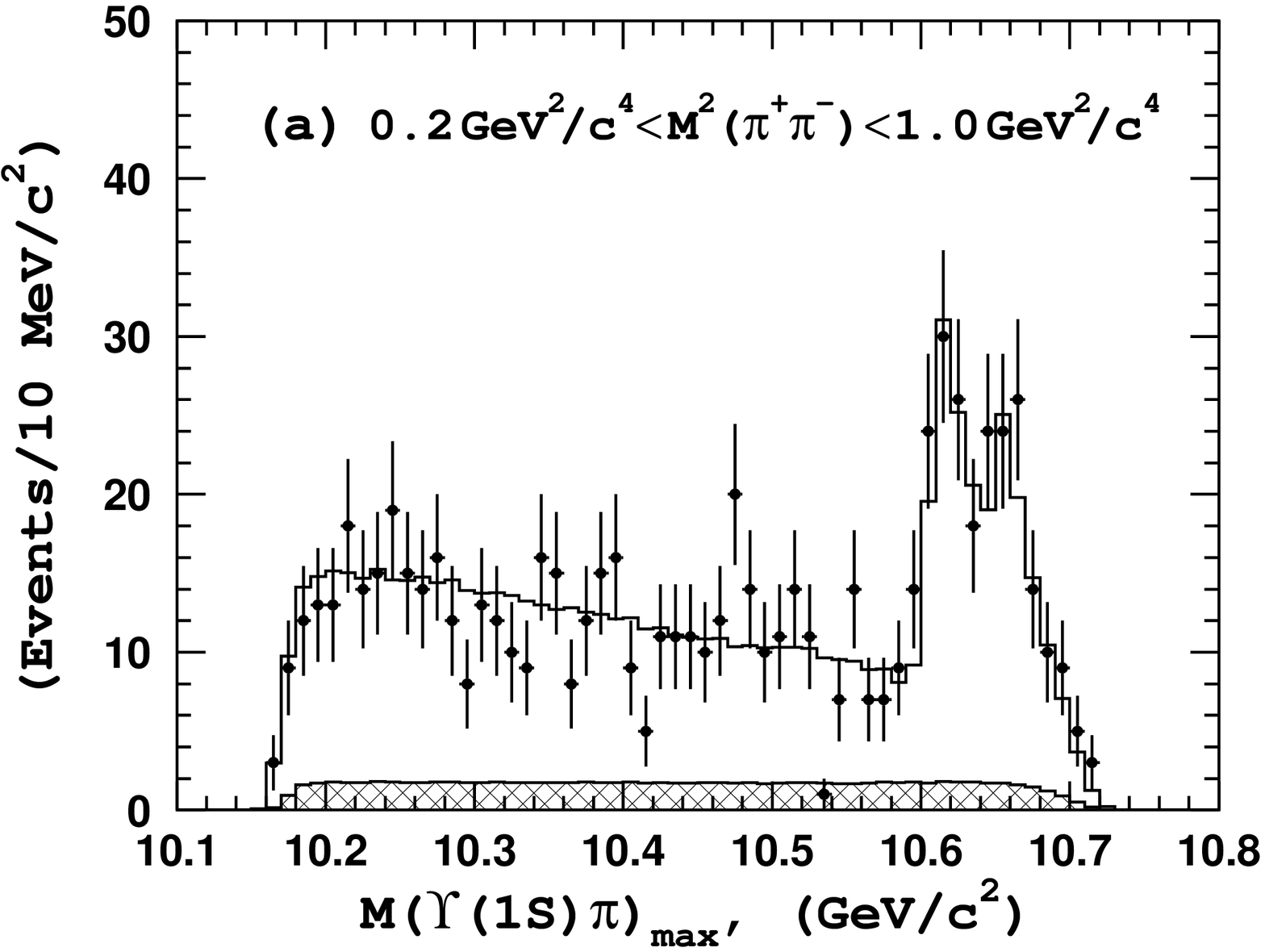} \hfill
  \includegraphics[width=0.32\textwidth]{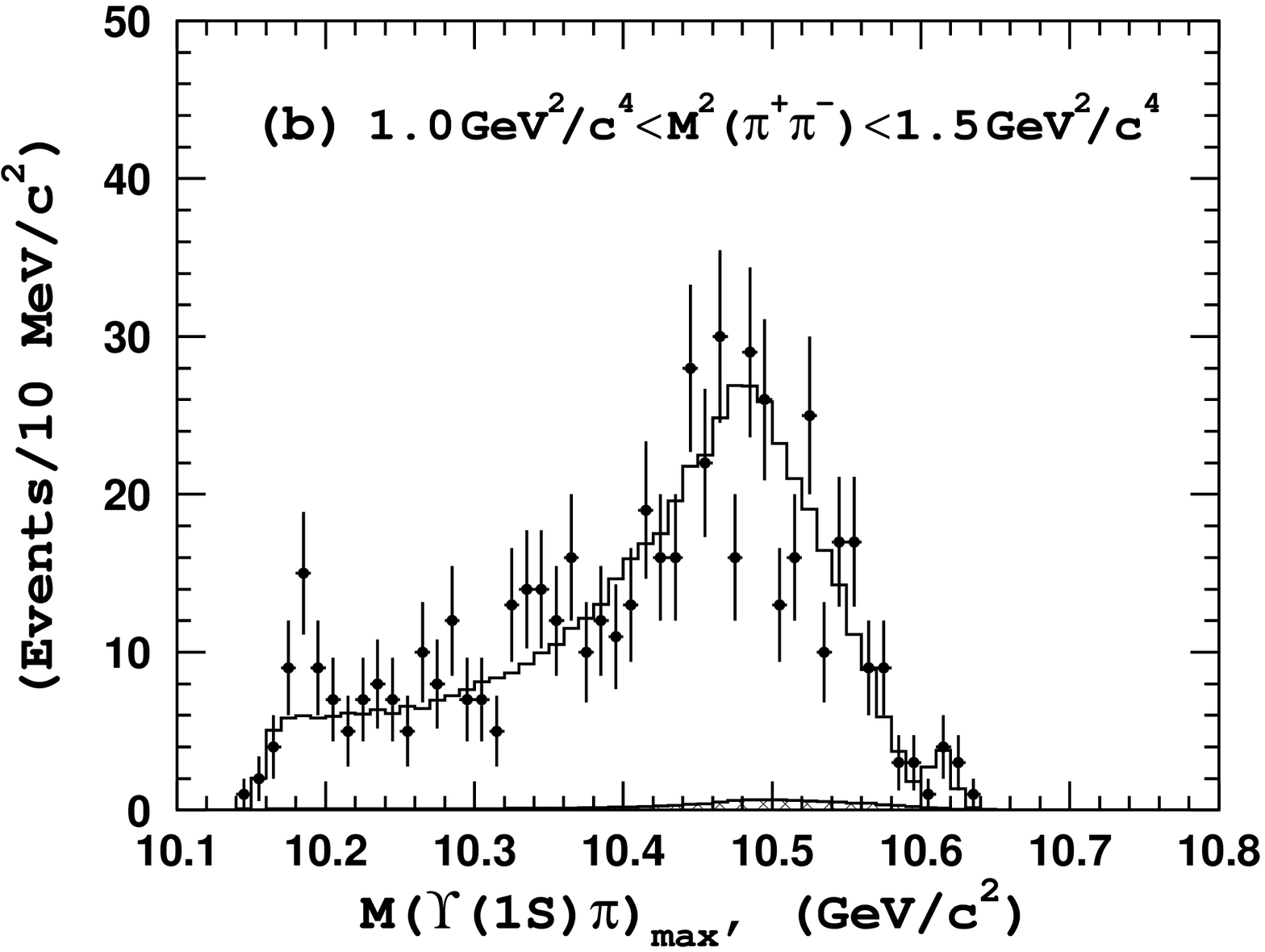} \hfill
  \includegraphics[width=0.32\textwidth]{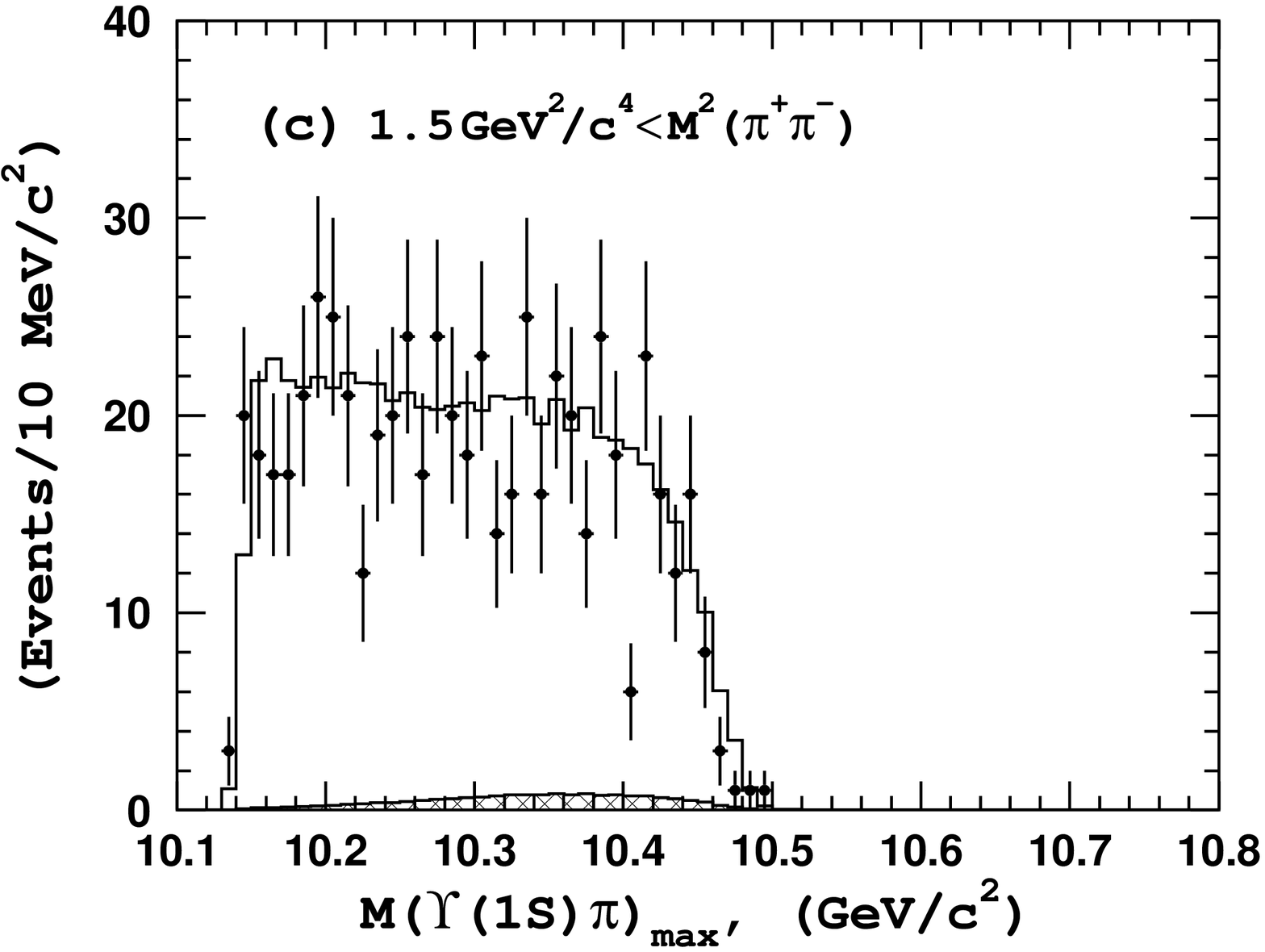} \\
  \includegraphics[width=0.32\textwidth]{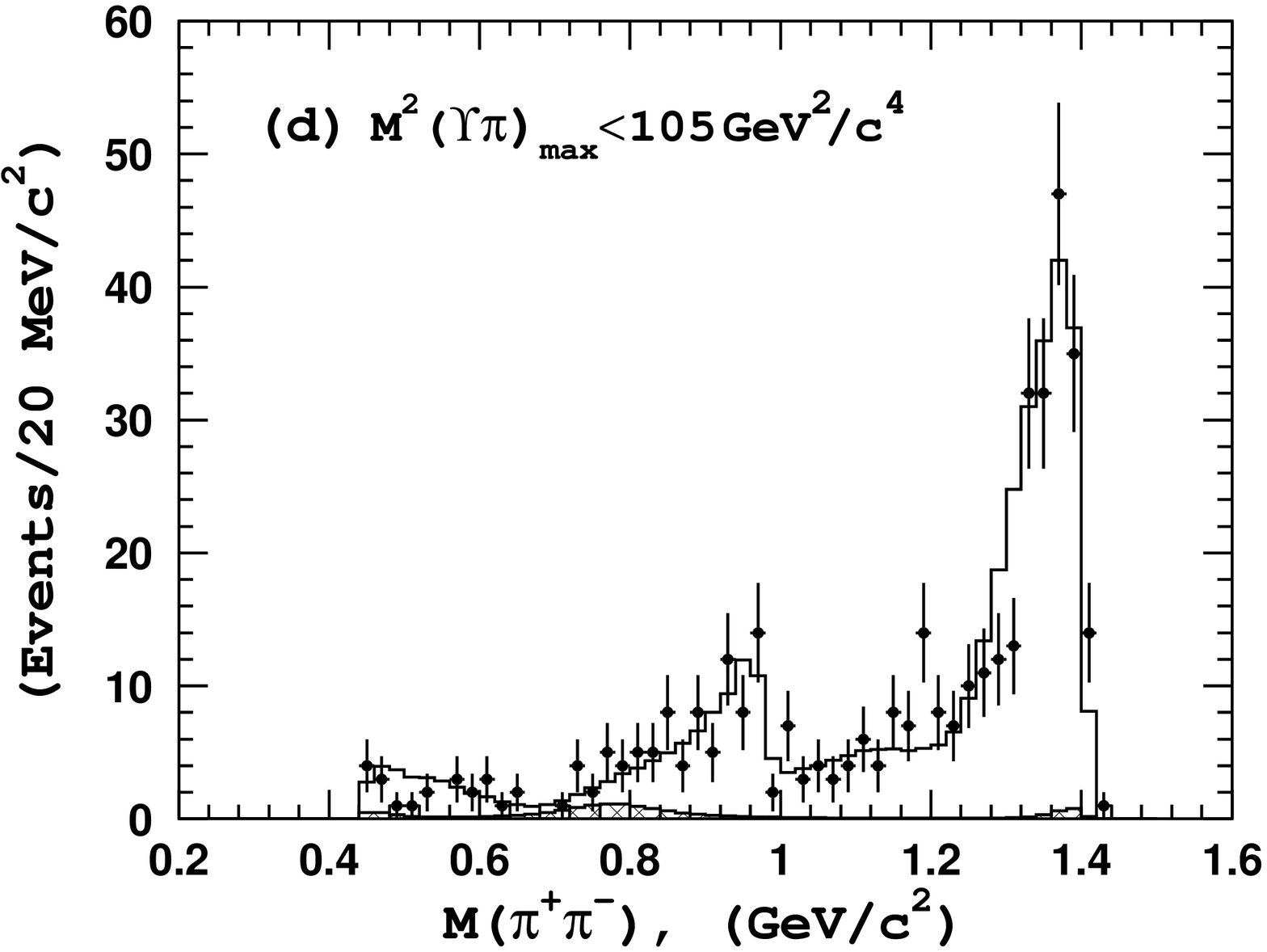} \hfill
  \includegraphics[width=0.32\textwidth]{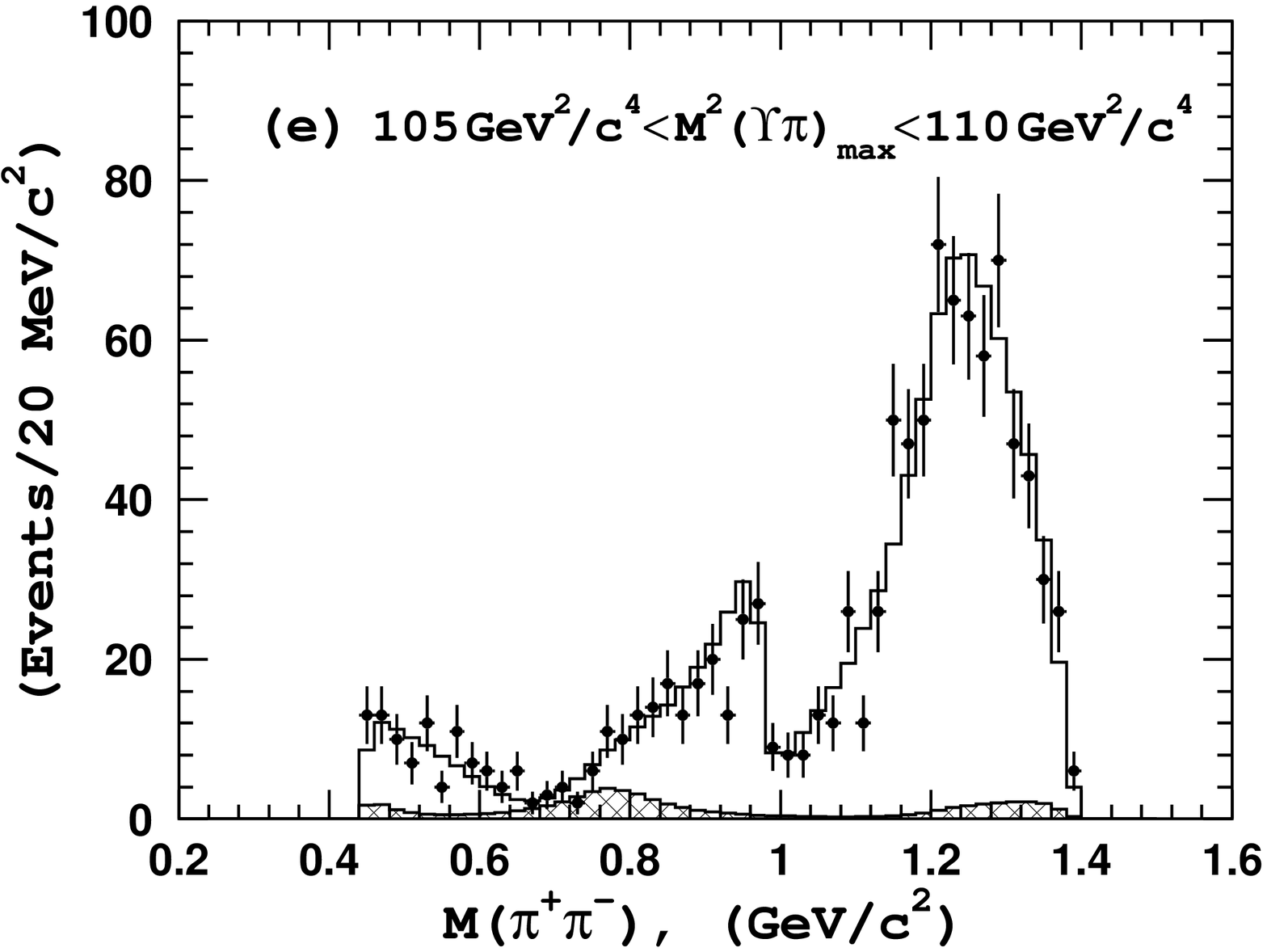} \hfill
  \includegraphics[width=0.32\textwidth]{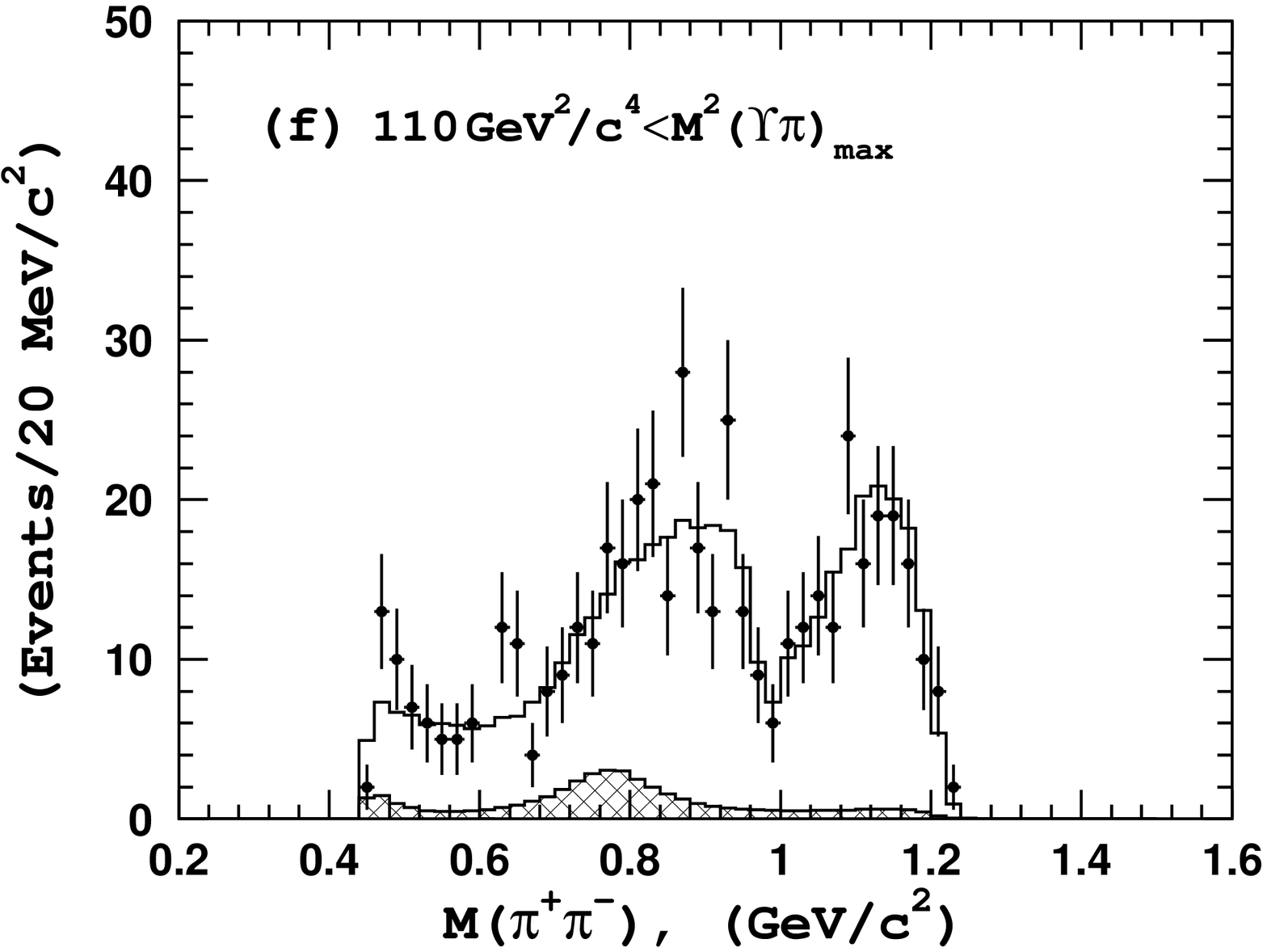}
  \caption{A detailed comparison of fit results with the nominal model 
           (open histogram) with the data (points with error bars) for 
           events in the $\Uu\pp$ signal region. 
           Hatched histograms show the estimated background components.}
\label{fig:y1spp-reg}
\end{figure*}
%
%======================================================================
%
\begin{figure*}[!t]
  \centering
  \includegraphics[width=0.24\textwidth]{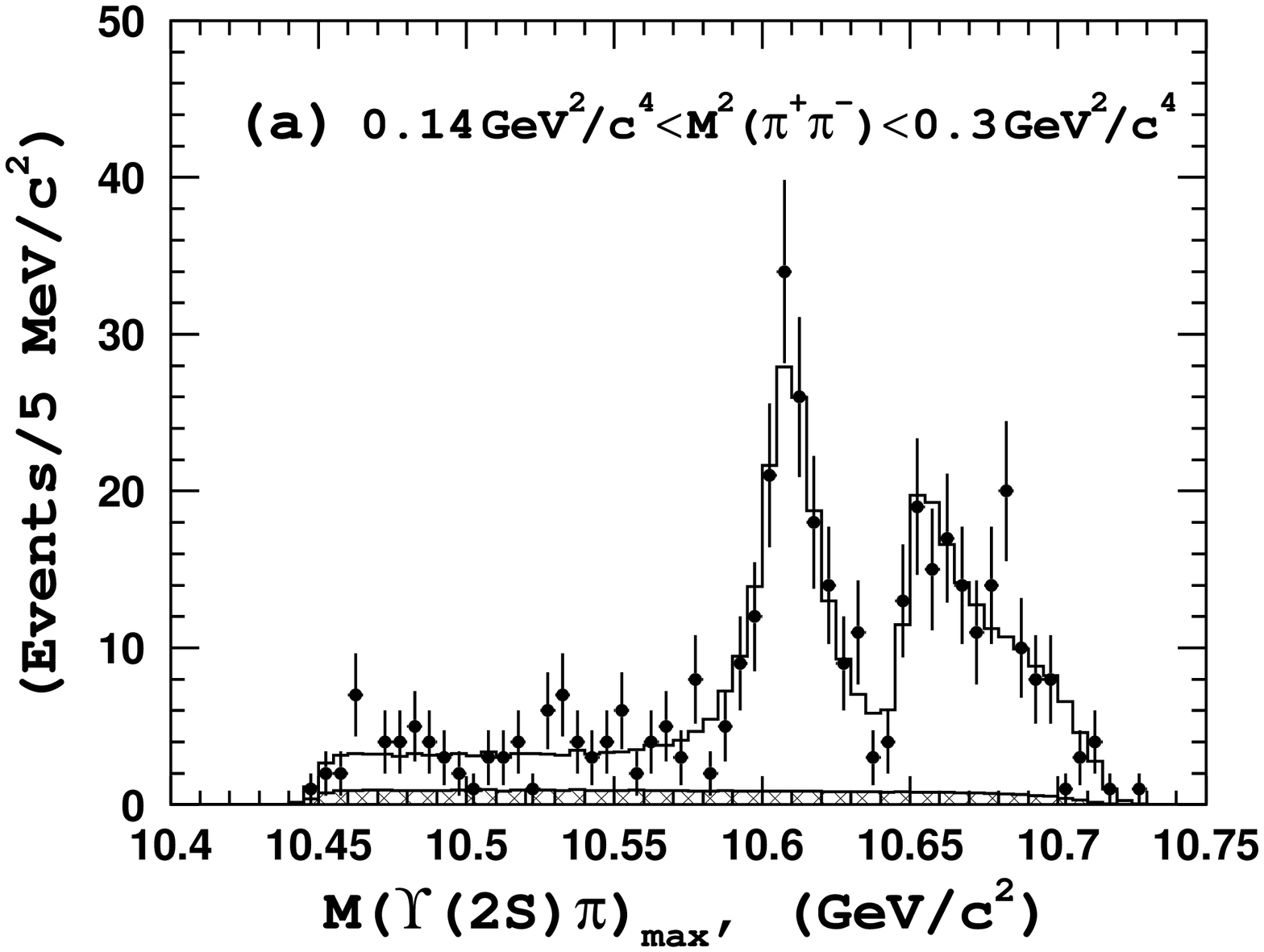} \hfill
  \includegraphics[width=0.24\textwidth]{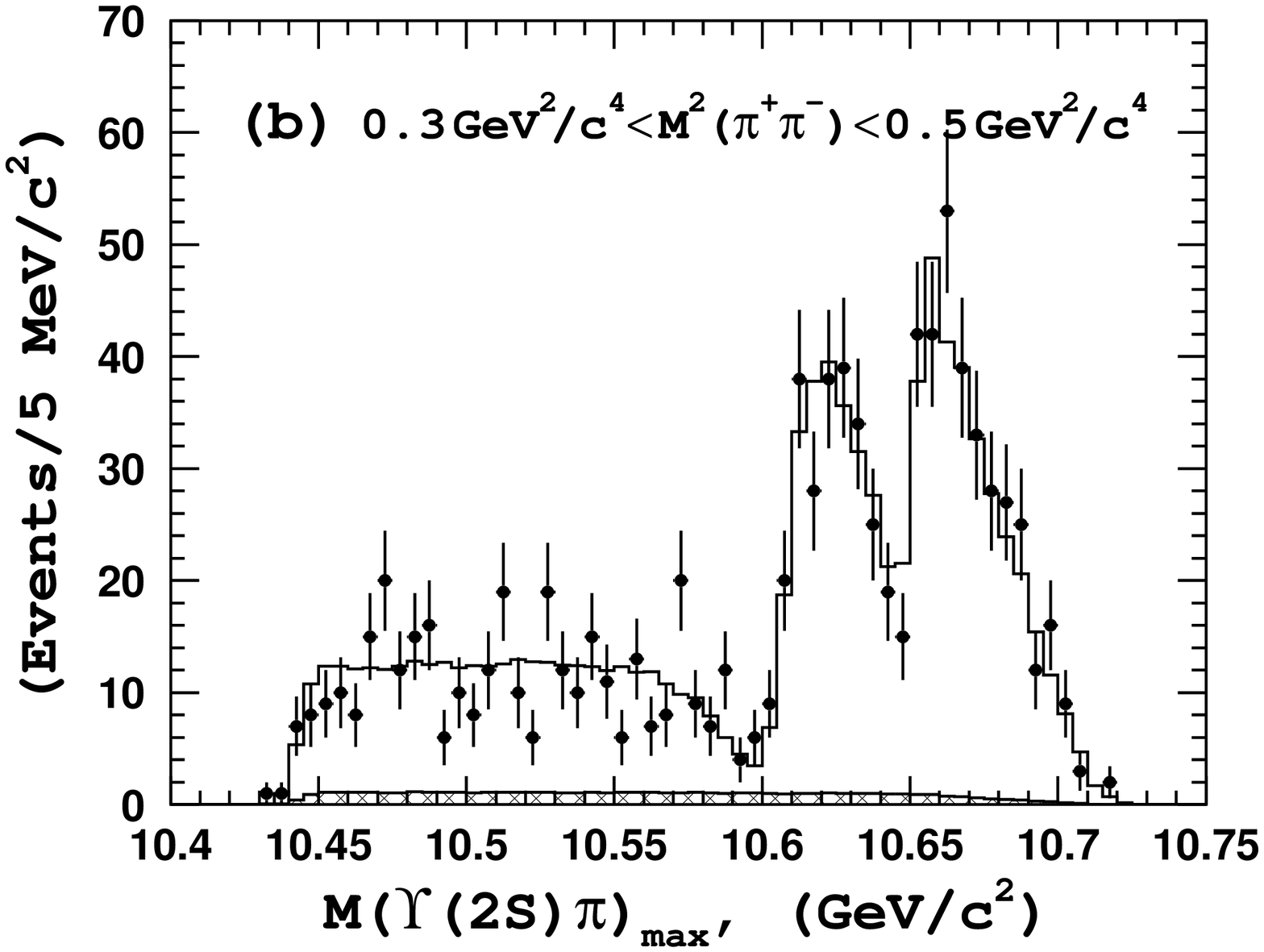} \hfill
  \includegraphics[width=0.24\textwidth]{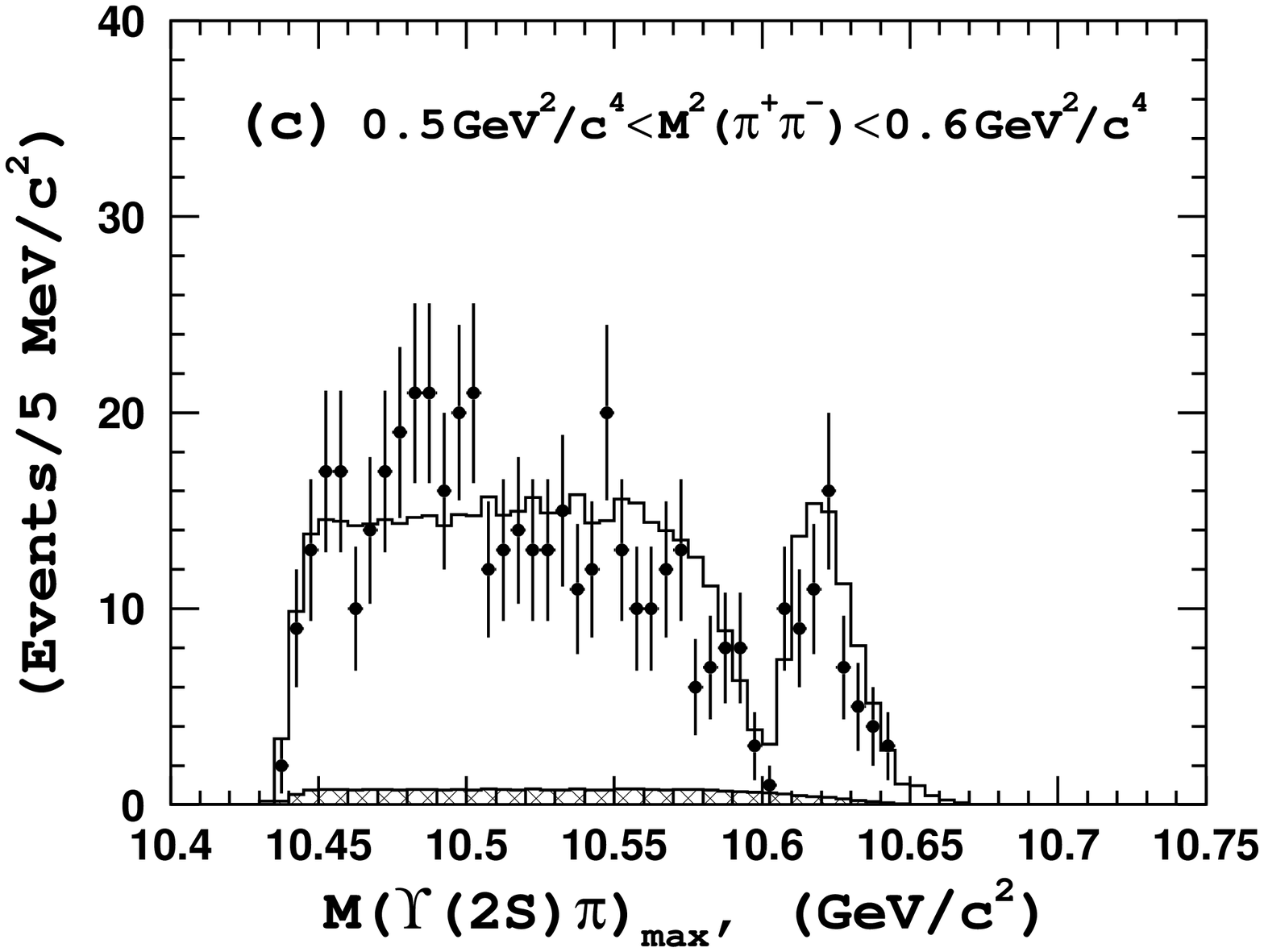} \hfill
  \includegraphics[width=0.24\textwidth]{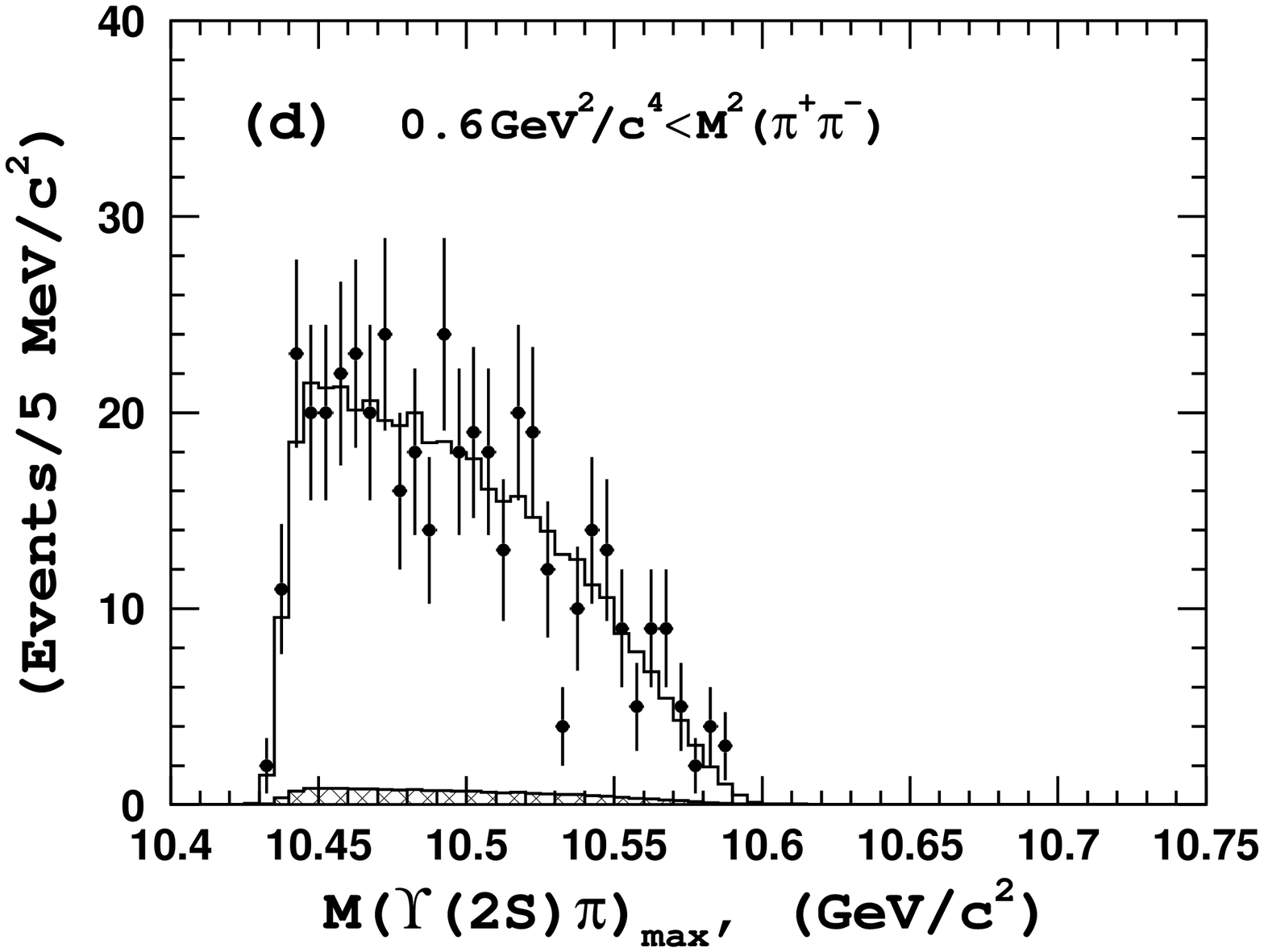} \\
  \includegraphics[width=0.24\textwidth]{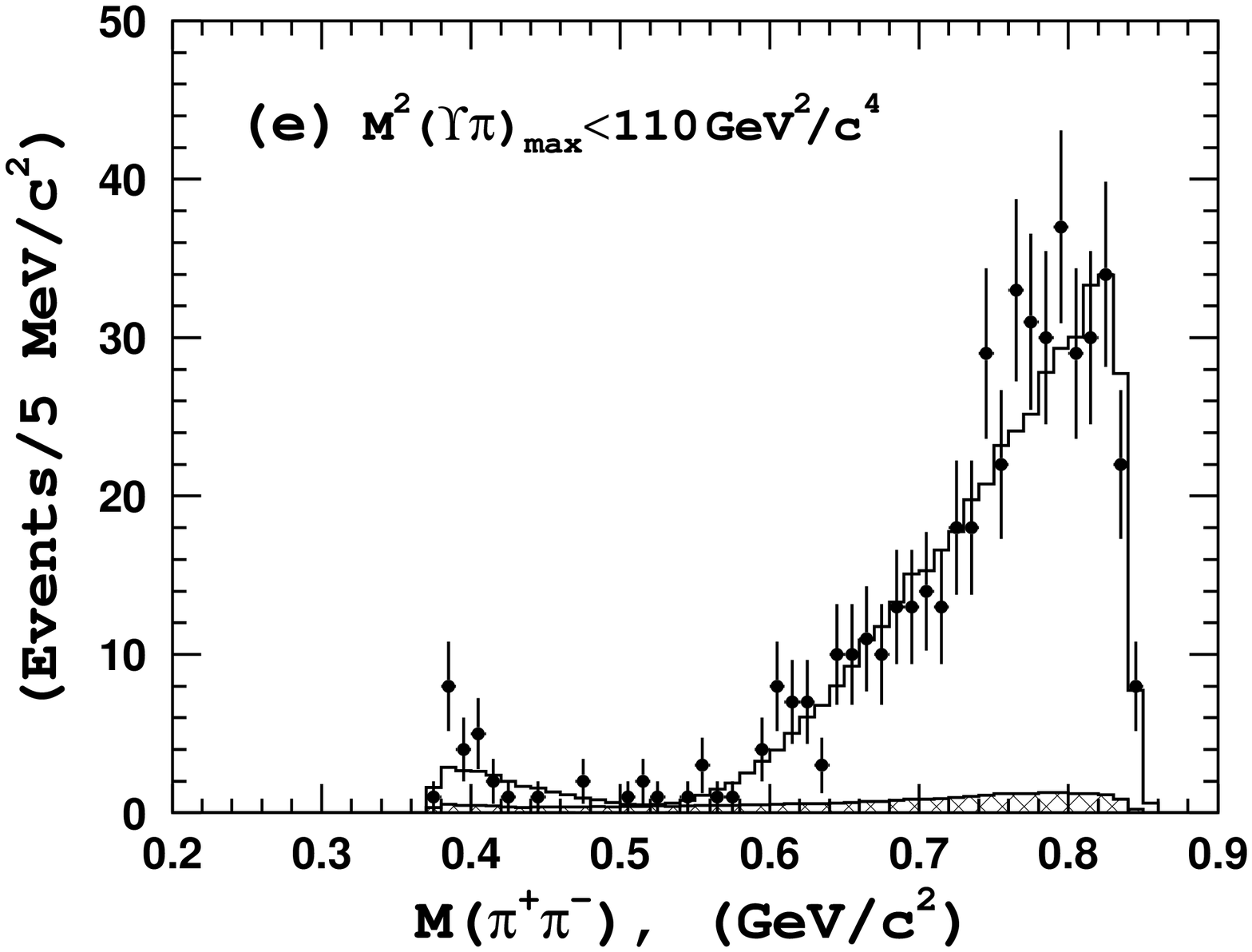} \hfill
  \includegraphics[width=0.24\textwidth]{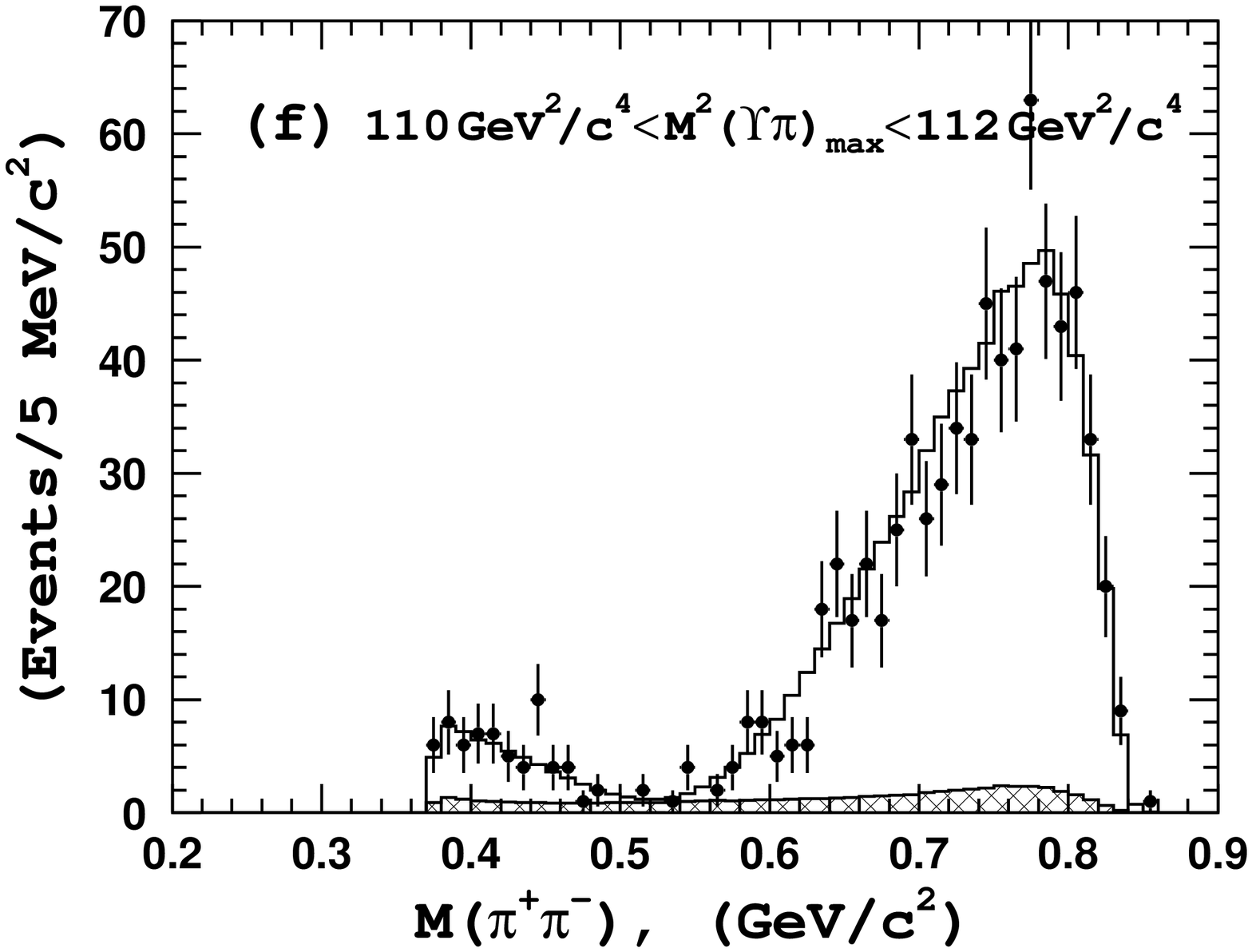} \hfill
  \includegraphics[width=0.24\textwidth]{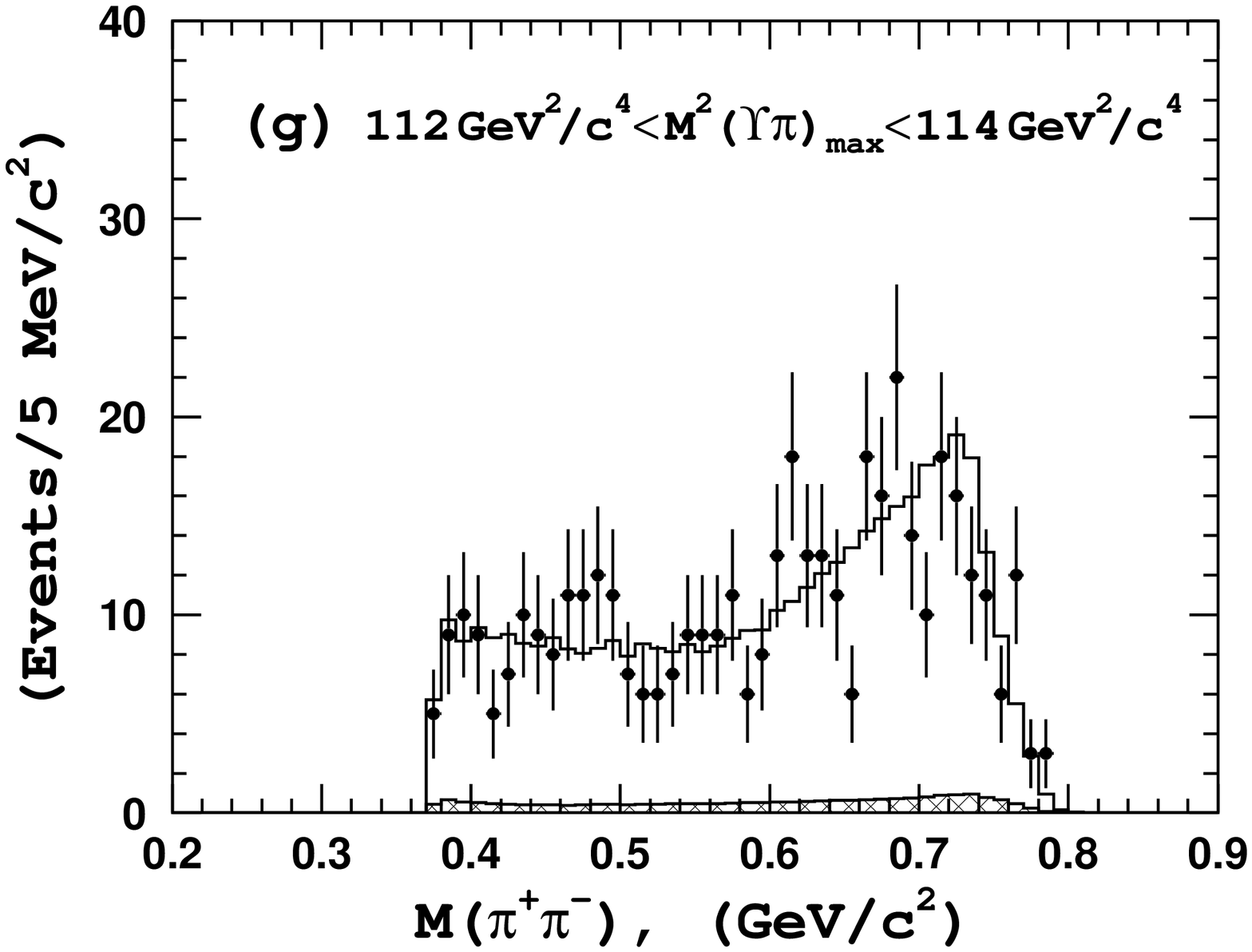} \hfill
  \includegraphics[width=0.24\textwidth]{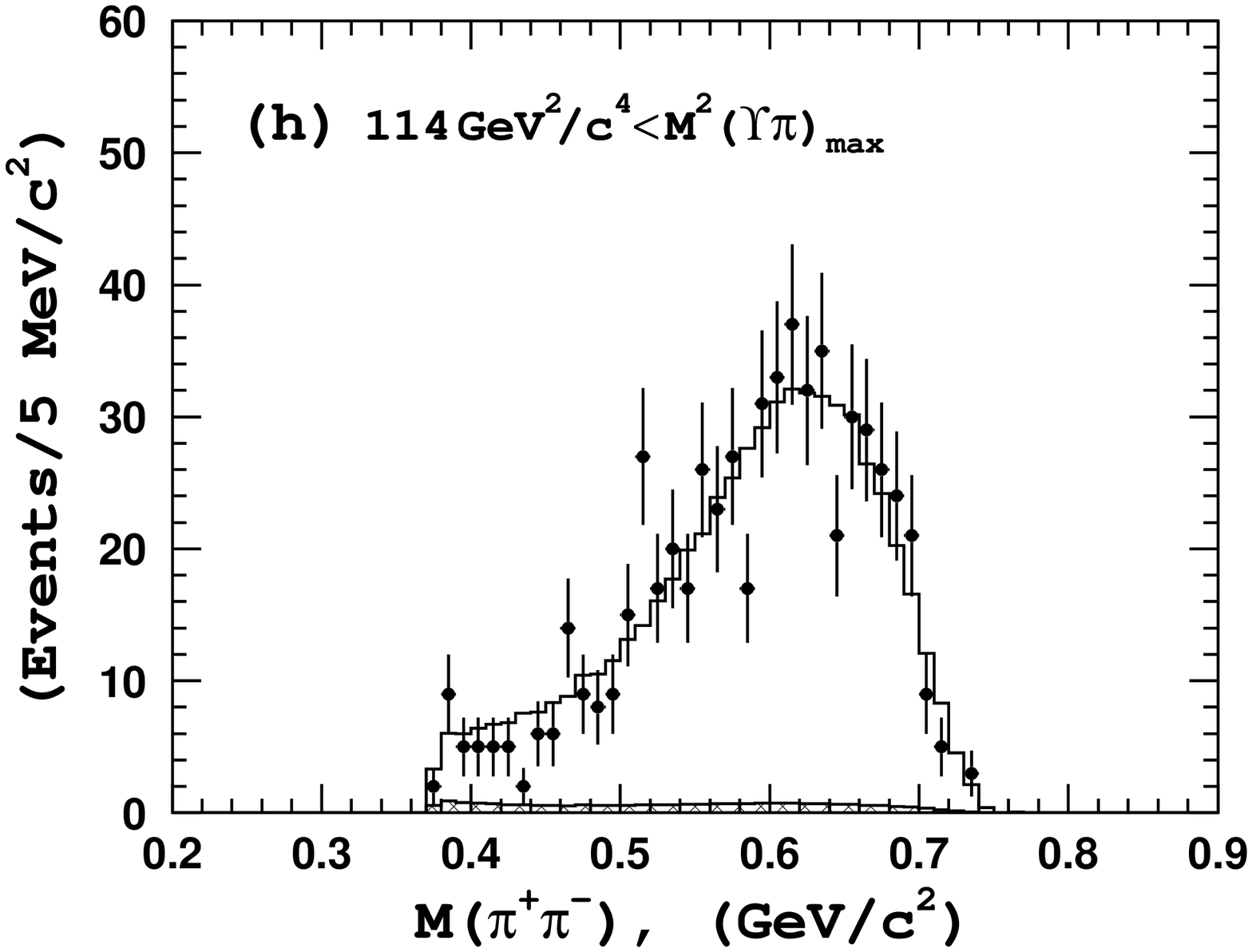}
  \caption{A detailed comparison of fit results with the nominal model 
           (open histogram) with the data (points with error bars) for 
           events in the $\Ud\pp$ signal region. 
           Hatched histograms show the estimated background components.}
\label{fig:y2spp-reg}
\end{figure*}

The logarithmic likelihood function ${\cal{L}}$ is 
\begin{equation}
{\cal{L}} = -2\sum_{\rm events}{\ln(f_{\rm sig}S + (1-f_{\rm sig})B)},
\label{eq:logl}
\end{equation}
where the summation is performed over all selected candidate events.
The $S$ term in Eq.~\ref{eq:logl} is formed from $|M_{\Upsilon\pi\pi}|^2$ 
(see the Appendix) convolved with the detector resolution, $f_{\rm sig}$ 
is the fraction of signal events in the data sample 
(see Table~\ref{tab:ynspp_sfrac}), and $B$ is a background density 
function determined from the fit to the sideband events. Both $S$ and 
$B$ are normalized to unity. 

For normalization, we use a large sample of signal $\ee\to\Un\pp\to\uu\pp$ 
MC events generated with a uniform distribution over the phase space and 
processed through the full detector simulation. The simulation also 
accounts for the beam energy spread of $\sigma=5.3$~MeV and c.m.\ energy 
variations throughout the data taking period. The use of the full MC 
events for the normalization allows us to account for variations of the 
reconstruction efficiency over the phase space. More details can be found
in Ref.~\cite{Belle_hhh}. Results of fits to $\Un\pp$ events in the signal
regions with the nominal model are shown in Fig.~\ref{fig:ynspp-f-hh}, 
where one-dimensional projections of the data and fits are presented. In 
order to combine $Z^+_b$ and $Z^-_b$ signals, we plot the 
$M(\Un\pi)_{\rm max}$ distribution rather than individual $M(\Un\pi^+)$ 
and $M(\Un\pi^-)$ spectra. To quantify the goodness of fits, we calculate
$\chi^2$ values for one-dimensional projections shown in 
Fig.~\ref{fig:ynspp-f-hh}, combining any bin with fewer than nine events
with its neighbor. A $\chi^2$ variable for the multinomial distribution 
is then calculated as
\begin{equation}
   \chi^2 = -2\sum^{N_{\rm bins}}_{i=1}n_i\ln\left(\frac{p_i}{n_i}\right),
\end{equation}
where $n_i$ is the number of events observed in the $i$-th bin and $p_i$
is the number of events expected from the model. For a large number of 
events, this formulation becomes equivalent to the standard $\chi^2$ 
definition. Since we are minimizing the unbinned likelihood function, 
such a constructed $\chi^2$ variable does not asymptotically follow a 
typical $\chi^2$ distribution but is rather bounded by two $\chi^2$ 
distributions with ($N_{\rm bins}-1$) and ($N_{\rm bins}-k-1$) degrees of
freedom~\cite{kendal}, where $k$ is the number of fit parameters. 
Because it is bounded by two $\chi^2$ distributions, it remains a useful 
statistic to estimate the goodness of the fits. Results are presented in
Table~\ref{tab:ynspp-chi2}. For all final states, the nominal model 
provides a good description of the data.
\begin{table}[!b]
  \caption{Results of the $\chi^2/n_{\rm bins}$ calculations for 
      one-dimensional projections shown in Fig.~\ref{fig:ynspp-f-hh}.}
  \medskip
  \label{tab:ynspp-chi2}
\centering
  \begin{tabular}{lccc} \hline \hline
                & ~$\Uu\pp$~         &
                  ~$\Ud\pp$~         &
                  ~$\Ut\pp$~    
\\ \hline
  $M(\Upsilon\pi)_{\rm max}$   & 61.5/53 & 46.6/54 & 12.0/20  \\ 
  $\mpp$               & 68.3/49 & 45.1/48 & 18.6/20  \\
\hline \hline
\end{tabular}
\end{table}

\begin{figure*}[!t]
  \centering
  \includegraphics[width=0.32\textwidth]{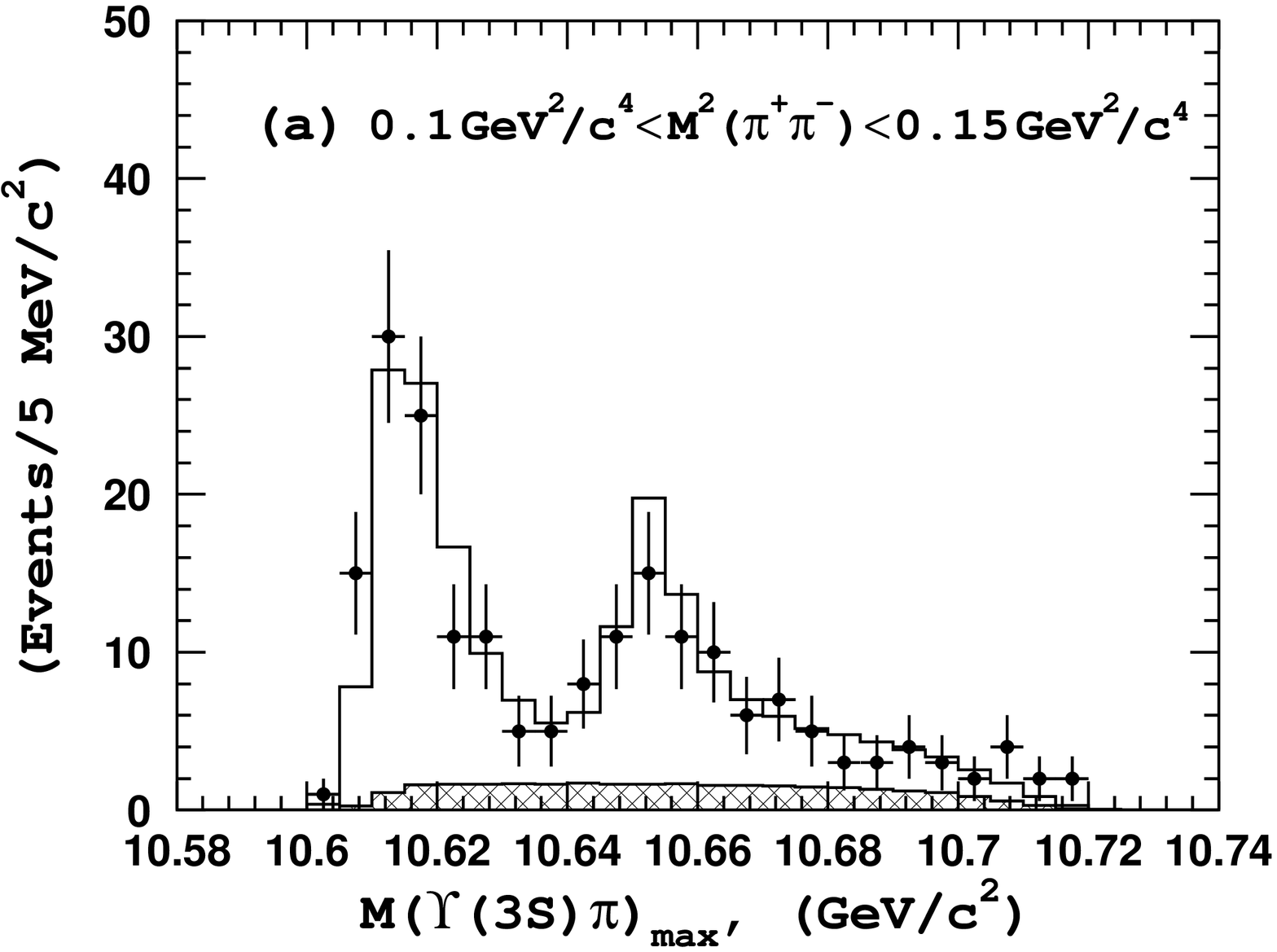} \hfill
  \includegraphics[width=0.32\textwidth]{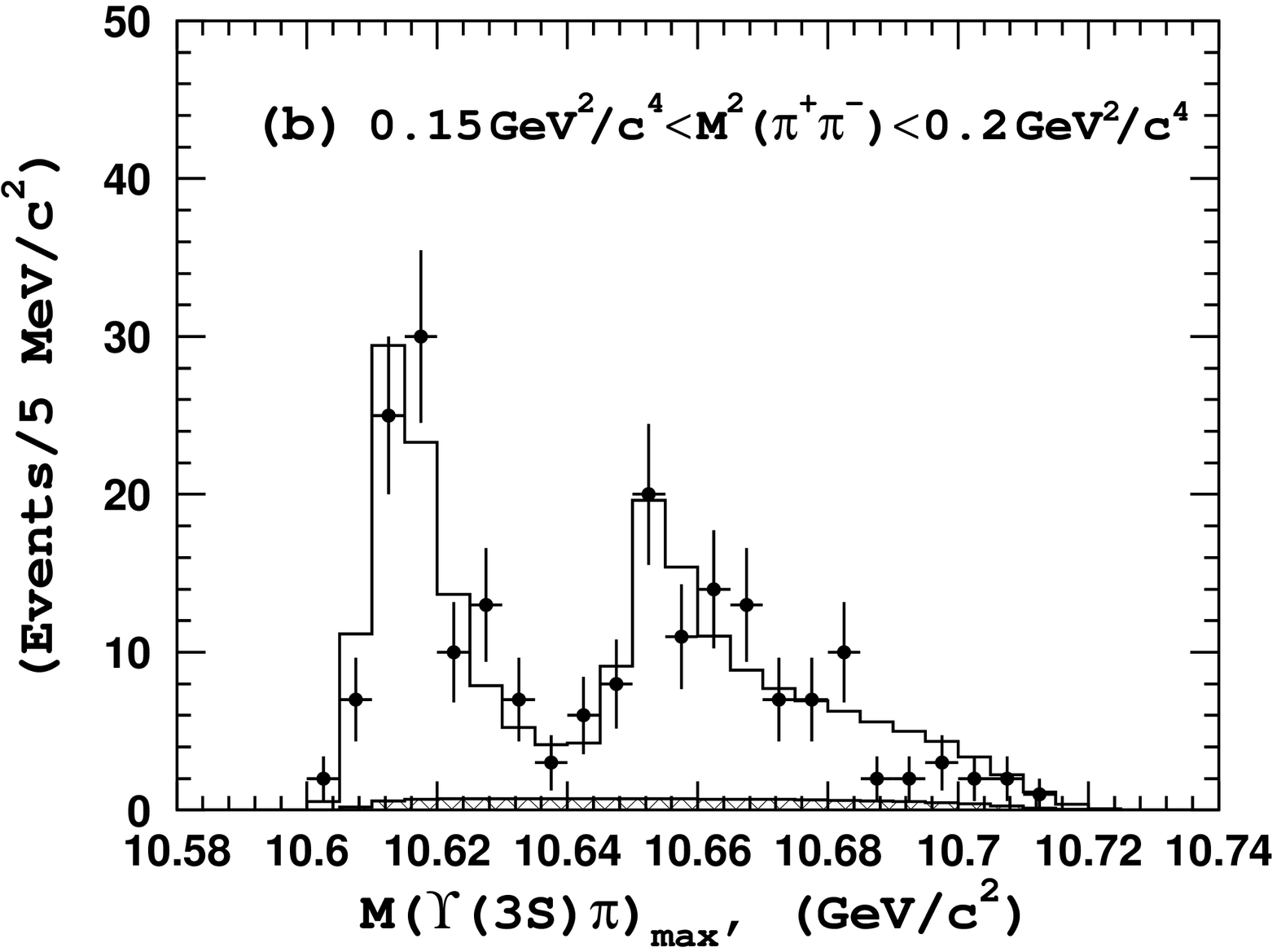} \hfill
  \includegraphics[width=0.32\textwidth]{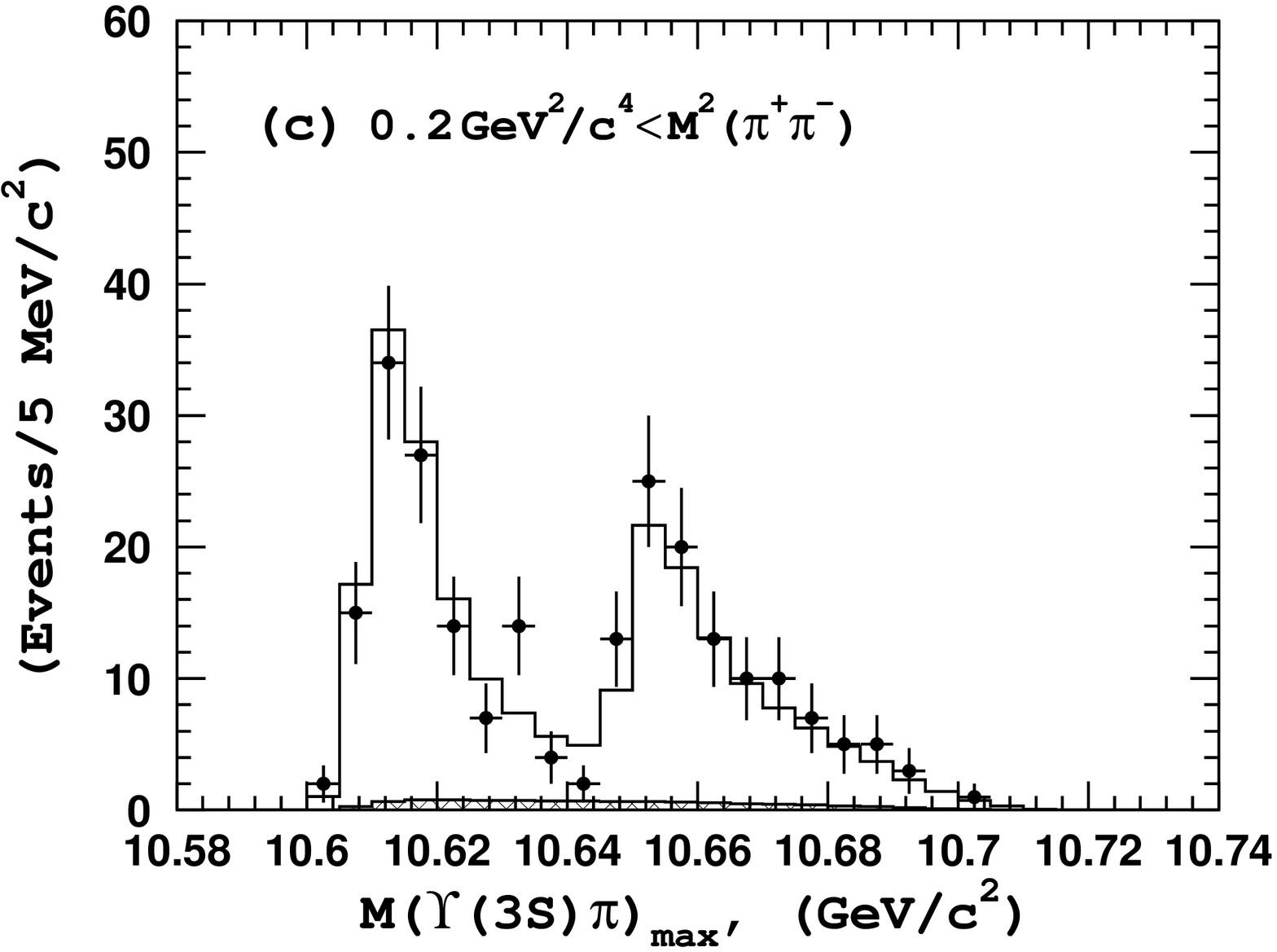} \\
  \includegraphics[width=0.32\textwidth]{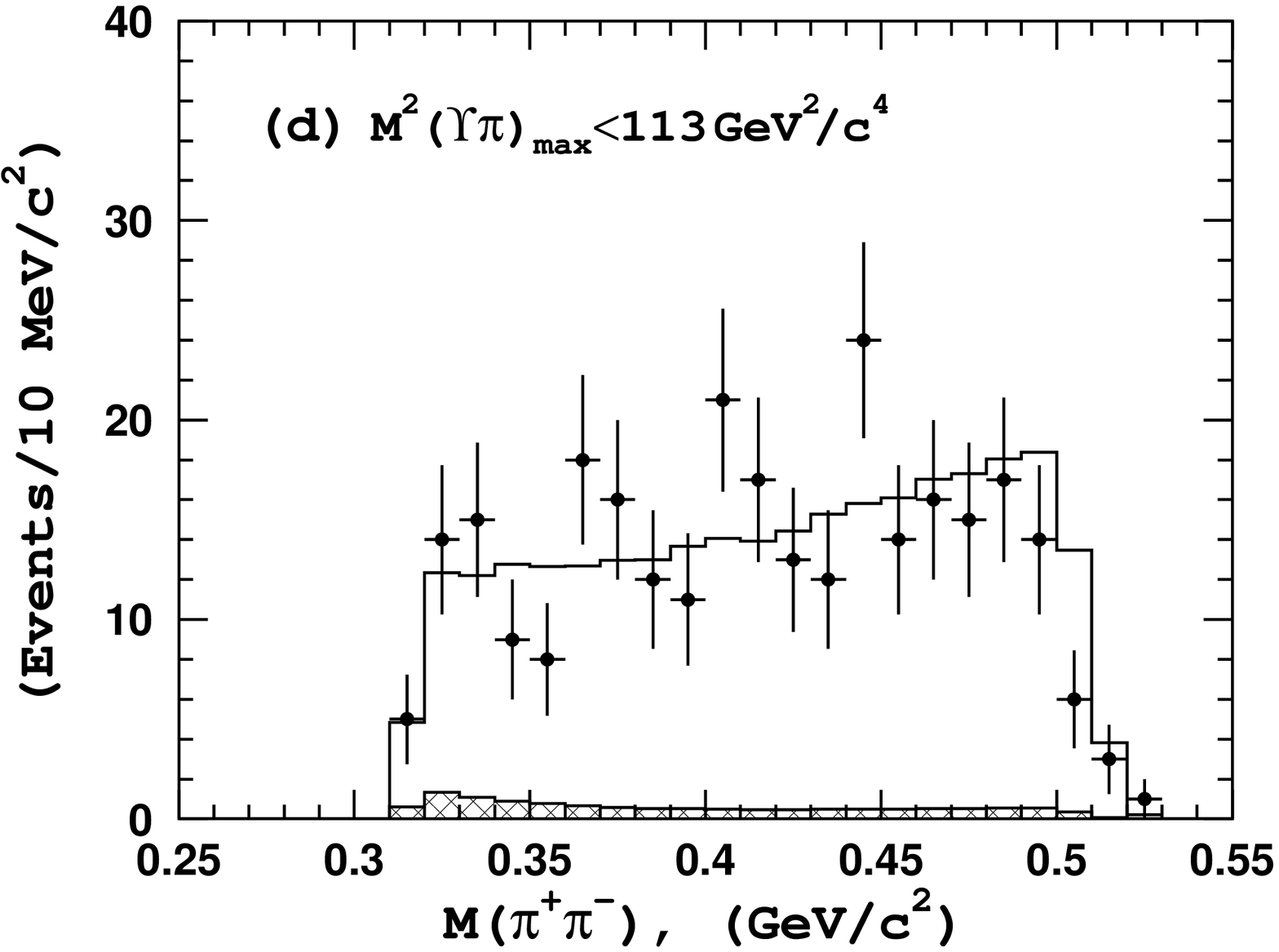} \hfill
  \includegraphics[width=0.32\textwidth]{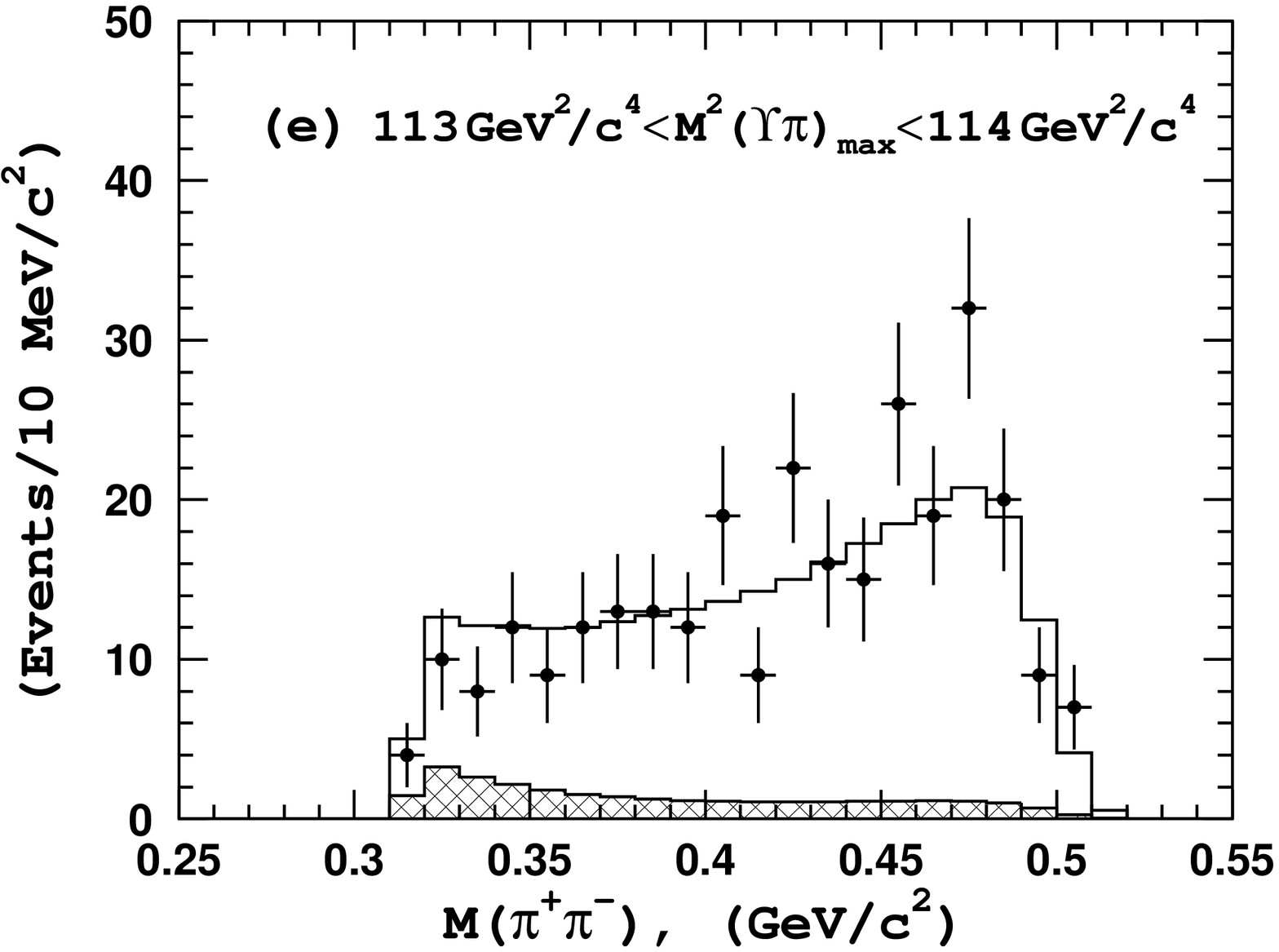} \hfill
  \includegraphics[width=0.32\textwidth]{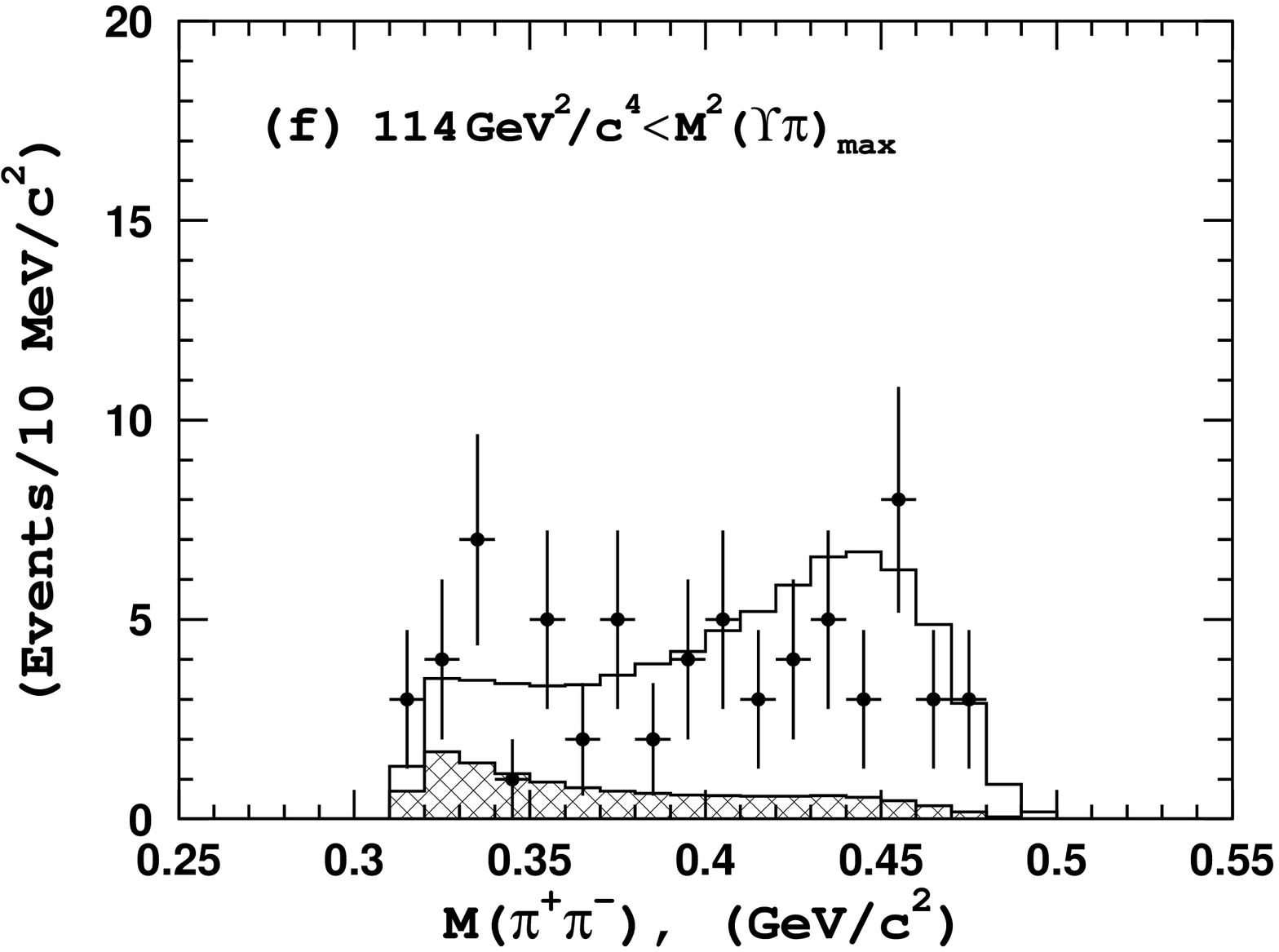} \\
  \caption{A detailed comparison of fit results with the nominal model 
           (open histogram) with the data (points with error bars) for 
           events in the $\Ut\pp$ signal region. 
           Hatched histograms show the estimated background components.}
\label{fig:y3spp-reg}
\end{figure*}

\begin{figure*}[!t]
  \centering
  \includegraphics[width=0.24\textwidth]{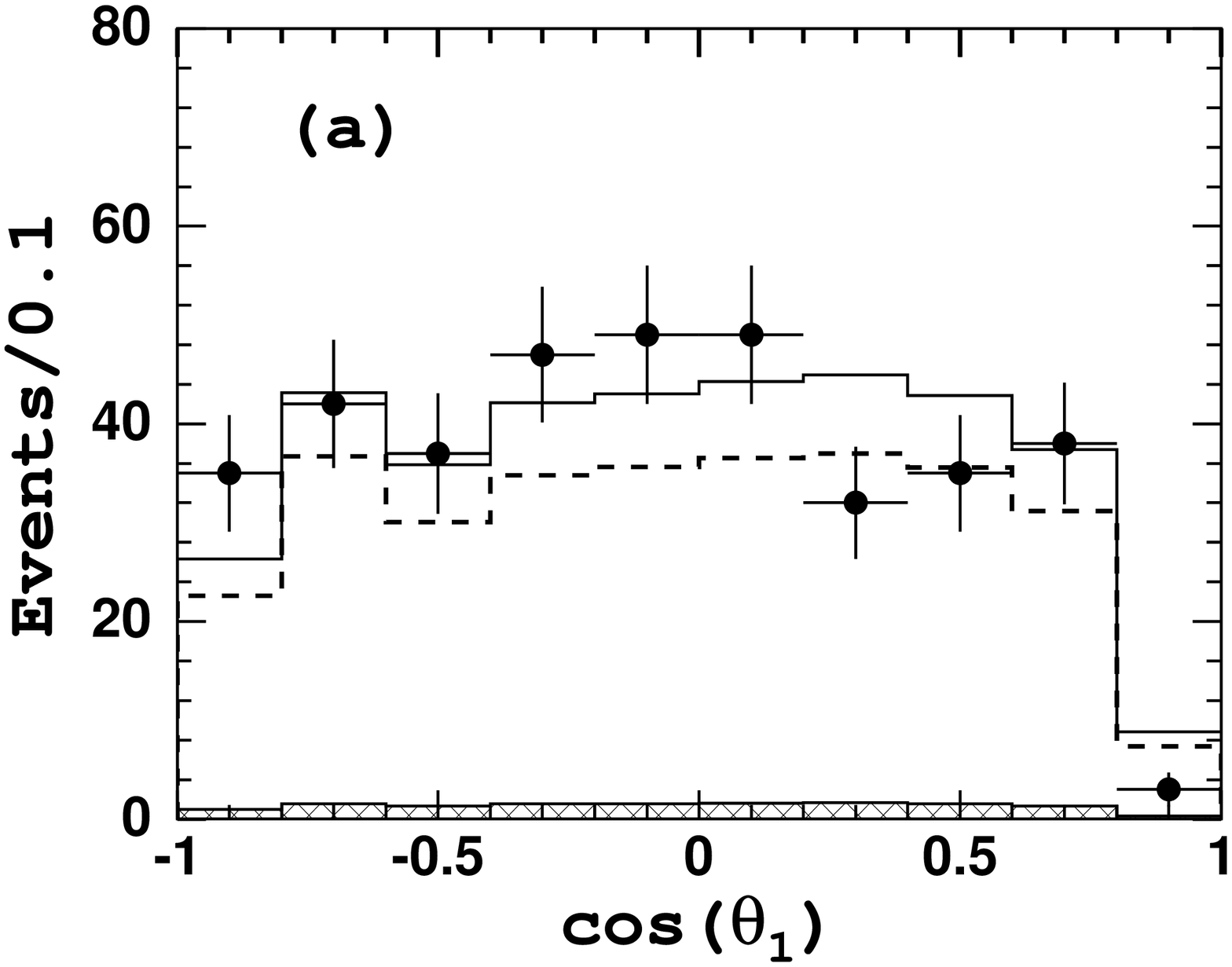} \hfill
  \includegraphics[width=0.24\textwidth]{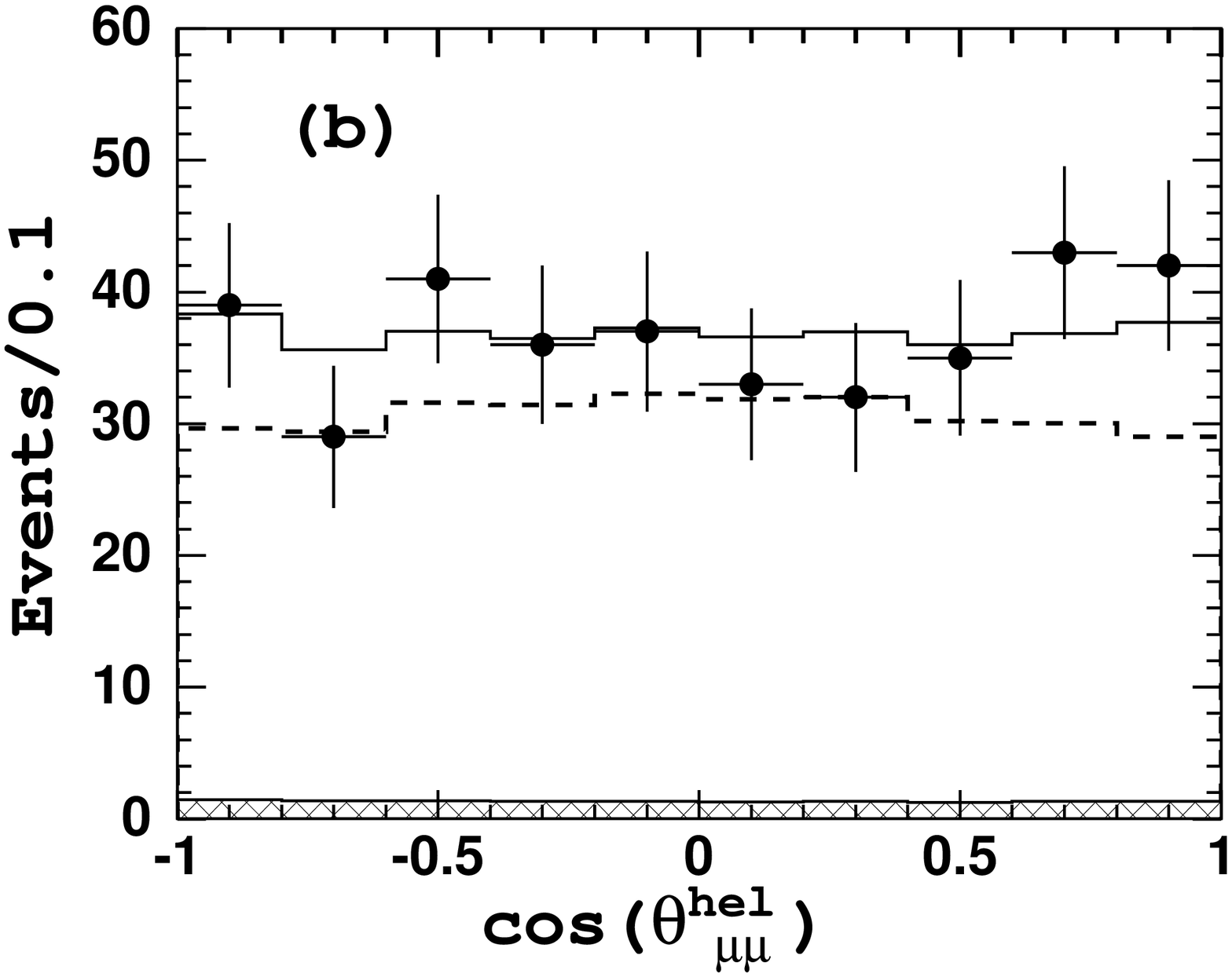} \hfill
  \includegraphics[width=0.24\textwidth]{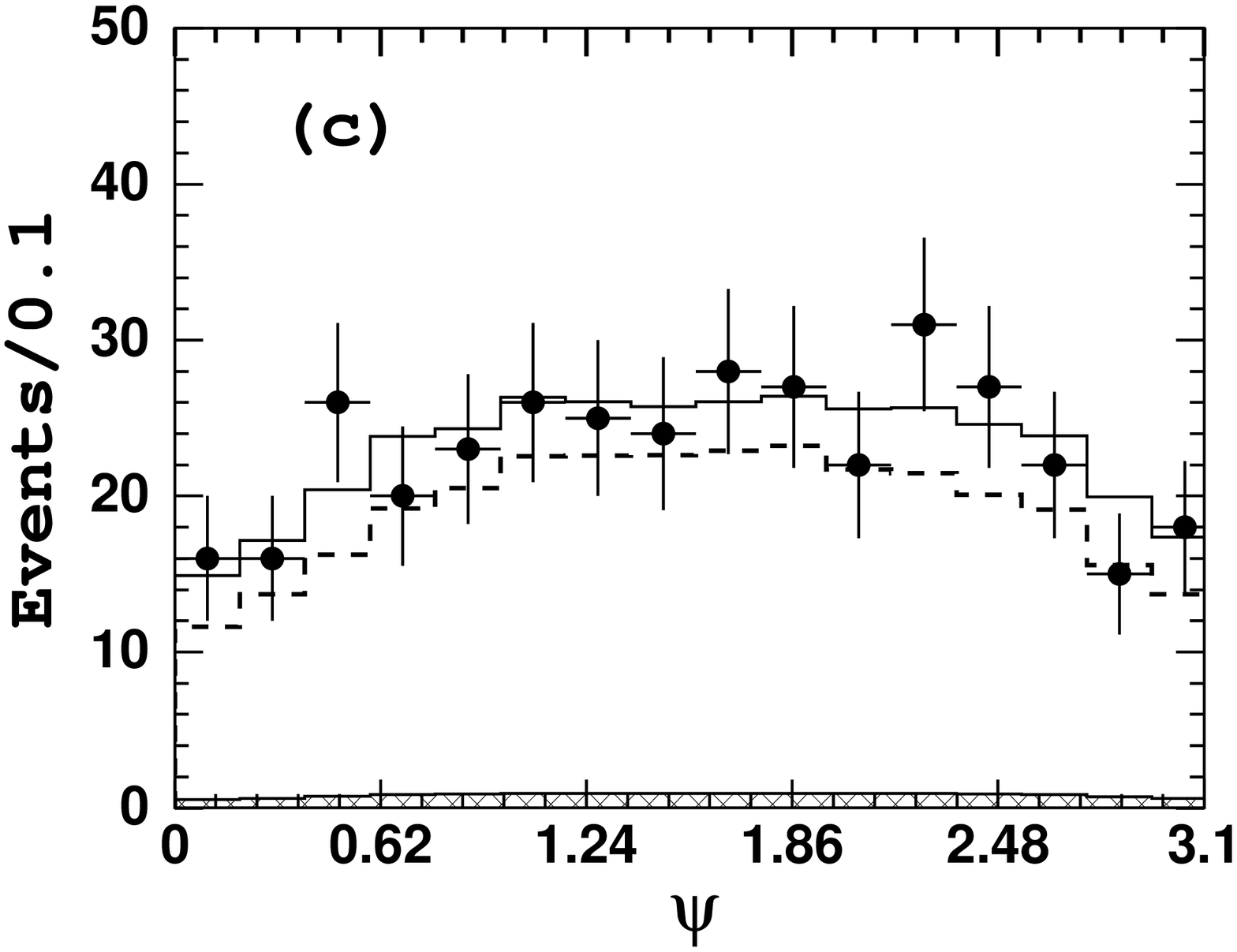} \hfill
  \includegraphics[width=0.24\textwidth]{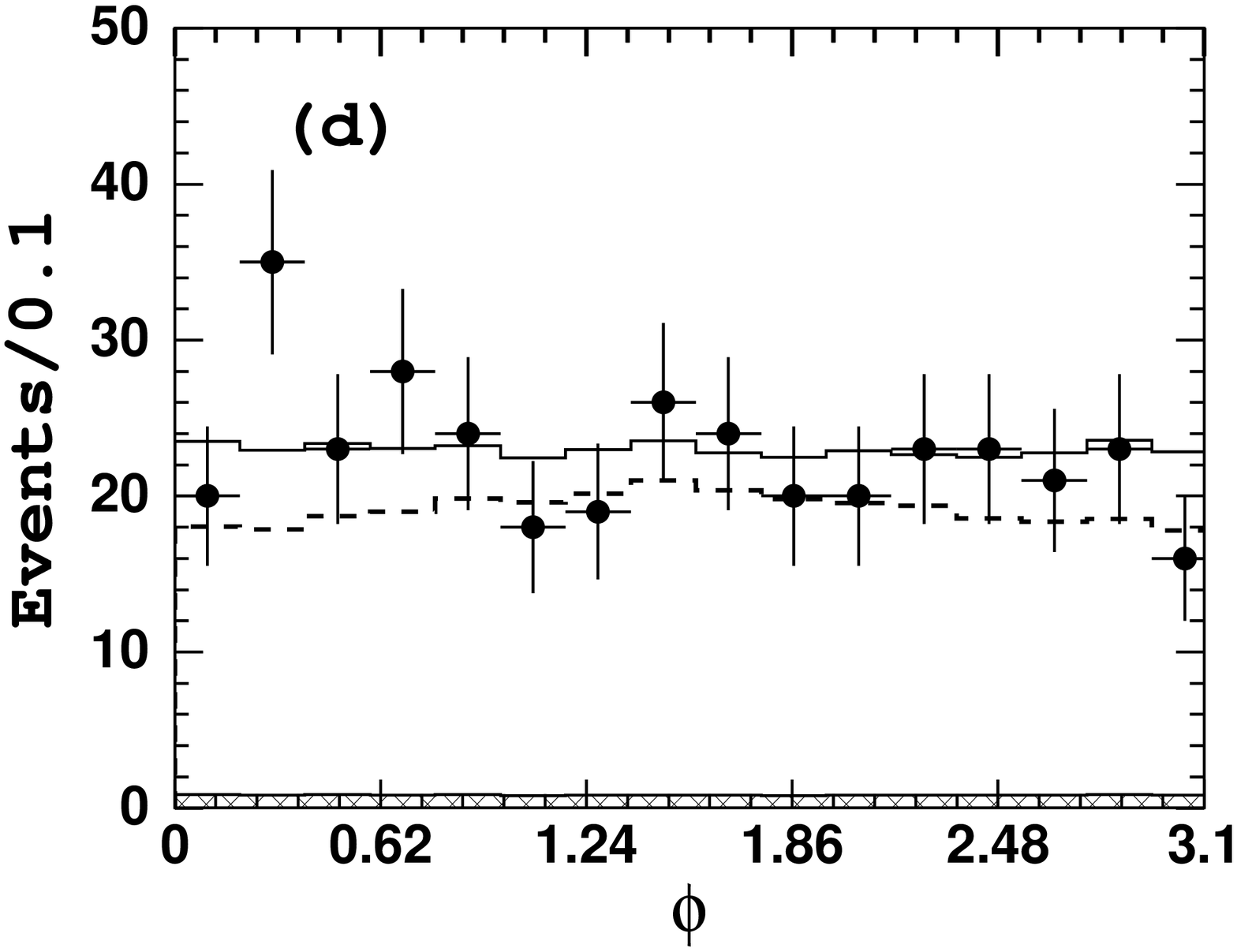} \\
  \includegraphics[width=0.24\textwidth]{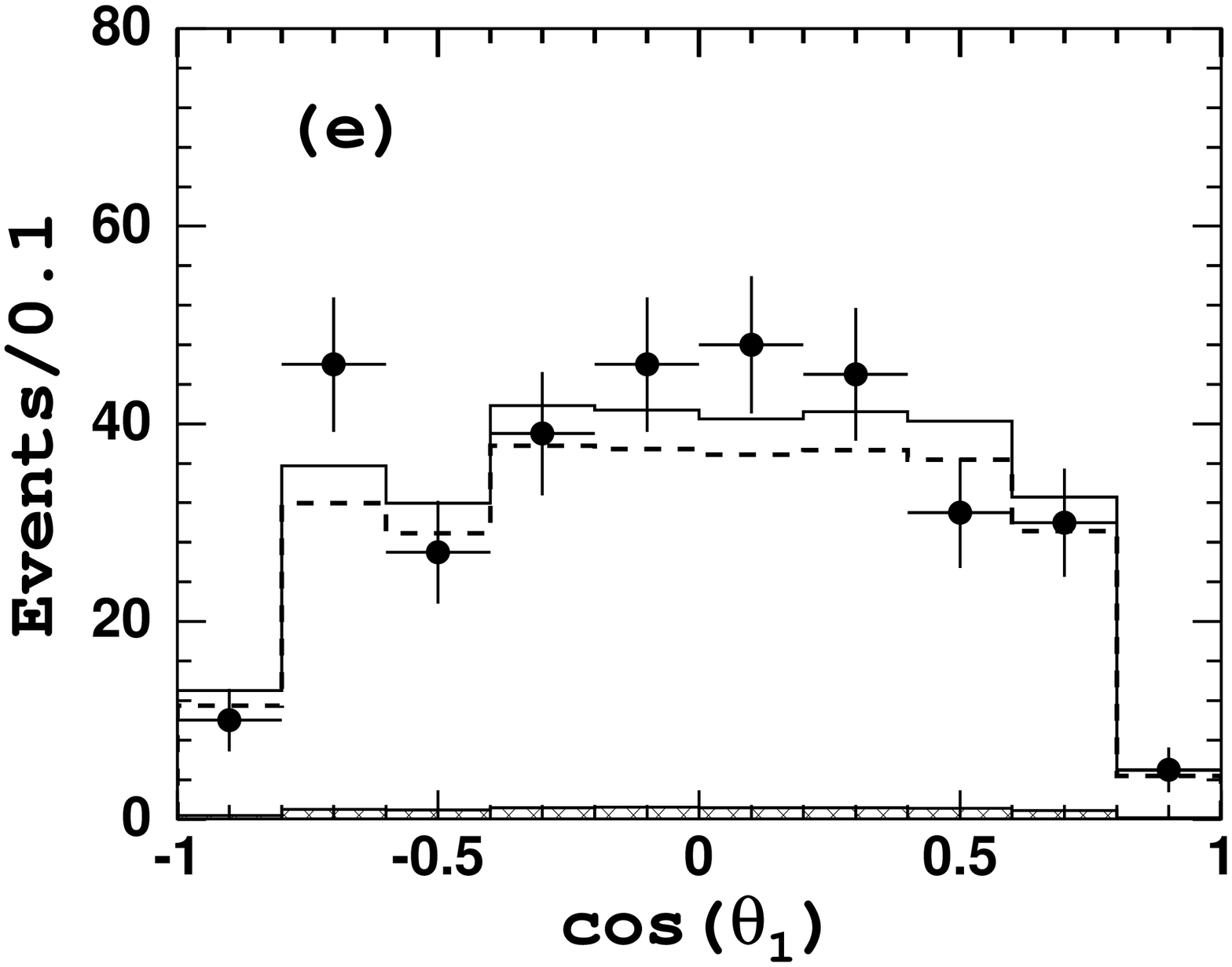} \hfill
  \includegraphics[width=0.24\textwidth]{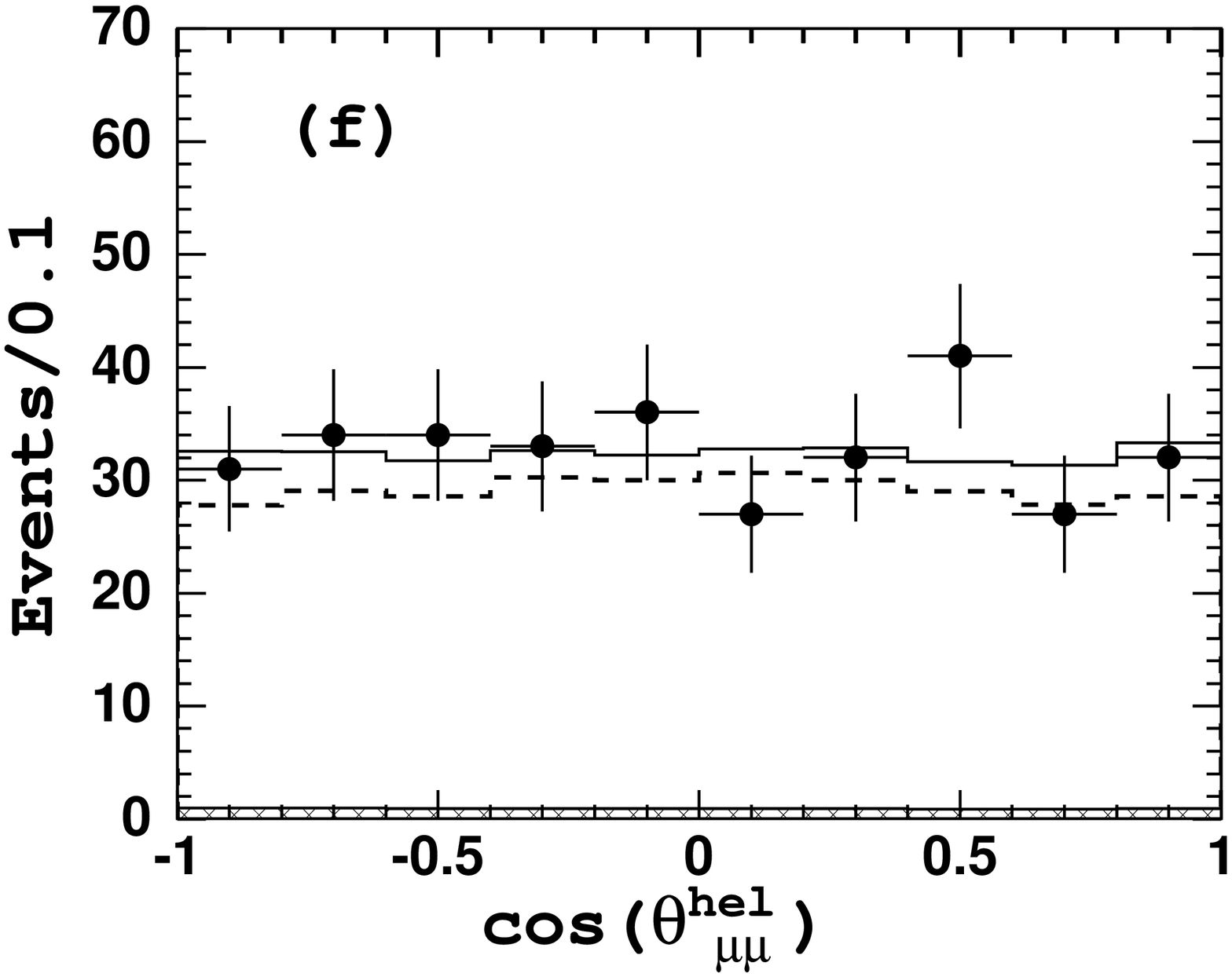} \hfill
  \includegraphics[width=0.24\textwidth]{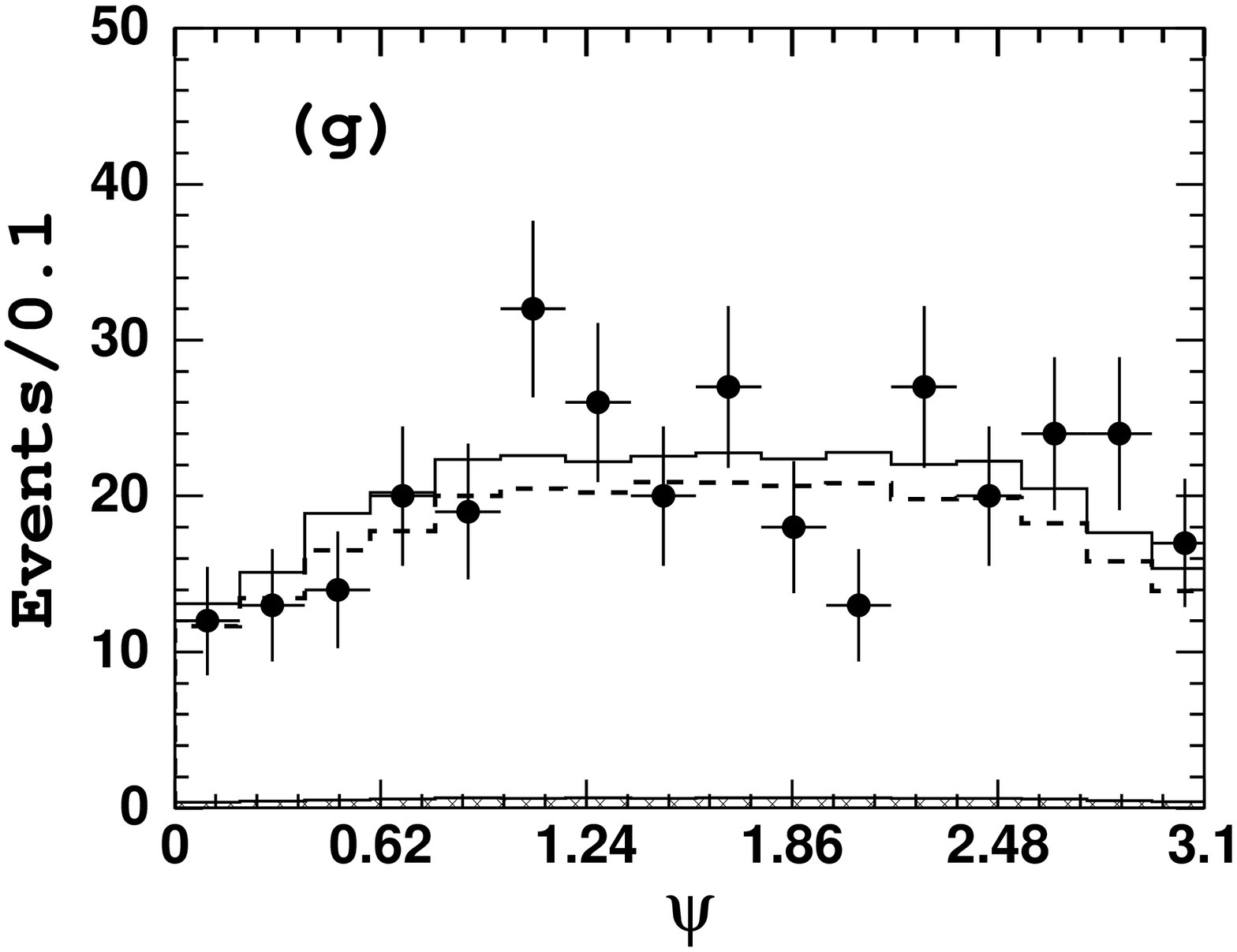} \hfill
  \includegraphics[width=0.24\textwidth]{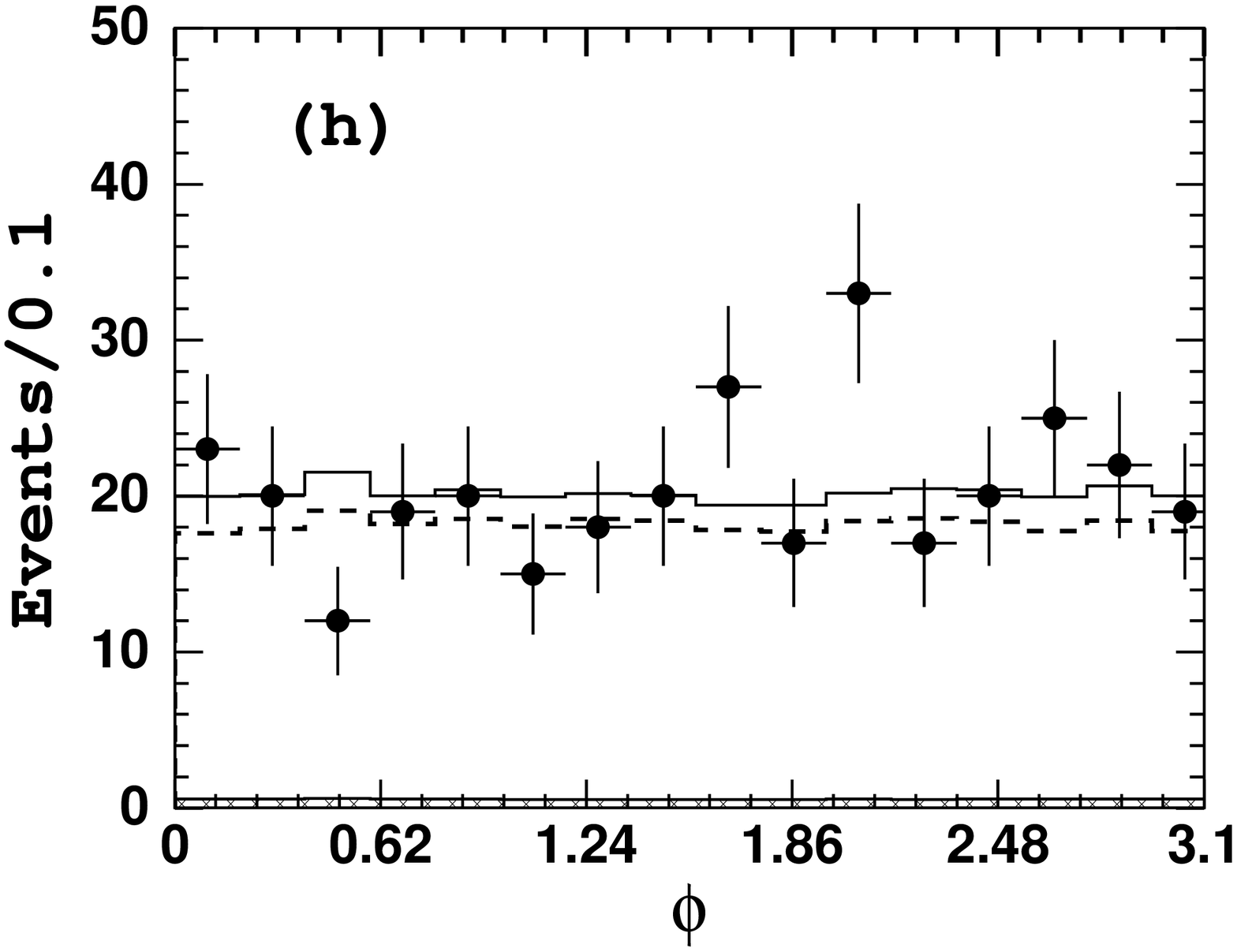} \\
  \includegraphics[width=0.24\textwidth]{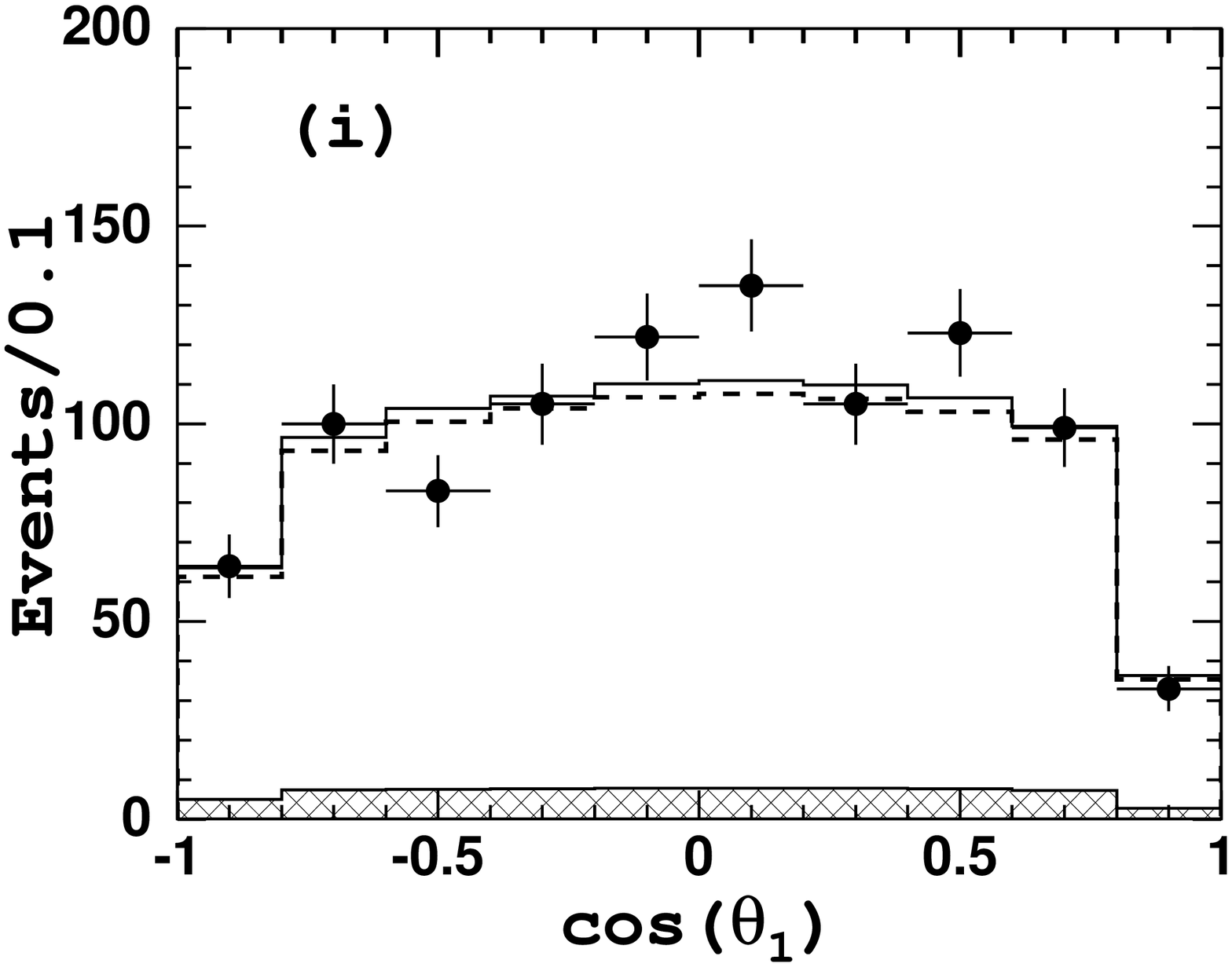} \hfill
  \includegraphics[width=0.24\textwidth]{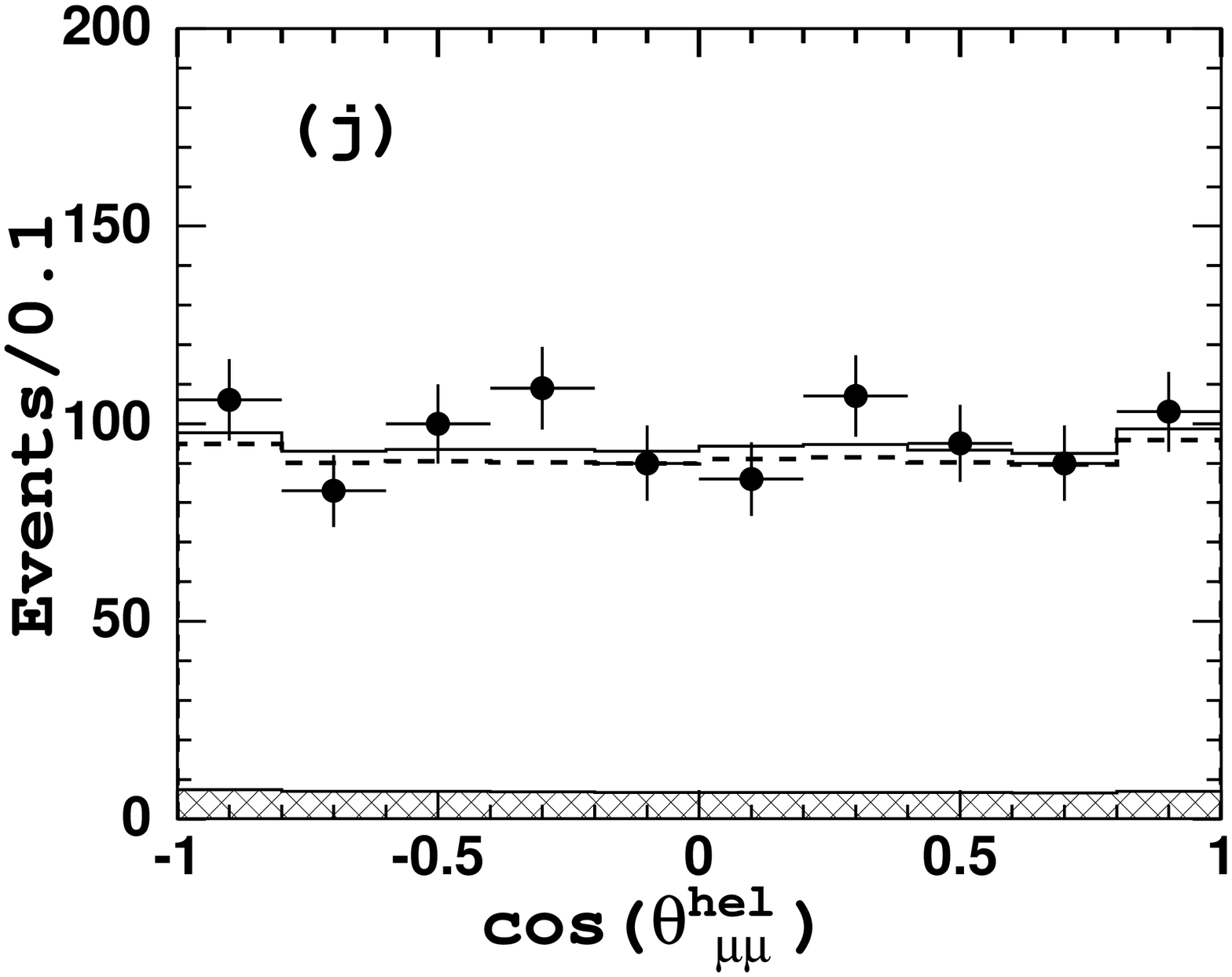} \hfill
  \includegraphics[width=0.24\textwidth]{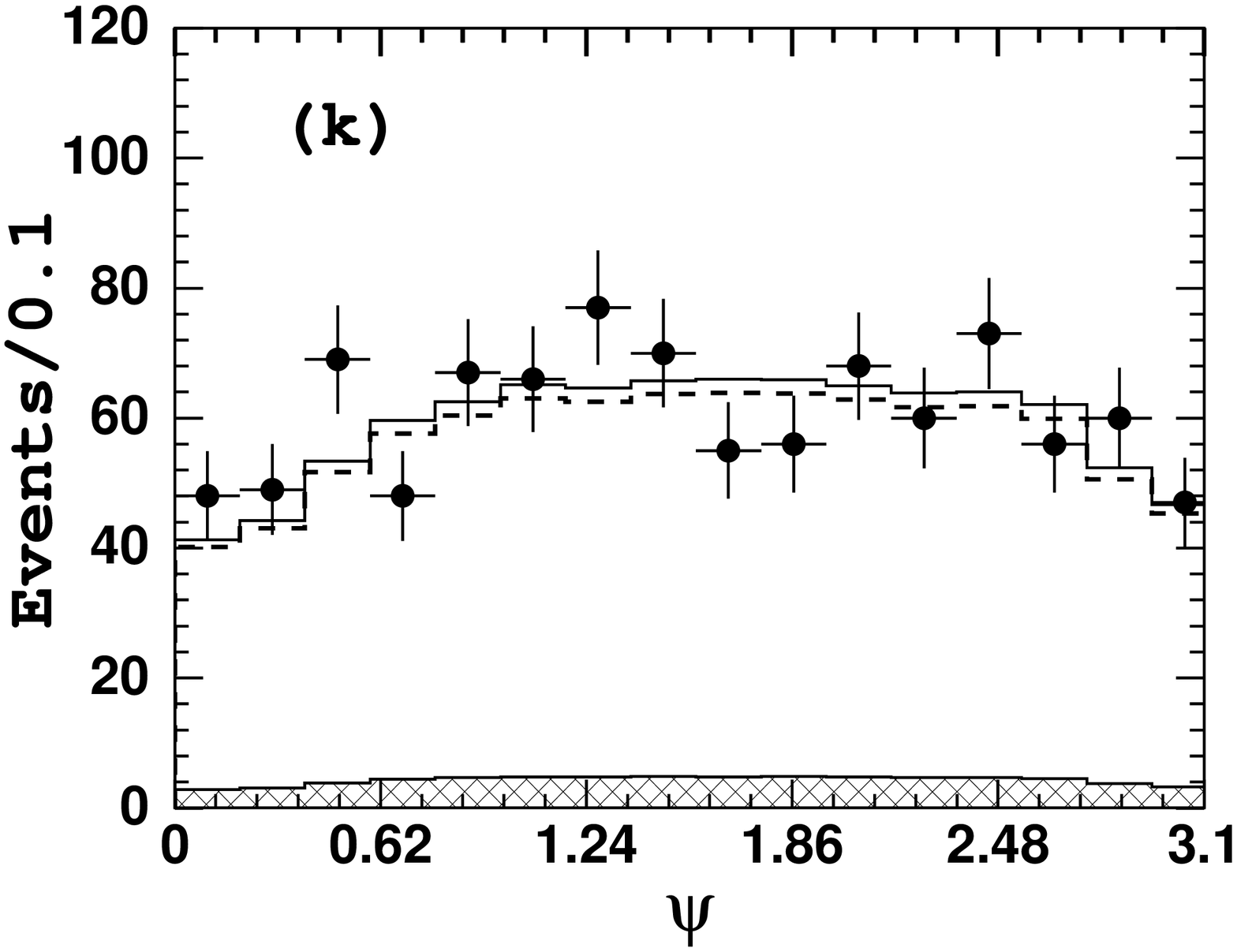} \hfill
  \includegraphics[width=0.24\textwidth]{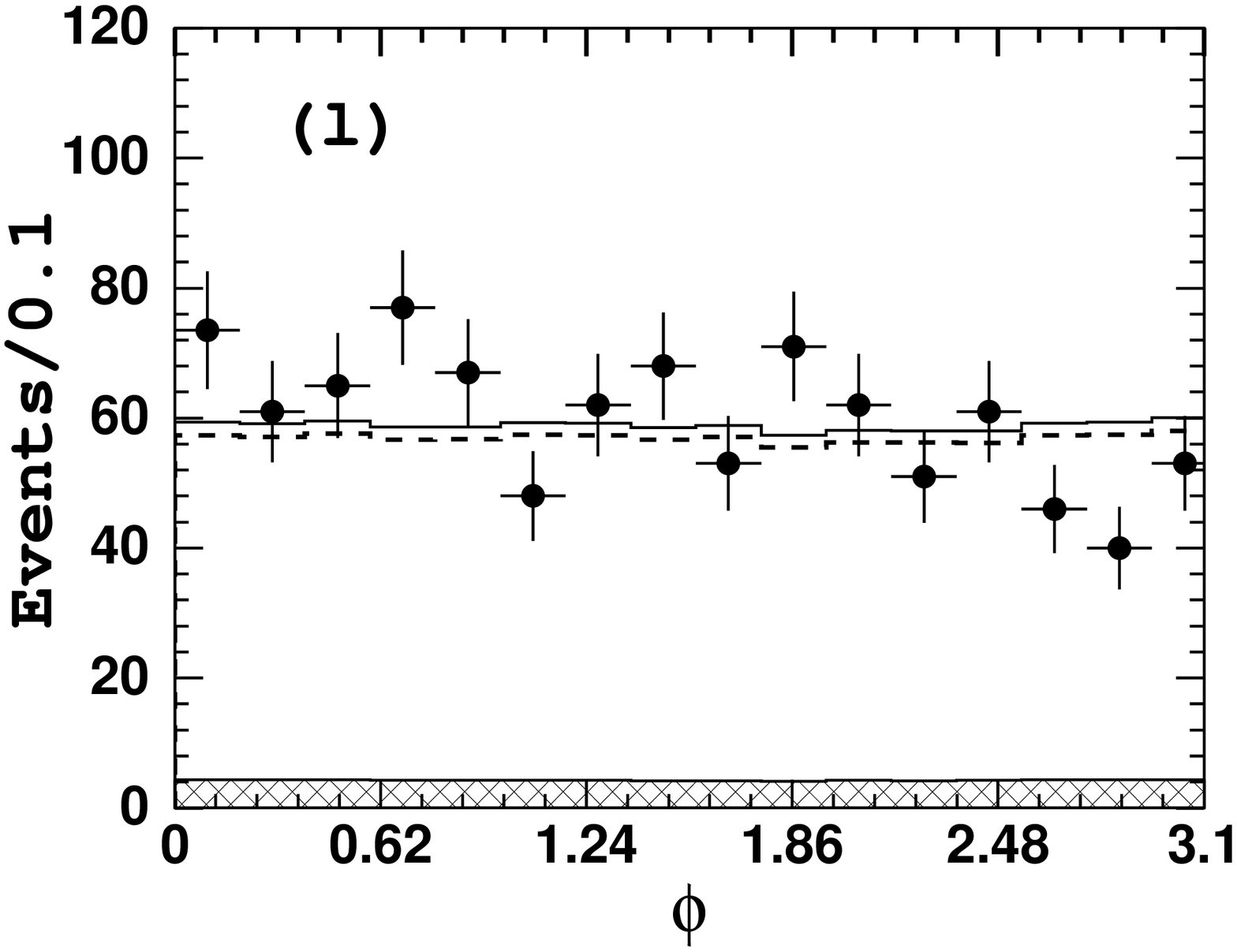} 
  \caption{Comparison of angular distributions for signal $\Ud\pp$ events
           in data (points with error bars), fit with the nominal model 
           with $J^P=1^+$ (open histogram), and fit with $J^P=2^+$ the 
           model (dashed histogram).
           Hatched histograms show the estimated background components.
           The top row is for the $Z_b(10610)$ region, the middle row is
           for the $Z_b(10650)$ region and the bottom row is for the 
           non-resonant region. See text for details.}
\label{fig:y2spp-ang}
\end{figure*}

\begin{figure*}[!t]
  \centering
  \includegraphics[width=0.24\textwidth]{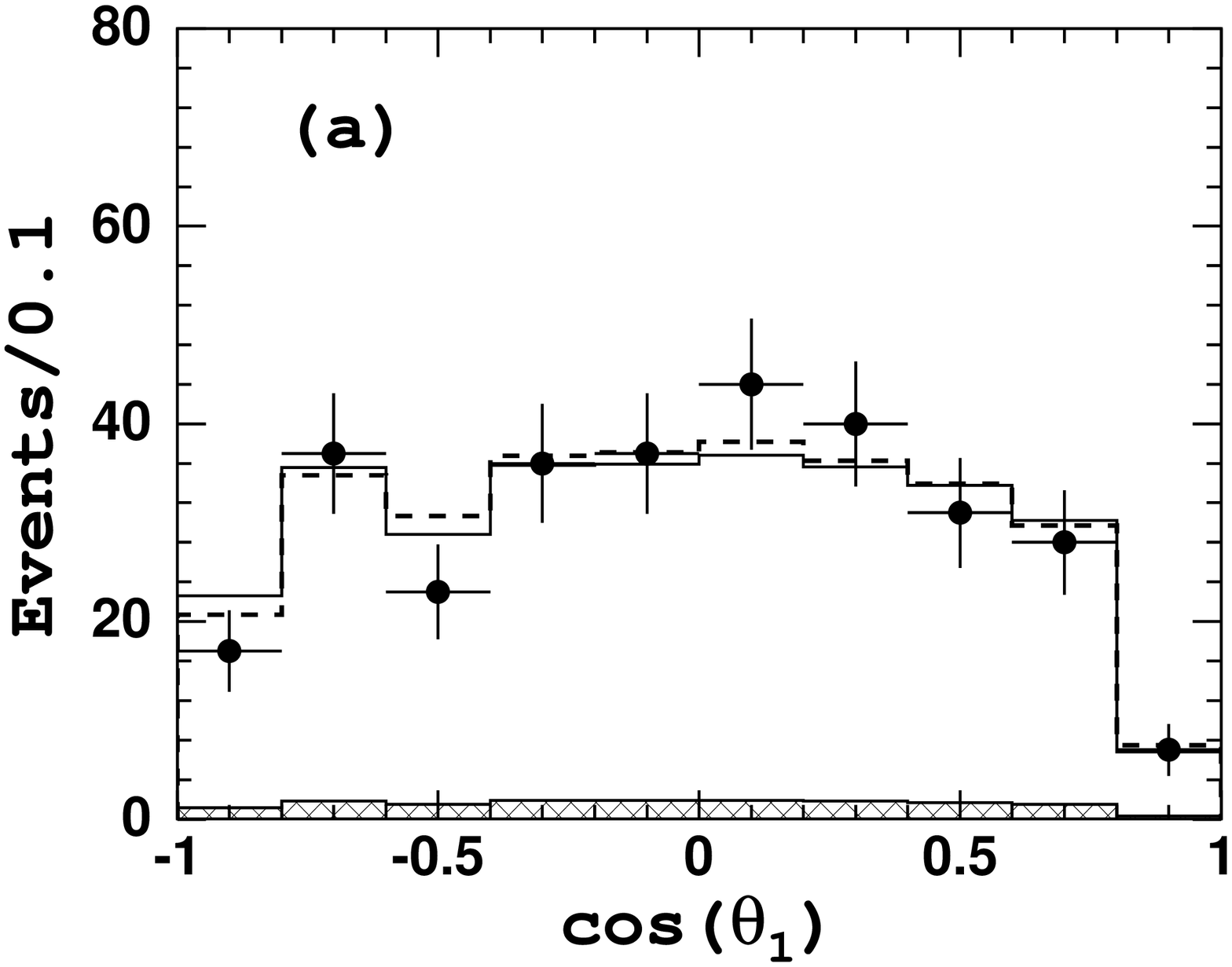} \hfill
  \includegraphics[width=0.24\textwidth]{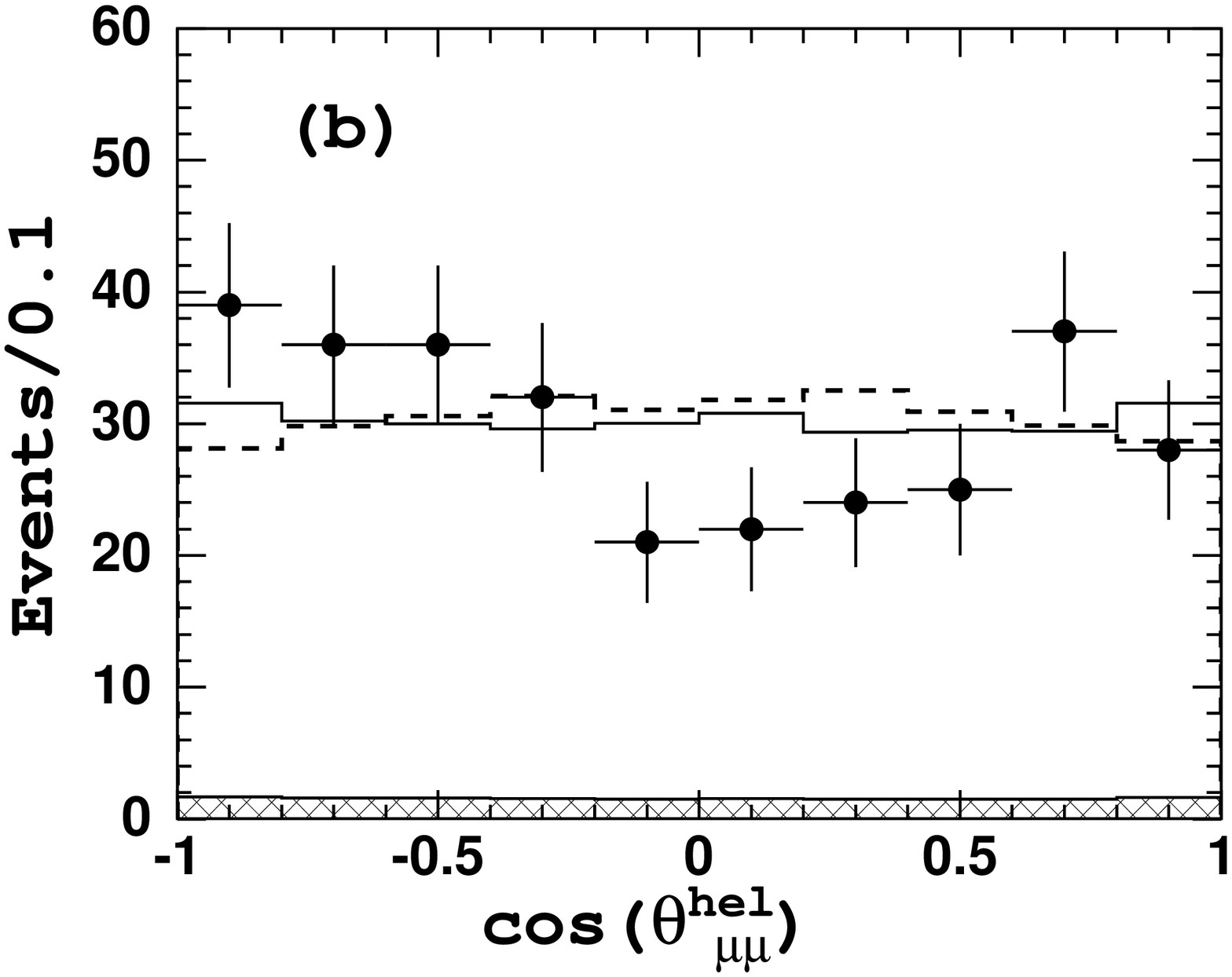} \hfill
  \includegraphics[width=0.24\textwidth]{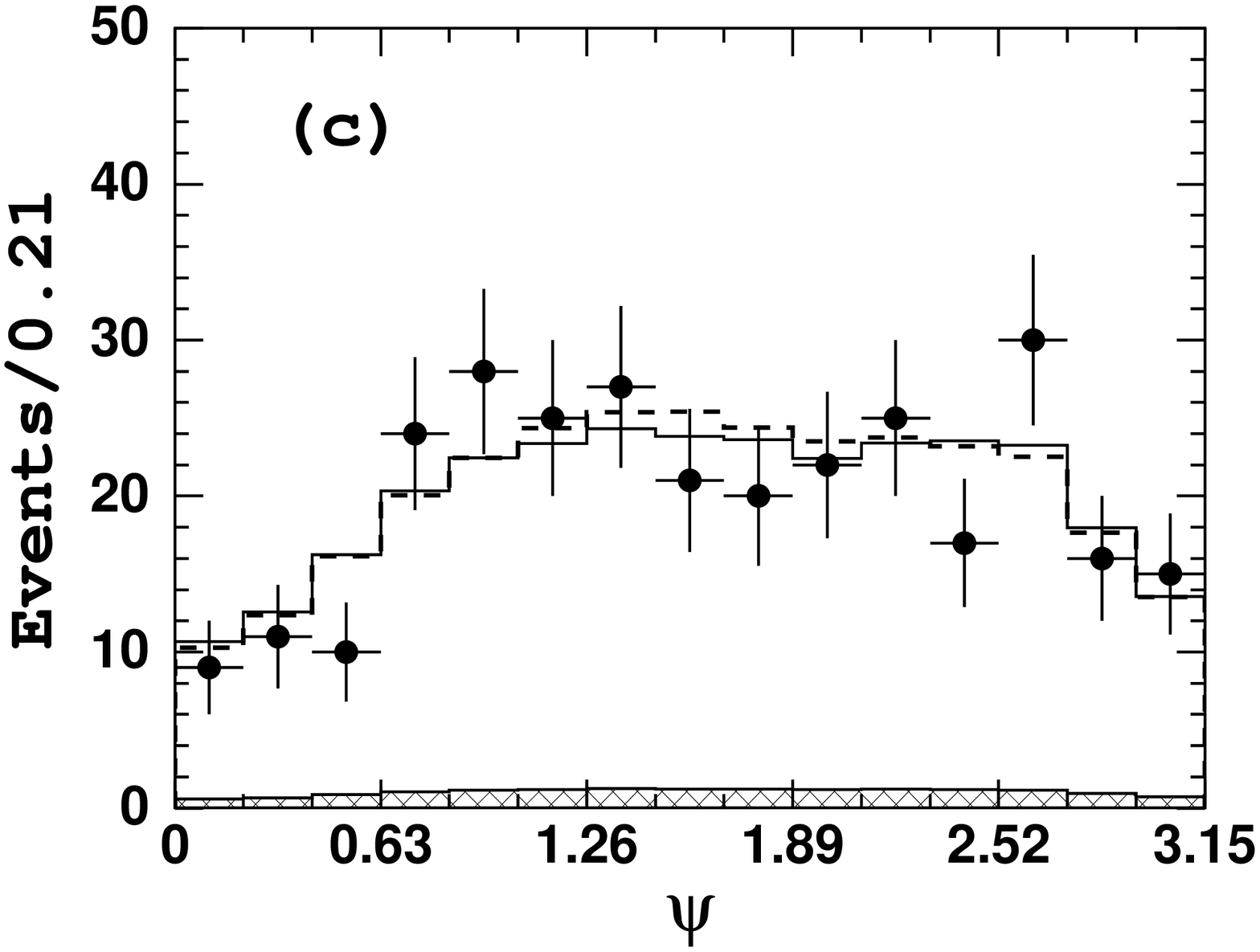} \hfill
  \includegraphics[width=0.24\textwidth]{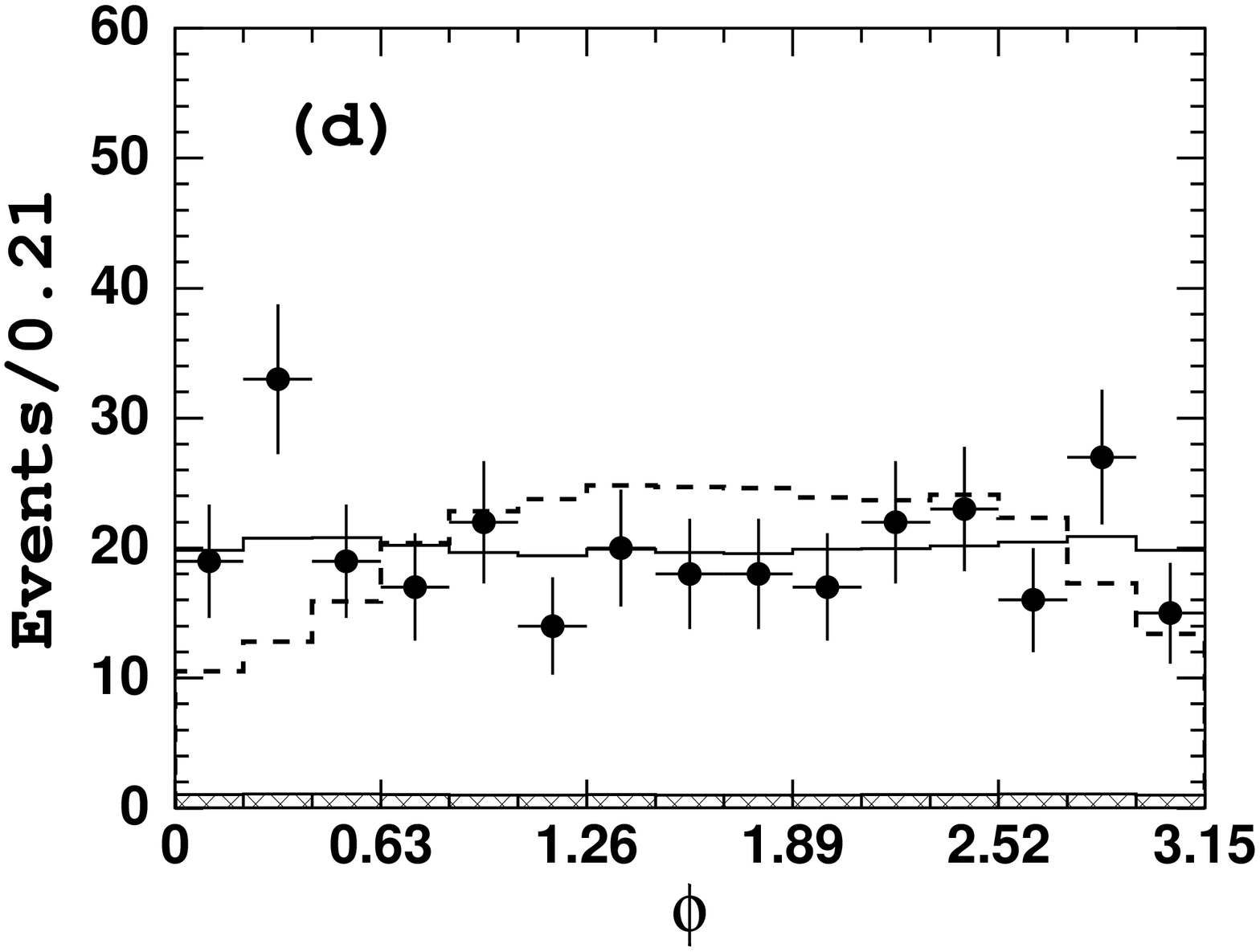} 
  \includegraphics[width=0.24\textwidth]{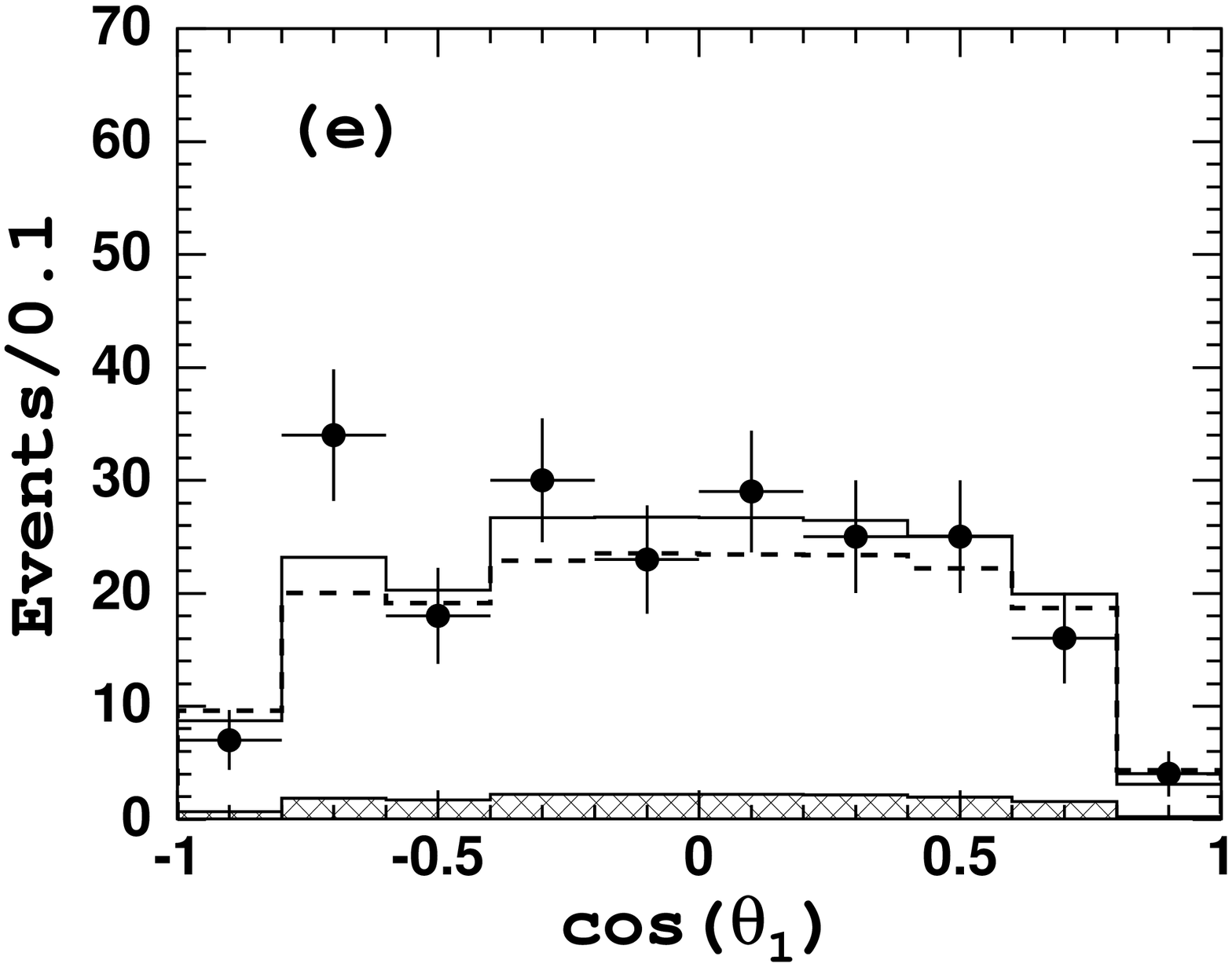} \hfill
  \includegraphics[width=0.24\textwidth]{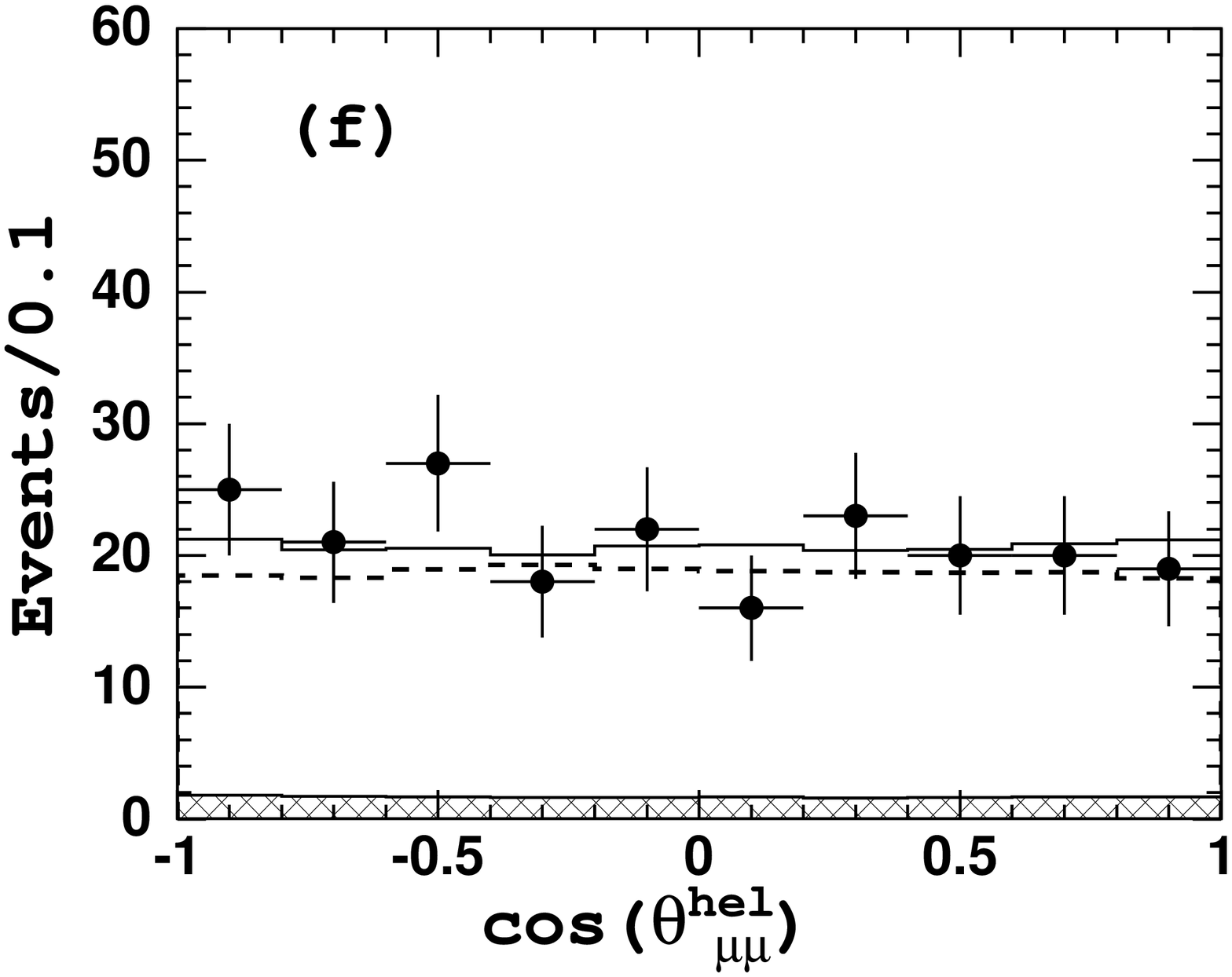} \hfill
  \includegraphics[width=0.24\textwidth]{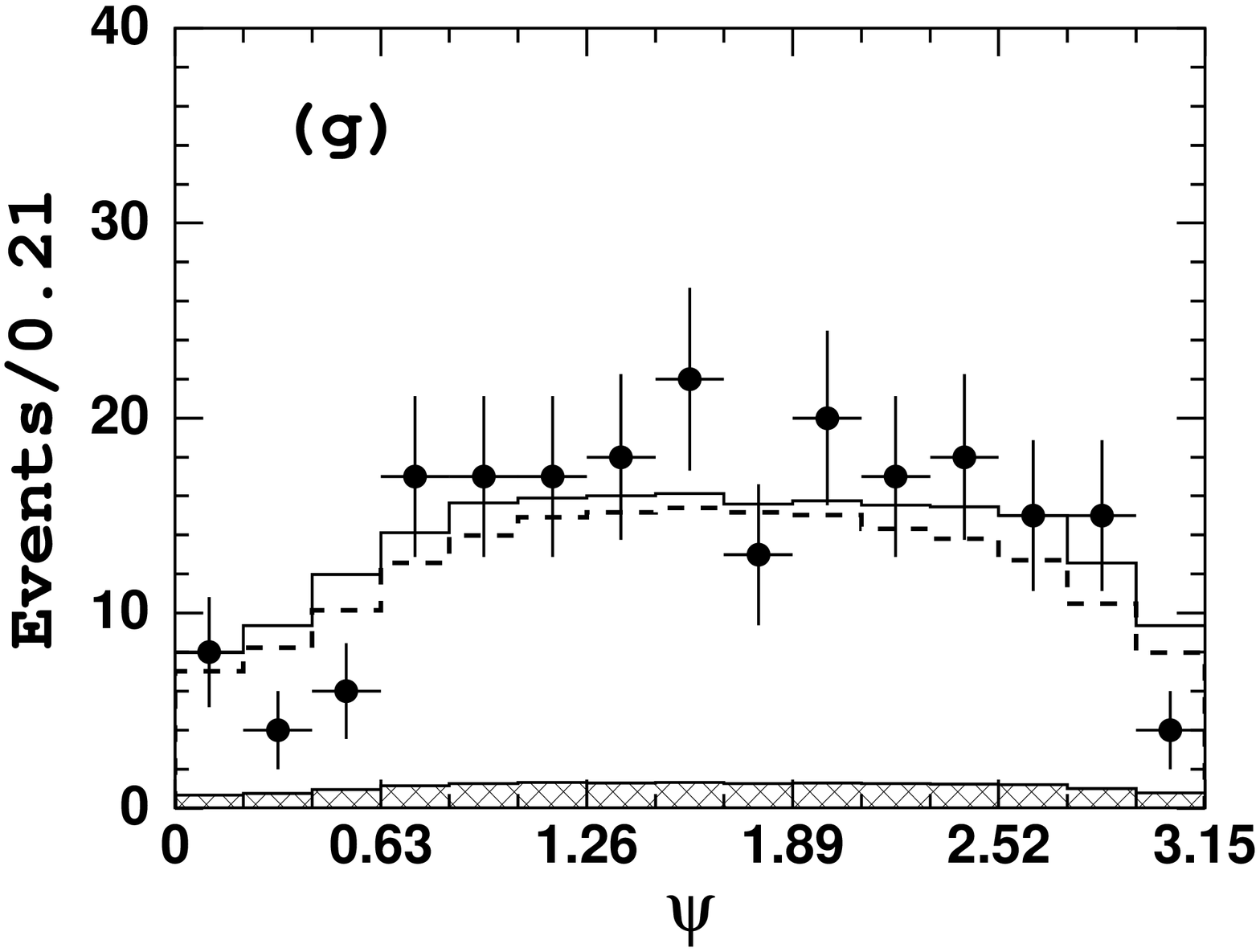} \hfill
  \includegraphics[width=0.24\textwidth]{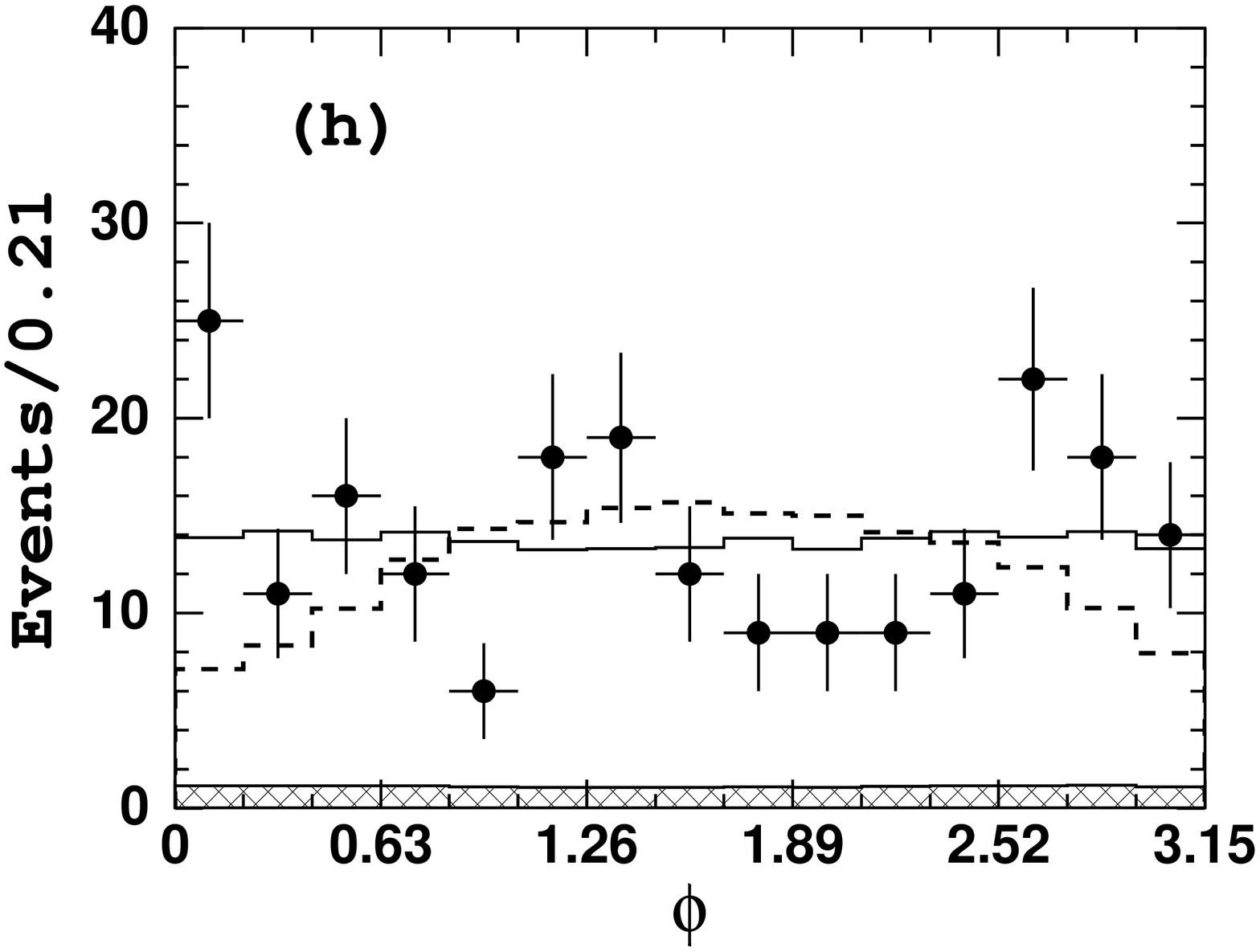} 
  \caption{Comparison of angular distributions for signal $\Ut\pp$ events
           in data (points with error bars), fit with the nominal model 
           with $J^P=1^+$ (open histogram), and fit with the $J^P=2^+$ 
           model (dashed histogram).
           Hatched histograms show the estimated background components.
           The top row is for the $Z_b(10610)$ region and the bottom row 
           is for the $Z_b(10650)$ region. See text for details.}
\label{fig:y3spp-ang}
\end{figure*}

A more detailed comparison of the fit results and the data is shown in 
Figs.~\ref{fig:y1spp-reg}-\ref{fig:y3spp-reg}, where mass projections
for various regions of the Dalitz plots are presented. In addition, 
comparison of the angular distributions for the $Z_b(10610)$ region 
($10605 {\rm ~MeV}/c^2<M(\Un\pi)_{\rm max}<10635 {\rm ~MeV}/c^2$), 
the $Z_b(10650)$ region 
($10645 {\rm ~MeV}/c^2<M(\Un\pi)_{\rm max}<10675 {\rm ~MeV}/c^2$), and the
non-resonant region ($M(\Un\pi)_{\rm max}<10570 {\rm ~MeV}/c^2$)
is presented for the $\Ud\pp$ final state in Fig.~\ref{fig:y2spp-ang} 
and for the $\Ut\pp$ final state in Fig.~\ref{fig:y3spp-ang}.
Here, $\theta_1$ is the angle between the prompt pion and the beam axis in 
the c.m.\ frame, $\theta^{\rm hel}_{\mu\mu}$ is the angle between the $Z_b$ 
and the $\mu^+$ momenta calculated in the $\Un$ rest frame (that is, the
$\Un\to\uu$ helicity angle), $\phi$ is the angle between the plane formed 
by the $\pp$ system and the $\Un$ decay plane in the $Z_b$ rest frame, and, 
finally, $\psi$ is the angle between the plane formed by the prompt 
pion and the beam axis and the $\Un$ decay plane calculated in the $Z_b$ 
rest frame.

We find that the model with $J^P=1^+$ assigned to both $Z_b$ states 
provides the best description of the data for all three three-body final 
states. Fits to the data with alternative $J^P$ values assigned to the 
two $Z_b$ states are compared with the nominal one in terms of the 
likelihood values returned by the fits. For each model, we calculate 
$\Delta{\cal{L}}={\cal{L}}(J^P)-{\cal{L}}_0$ that is the difference in the 
likelihood values returned by the fit to a model with an alternative $J^P$ 
assignment and the nominal one. Results of this study for the $\Ud\pp$ 
and $\Ut\pp$ modes (where the $Z_b\pi$ signal comprises a significant 
fraction of the three-body signal) are summarized in Table~\ref{tab:zjp}.
For the $\Uu\pp$ mode, we fit the data only to models with the same $J^P$ 
assigned to both $Z_b$ states. The obtained $\Delta{\cal{L}}$ values are 
64, 41, and 59 for the $J^P=1^-$, $2^+$, and $2^-$ models, respectively.

\begin{table}[!b]
  \caption{Results of the fit to $\Ud\pp$ ($\Ut\pp$) events with different
$J^P$ values assigned to the $Z_b(10610)$  and $Z_b(10650)$ states. 
Shown in the table is the difference in ${\cal{L}}$ values for fits to an 
alternative model and the nominal one.}
  \medskip
  \label{tab:zjp}
  \centering
  \begin{tabular}{lcccc} \hline \hline
 ~~~~~~~~~~$Z_b(10650)$   &
                     ~$1^+$~             &
                     ~$1^-$~             &
                     ~$2^+$~             &
                     ~$2^-$~            \\
~~$Z_b(10610)$   & \\ \hline
~~~~~~$1^+$~  & $  0~(0)$   & $ 60~(33)$   & $ 42~(33)$   & $ 77~(63)$  \\
~~~~~~$1^-$~  & $226~(47)$  & $264~(73)$   & $224~(68)$   & $277~(106)$ \\
~~~~~~$2^+$~  & $205~(33)$  & $235~(104)$  & $207~(87)$   & $223~(128)$ \\
~~~~~~$2^-$~  & $289~(99)$  & $319~(111)$  & $321~(110)$  & $304~(125)$ \\
\hline  \hline
  \end{tabular}
\end{table}

\begin{table*}[!t]
  \caption{Results on cross sections for three-body $\ee\to\Un\pp$ transitions.
 The first quoted error is statistical and the second is systematics.
 The last line quotes results from our previous publication for comparison.}
  \medskip
  \label{tab:ypp-frac}
\centering
  \begin{tabular}{lccc} \hline \hline
 Final state    &  ~~~~~~~$\Uu\pp$~~~~~~~    &
                   ~~~~~~~$\Ud\pp$~~~~~~~    &
                   ~~~~~~~$\Ut\pp$~~~
\\ \hline
Signal yield    &  $2090\pm115$  &  $2476\pm97$  & $628\pm41$    \\
Efficiency, \%  &    $45.9$     &    $39.0$     &   $24.4$      \\
${\cal{B}}_{\Un\to\mu^+\mu^-}$, \%~\cite{PDG}
                & $2.48\pm0.05$ & $1.93\pm0.17$ & $2.18\pm0.21$ \\
% $f_{\rm corr}$   &  $1.0$ & $1.0$ & $1.0$ \\
\hline
$\sigma^{\rm vis}_{\ee\to\Un\pp}$, pb
                & $1.51\pm0.08\pm0.09$  
                & $2.71\pm0.11\pm0.30$ 
                & $0.97\pm0.06\pm0.11$ \\ 
$\sigma_{\ee\to\Un\pp}$, pb
                & $2.27\pm0.12\pm0.14$  
                & $4.07\pm0.16\pm0.45$ 
                & $1.46\pm0.09\pm0.16$ \\ \hline
$\sigma^{\rm vis}_{\ee\to\Un\pp}$, pb~\cite{Belle_ypp}
                & $1.61\pm0.10\pm0.12$
                & $2.35\pm0.19\pm0.32$
                & $1.44^{+0.55}_{-0.45}\pm0.19$  \\
\hline \hline
\end{tabular}
\end{table*}

The best discrimination is provided by the $\ee\to\Ud\pp$ channel, where 
the $Z_b$ and the underlying non-$Z_b$ amplitudes are comparable in size, 
thus maximizing the relative size of the interference term. To cross-check 
the separation power, we perform a MC study in which we generate a large 
number of $\Un\pp$ samples, each with statistics equivalent to the data 
and perform fits of each pseudo-experiment with different $J^P$ models. 
The obtained $\Delta{\cal{L}}$ distributions are fit to a Gaussian 
function (a bifurcated Gaussian function for asymmetric distributions) to 
estimate the probability of $\Delta{\cal{L}}>42$.
We find that alternative models with the same $J^P$ assigned to both 
$Z_b$ states are rejected at a level exceeding eight standard deviations.
The comparisons of the fit result where both $Z_b$ are assumed to be 
$J^P=2^+$ states (the next best hypothesis) and the data are shown in 
Figs.~\ref{fig:ynspp-f-hh},~\ref{fig:y2spp-ang} and~\ref{fig:y3spp-ang}. 

In fits with different $J^P$ values assigned to the $Z_b(10610)$ and 
$Z_b(10650)$ states, the smallest $\Delta{\cal{L}}$ value is provided
by the model with $Z_b(10610)$ assumed to be an $1^+$ state and $Z_b(10650)$
a $2^+$ state, as shown in Table~\ref{tab:zjp}. A similar study with 
MC pseudo-experiments shows that this alternative hypothesis is rejected 
at a level exceeding six standard deviations.

Finally, we note that multiple solutions are found in the fit to the 
$\Uu\pp$ and $\Ud\pp$ final states. This is due to the presence of several
$S$-wave components in the three-body amplitudes for these modes. While 
the overall fraction of the $S$-wave contribution is a well defined 
quantity, the individual components are strongly correlated and thus
poorly separated by the fit. Because of this effect, we do not present
relative phases and fractions of individual $S$-wave contributions 
except for the $\Uu f_0(980)$ mode, whose parameters are well defined.

%======================================================================
%======================================================================
%======================================================================

\section{Cross sections}

The cross sections of the three-body $\ee\to\Un\pp$ processes are
calculated using the following formula:
\begin{eqnarray}
\sigma_{\ee\to\Un\pp} = 
\frac{\sigma^{\rm vis}_{\ee\to\Un\pp}}{1+\delta_{\rm ISR}} =  \nonumber \\
\frac{N_{\Un\pp}}
{L\cdot {\cal{B}}_{\Un\to\uu}\cdot\varepsilon_{\Un\pp}(1+\delta_{\rm ISR})}, 
\label{eq:csec}
\end{eqnarray}
where $\sigma_{\rm vis}$ is the visible cross section. The initial state
radiation (ISR) correction factor $(1+\delta_{\rm ISR})=0.666\pm0.013$ is 
determined using formulae given in Ref.~\cite{radcorr}, where we use 
the energy dependence of the $\ee\to\Un\pp$ cross section measured in 
Ref.~\cite{ypp_scan}. The integrated luminosity is measured to be 
$L=121.4$~fb$^{-1}$, and the reconstruction efficiency $\varepsilon_{\Un\pp}$ 
(including trigger efficiency and final state radiation) is determined 
from the signal MC events generated according to the nominal model from 
the amplitude analysis. For the branching fractions of the $\Un\to\uu$ 
decays, the world average values are used~\cite{PDG}. Results of the 
calculations are summarized in Table~\ref{tab:ypp-frac}. The Born cross 
section can be obtained by multiplying Eq.~\ref{eq:csec} by the vacuum 
polarization correction factor, $|1-\Pi|^2=0.9286$~\cite{vacpolar}. 
The $\Uf\to\Un\pp$ branching fractions listed in Ref.~\cite{PDG} can be 
obtained by dividing our results for $\sigma^{\rm vis}$ in 
Table~\ref{tab:ypp-frac} by the $\ee\to b\bar{b}$ cross section measured 
at the $\Uf$ peak,
$\sigma_{\ee\to b\bar{b}}(\sqrt{s}=10860) = 0.340\pm0.016$~nb~\cite{ee2bb}.

\begin{table}[!b]
  \caption{List of dominant sources of systematic uncertainties (in percent)
contributing to the measurement of three-body $\ee\to\Un\pp$ cross sections.}
  \medskip
  \label{tab:ypp-syst}
\centering
  \begin{tabular}{lccc} \hline \hline
 Final state           & $\Uu\pp$ & $\Ud\pp$ & $\Ut\pp$
\\ \hline
${\cal{B}}_{\Un\to\uu}$,~\cite{PDG}
                       &    $2.0$   &    $8.8$   &   $9.6$     \\
Signal yield           &    $4.5$   &    $5.3$   &   $4.9$     \\
Muon ID                &    $2.0$   &    $2.0$   &   $2.0$     \\
Tracking               &    $2.7$   &    $2.7$   &   $2.7$     \\
ISR correction         &    $2.0$   &    $2.0$   &   $2.0$     \\
Luminosity             &    $1.4$   &    $1.4$   &   $1.4$     \\ 
\hline 
Total                  &    $6.2$   &   $10.9$   &  $11.4$     \\
\hline \hline
\end{tabular}
\end{table}

\begin{table*}[!t]
  \caption{Summary of results of fits to 
           $\Un\pp$ events in the signal regions.}
  \medskip
  \label{tab:results}
\centering
  \begin{tabular}{lccc} \hline \hline
Parameter & ~~~~$\Uu\pp$~~~~   &
              ~~~~$\Ud\pp$~~~~   &
              ~~~~$\Ut\pp$~~~~
\\ \hline
%    $f_{Z^\mp_b(10610)\pi^\pm}{\cal{B}}(Z^\mp_b(10610)\to\Un\pi^\mp)$, \%  &
           $f_{Z^\mp_b(10610)\pi^\pm}$, \%  &
           $4.8\pm1.2^{+1.5}_{-0.3}$   &
           $18.1\pm3.1^{+4.2}_{-0.3}$   &
           $30.0\pm6.3^{+5.4}_{-7.1}$  
\\
           $Z_b(10610)$ mass, MeV/$c^2$  &
         ~~$10608.5\pm3.4^{+3.7}_{-1.4}$~~     &
         ~~$10608.1\pm1.2^{+1.5}_{-0.2}$~~     &
         ~~$10607.4\pm1.5^{+0.8}_{-0.2}$~~    
\\
           $Z_b(10610)$ width, MeV/$c^2$  &
         ~~$18.5\pm5.3^{+6.1}_{-2.3}$~~    &
         ~~$20.8\pm2.5^{+0.3}_{-2.1}$~~    &
         ~~$18.7\pm3.4^{+2.5}_{-1.3}$~~    
\\
%    $f_{Z^\mp_b(10650)\pi^\pm}{\cal{B}}(Z^\mp_b(10650)\to\Un\pi^\mp)$, \%  &
           $f_{Z^\mp_b(10650)\pi^\pm}$, \%  &
           $0.87\pm0.32^{+0.16}_{-0.12}$    &
           $4.05\pm1.2^{+0.95}_{-0.15}$    &
           $13.3\pm3.6^{+2.6}_{-1.4}$  
\\
           $Z_b(10650)$ mass, MeV/$c^2$  &
         ~~$10656.7\pm5.0^{+1.1}_{-3.1}$~~     &
         ~~$10650.7\pm1.5^{+0.5}_{-0.2}$~~     &
         ~~$10651.2\pm1.0^{+0.4}_{-0.3}$~~    
\\
           $Z_b(10650)$ width, MeV/$c^2$  &
         ~~$12.1^{+11.3+2.7}_{-4.8-0.6}$~~    &
         ~~$14.2\pm3.7^{+0.9}_{-0.4}$~~    &
         ~~$ 9.3\pm2.2^{+0.3}_{-0.5}$~~    
\\
           $\phi_{Z}$, degrees      &
           $67\pm36^{+24}_{-52}$     &
           $-10\pm13^{+34}_{-12}$    &
           $-5\pm22^{+15}_{-33}$  
\\
           $c_{Z_b(10650)}/c_{Z_b(10610)}$   &
           $0.40\pm0.12^{+0.05}_{-0.11}$  &
           $0.53\pm0.07^{+0.32}_{-0.11}$  &
           $0.69\pm0.09^{+0.18}_{-0.07}$
\\
%           $f_{\Un f_2(1270)}{\cal{B}}(f_2(1270)\to\pp)$, \%  &
           $f_{\Un f_2(1270)}$, \%  &
           $14.6\pm1.5^{+6.3}_{-0.7}$      &
           $4.09\pm1.0^{+0.33}_{-1.0}$     &
           $-$  
\\
           $f_{\Un(\pp)_S}$, \%  &
           $86.5\pm3.2^{+3.3}_{-4.9}$     &
           $101.0\pm4.2^{+6.5}_{-3.5}$    &
           $44.0\pm6.2^{+1.8}_{-4.3}$  
\\
%           ~~~$f_{\Un f_0(980)}{\cal{B}}(f_0(980)\to\pp)$, \%  &
           ~~~$f_{\Un f_0(980)}$, \%  &
           $6.9\pm1.6^{+0.8}_{-2.8}$     &
           $-$    &
           $-$  
\\
\hline \hline
\end{tabular}
\end{table*}

The dominant sources of systematic uncertainties contributing
to the measurements of cross sections for the three-body $\ee\to\Un\pp$ 
transitions are given in Table~\ref{tab:ypp-syst}.
The uncertainty in the signal yield is estimated by varying fit parameters
within one standard deviations one by one and repeating the fit to the 
corresponding $\mmpp$ distribution.
The uncertainty in the muon identification is determined using a large 
sample of $J/\psi\to\uu$ events in data and MC and found to be 1\% per muon.
The uncertainty in tracking efficiency is estimated using partially 
reconstructed $D^{*-}\to\pi^- D^0[K^0_S\pp]$ events and found to be 0.35\% 
per a high momentum track (muons from $\Un\to\uu$ decays) and 1\% per 
a lower momentum track (pions).
The uncertainty in the radiative correction factor is determined from a 
dedicated study. It is found to be due mainly to the uncertainty in the 
parametrization of the energy dependence of the $\ee\to\Un\pp$ cross 
section, the uncertainty in the c.m.\ energy and the selection criteria.
All contributions are added in quadrature to obtain the overall systematic
uncertainty of 6.2\%, 10.9\%, and 11.4\% for $n=1,2$ and $3$, respectively. 
Our results for $\sigma_{\rm vis}(\ee\to\Un\pp)$ may be compared with the 
previous measurements by Belle performed with a data sample of 
21~fb$^{-1}$~\cite{Belle_ypp} (see last line in Table~\ref{tab:ypp-frac}). 
We find the two sets of measurements are consistent within uncertainties.

Results of the amplitude analysis are summarized in 
Table~\ref{tab:results}, where fractions of individual quasi-two-body
modes, masses and widths of the two $Z_b$ states, the relative 
phase, $\phi_Z$, between the two $Z_b$ amplitudes and fraction 
$c_{Z_{10610}}/c_{Z_{10650}}$ of their amplitudes are given.
The fraction $f_X$ of the total three-body signal attributed to a 
particular quasi-two-body intermediate state is calculated as
\begin{equation}
 f_X = \frac{\int |{\cal A}_X|^2\, d\Omega}
            {\int |{\cal M}_{\Un\pi\pi}|^2\, d\Omega},
\label{eq:fraction}
\end{equation}
where ${\cal A}_X$ is the amplitude for a particular component $X$ of the 
three-body amplitude ${\cal M}_{\Un\pi\pi}$, defined in the Appendix.
For amplitudes where the $\pp$ system is in an $S$-wave, we do not 
calculate individual fractions for every component but present the result 
only for the combination $\Un(\pp)_S$ of all such components. The only 
exception is the $\Uu f_0(980)$ component. The statistical significance
of this signal, determined as $\sqrt{{\cal{L}}_{f_0}-{\cal{L}}_0}$, where
${\cal{L}}_{f_0}$ is the likelihood value with $f_0(980)$ amplitude fixed
at zero, exceeds eight standard deviations. Note, that the sum of the fit 
fractions for all components is not necessarily unity because of the 
interference. Statistical uncertainties for relative fractions of 
intermediate channels quoted in Table~\ref{tab:results} are determined 
utilizing a MC pseudo-experiment technique. For each three-body final 
state, we generate a large number of MC samples, each with statistics 
equivalent to the experimental data (including background) and with a 
phase space distribution according to the nominal model. Each MC sample 
is then fit to the nominal model and fractions $f_i$ of contributing 
submodes are determined. The standard deviation of the $f_i$ distribution 
is then taken as the statistical uncertainty for the fraction of the 
corresponding submode; see Table~\ref{tab:results}.

Combining results for the three-body cross sections from 
Table~\ref{tab:ypp-frac} with the results of the amplitude analysis 
from Table~\ref{tab:results}, we calculate the product
$\sigma_{Z^\pm_b\pi^\mp}\times{\cal{B}}_{\Un\pi^\mp}$, where 
$\sigma_{Z^\pm_b\pi^\mp}$ is the cross section of the $\ee$ annihilation to 
$Z^\pm_b\pi^\mp$ and ${\cal{B}}_{\Un\pi^\mp}$ is the branching fraction 
of $Z^\pm_b$ decay to $\Un\pi^\pm$. For the $Z^\pm_b(10610)$, we obtain
\begin{eqnarray}
  \sigma_{Z^\pm_b(10610)\pi^\mp}\times{\cal{B}}_{\Uu\pi^\mp} = & 
109 \pm 27  {}^{+35}_{-10} & {\rm fb} \nonumber \\
  \sigma_{Z^\pm_b(10610)\pi^\mp}\times{\cal{B}}_{\Ud\pi^\mp} = & 
737 \pm 126 {}^{+188}_{-85} & {\rm fb} \nonumber \\
  \sigma_{Z^\pm_b(10610)\pi^\mp}\times{\cal{B}}_{\Ut\pi^\mp} = & 
438 \pm 92  {}^{+92}_{-114} & {\rm fb},
\end{eqnarray}
and for the $Z^\pm_b(10650)$, we obtain
\begin{eqnarray}
  \sigma_{Z^\pm_b(10650)\pi^\mp}\times{\cal{B}}_{\Uu\pi^\mp} = & 
20 \pm   7 {}^{+4}_{-3}  & {\rm fb} \nonumber \\
  \sigma_{Z^\pm_b(10650)\pi^\mp}\times{\cal{B}}_{\Ud\pi^\mp} = & 
165 \pm 49 {}^{+43}_{-20}  & {\rm fb} \nonumber \\
  \sigma_{Z^\pm_b(10650)\pi^\mp}\times{\cal{B}}_{\Ut\pi^\mp} = & 
194 \pm 53 {}^{+43}_{-25}  & {\rm fb}.
\end{eqnarray}

The main sources of systematic uncertainties in the amplitude analysis are
\begin{itemize}
  \item{the uncertainty in parametrization of the transition amplitude. 
To estimate this uncertainty, we use various modifications of the nominal 
model and repeat the fit to the data. In particular, for the $\Uu\pp$ and
$\Ud\pp$ channels, we modify the parametrization of the non-resonant 
amplitude, replacing the $s_{23}$ dependence from linear to a 
$\sqrt{s_{23}}$ form and replacing the $\Un f_2(1270)$ amplitude with a 
$D$-wave component in the non-resonant amplitude. For the $\Ut\pp$ 
channel, we modify the nominal model by adding various components of the 
amplitude initially fixed at zero: a $\Ut f_2(1270)$ component with an 
amplitude and phase fixed from the fit to the $\Uu\pp$ channel. We also 
fit the $\Ut\pp$ data with
the non-resonant amplitude set to be uniform. Variations in fit parameters
and fractions of contributing channels determined from fits with these 
models are taken as an estimation of the model related uncertainty;}
  \item{multiple solutions found for the $\Uu\pp$ and $\Ud\pp$ modes are
treated as model uncertainty, with variations in fit 
parameters included in the systematic uncertainty;}
  \item{uncertainty in the c.m.\ energy leads to uncertainty in the phase 
space boundaries. To estimate the associated effect on fit parameters, we 
generate a normalization phase space MC sample that corresponds to 
$E_{\rm cm}\pm 3$~MeV, where $E_{\rm cm}$ is the nominal c.m.\ energy, 
and refit;}
  \item{uncertainty in the fraction of signal events $f_{\rm sig}$ in the 
sample. To determine the associated uncertainties in fit parameters we 
vary  $f_{\rm sig}$ within its error and repeat the fit to the data. We also 
fit the data with $f_{\rm sig}$ relaxed;}
  \item{uncertainty in the parametrization of the distribution of 
background events. We repeat the fit the data with a background density 
set to be uniform over the phase space;}
  \item{uncertainty associated with the fitting procedure. 
This is estimated from MC studies.}
\end{itemize}
All the contributions are added in quadrature to obtain the overall 
systematic uncertainty. The size of the systematic uncertainty depends on 
the three-body $\Un\pp$ channel and on the particular decay submode.

%=============================================================================
%=============================================================================
%=============================================================================

\section{Conclusions}

In conclusion, we have performed a full amplitude analysis of three-body 
$\ee\to\Un\pp$ ($n=1,2,3$) transitions that allowed us to determine the 
relative fractions of various quasi-two-body components of the three-body 
amplitudes as well as the spin and parity of the two observed $Z_b$ states. 
The favored quantum numbers are $J^P=1^+$ for both $Z_b$ states while the
alternative $J^P=1^-$ and $J^P=2^\pm$ combinations are rejected at confidence
levels exceeding six standard deviations. This is a substantial improvement 
over the previous one-dimensional angular analysis reported in 
Ref.~\cite{Belle_ang}. This is 
due to the fact that the part of the amplitude most sensitive to the spin 
and parity of the $Z_b$ states is the interference term between the 
$Z_b\pi$ and the non-resonant amplitudes. Thus, the highest sensitivity is 
provided by the $\ee\to\Ud\pp$ transition, where the two amplitudes $Z_b\pi$ 
and the non-resonant one are comparable in size. The measured values of the 
spin and parity of the $Z_b$ states are in agreement with the expectations 
of the molecular model~\cite{Zb-molecular} yet do not contradict several 
alternative interpretations~\cite{Zb-other}.

We update the measurement of the three-body $\ee\to\Un\pp$ cross sections  
with significantly increased integrated luminosity compared to that in 
Ref.~\cite{Belle_ypp}. 
The results reported here supersede our measurements reported in 
Ref.~\cite{Belle_ypp}. We also report the first measurement of the relative 
fractions of the $\ee\to Z^\mp_b\pi^\pm$ transitions and the first observation
of the $\ee\to\Uu f_0(980)$ transition. Finally, we find a significant 
contribution from the $\ee\to\Uu(\pp)_{D\rm-wave}$ amplitude but cannot 
attribute it unambiguously to the $\Uu f_2(1270)$ channel: the data can be 
equally well described by adding a $D$-wave component to the non-resonant 
amplitude.

\section*{Acknowledgement}

We thank the KEKB group for the excellent operation of the
accelerator; the KEK cryogenics group for the efficient
operation of the solenoid; and the KEK computer group,
the National Institute of Informatics, and the 
PNNL/EMSL computing group for valuable computing
and SINET4 network support.  We acknowledge support from
the Ministry of Education, Culture, Sports, Science, and
Technology (MEXT) of Japan, the Japan Society for the 
Promotion of Science (JSPS), and the Tau-Lepton Physics 
Research Center of Nagoya University; 
the Australian Research Council and the Australian 
Department of Industry, Innovation, Science and Research;
Austrian Science Fund under Grant No. P 22742-N16;
the National Natural Science Foundation of China under Contracts 
No.~10575109, No.~10775142, No.~10825524, No.~10875115, No.~10935008 
and No.~11175187; 
the Ministry of Education, Youth and Sports of the Czech
Republic under Contract No.~LG14034;
the Carl Zeiss Foundation, the Deutsche Forschungsgemeinschaft
and the VolkswagenStiftung;
the Department of Science and Technology of India; 
the Istituto Nazionale di Fisica Nucleare of Italy; 
the WCU program of the Ministry Education Science and
Technology, National Research Foundation of Korea Grants
No.~2011-0029457, No.~2012-0008143, No.~2012R1A1A2008330,
No.~2013R1A1A3007772;
the BRL program under NRF Grant No.~KRF-2011-0020333,
No.~KRF-2011-0021196,
Center for Korean J-PARC Users, No.~NRF-2013K1A3A7A06056592; the BK21
Plus program and the GSDC of the Korea Institute of Science and
Technology Information;
the Polish Ministry of Science and Higher Education and 
the National Science Center;
%the Ministry of Education and Science of the Russian
%Federation and the Russian Federal Agency for Atomic Energy;
the Ministry of Education and Science of the Russian 
Federation, the Russian Federal Agency for Atomic Energy and 
the Russian Foundation for Basic Research grants RFBR 12-02-01296 and 
12-02-33015;
the Slovenian Research Agency;
the Basque Foundation for Science (IKERBASQUE) and the UPV/EHU under 
program UFI 11/55;
the Swiss National Science Foundation; the National Science Council
and the Ministry of Education of Taiwan; and the U.S.\
Department of Energy and the National Science Foundation.
This work is supported by a Grant-in-Aid from MEXT for 
Science Research in a Priority Area (``New Development of 
Flavor Physics'') and from JSPS for Creative Scientific 
Research (``Evolution of Tau-lepton Physics'').

%======================================================================
%======================================================================
%======================================================================

\section*{Appendix: The $\ee\to\Un\pp$ Amplitude}

Here, we present a Lorentz invariant form of the amplitude for the 
$\ee\to[\Un\pi_2]\pi_1$, $\Un\to\mu^+\mu^-$ transition. The amplitude 
might consist of several components, each describing a quasi-two-body 
process with a certain spin and parity of the intermediate state. 
The following symbols are used: $K_1$, $K_2$, $P_1$ and $P_2$ are 
4-momenta for the $\mu^+$, $\mu^-$, $\pi_1$ and $\pi_2$, respectively;
$Q_0=P_1+P_2$; $Q_1=Q_2+P_2$; $Q_2=K_1+K_2$; $P_0=Q_1+P_1$; and 
$\varepsilon_5$ and $\varepsilon_n$ are polarization vectors for the
virtual photon and $\Un$, ($n=1,2,3$), respectively. Greek indices denote 
4-momenta components and run from 0 to 3. The $\ee\to\Un\pp$ amplitude 
can be written as
\begin{eqnarray}
  {\cal M}_{\Un\pi\pi} = {\cal M}_{\ee\to\Un\pp}{\cal M}_{\Un\to\uu} = \nonumber \\
\varepsilon_5^\mu O_{\mu\nu} \varepsilon_n^{*\nu} \varepsilon_n^\alpha
(\bar{u}_1\gamma_\alpha u_2)
\end{eqnarray}
and
\begin{eqnarray}
  |{\cal M}_{\Un\pi\pi}|^2 = \hspace*{50mm} \nonumber  \\
\varepsilon_5^{\mu'}\varepsilon_5^\mu O_{\mu\nu} 
\varepsilon_n^{*\nu} \varepsilon_n^\alpha 
{\rm Sp}(K_1\gamma_\alpha K_2\gamma_{\alpha'})
\varepsilon_n^{*\alpha'} \varepsilon_n^{\nu'}  O_{\mu'\nu'}^*,
\end{eqnarray}
where $u_k$ are the muon spinors. Performing the summation over the 
repetitive Greek indices and neglecting the muon mass, one obtains
\begin{eqnarray}
  R^{\nu\nu'} = 
\varepsilon_n^{*\nu} \varepsilon_n^\alpha 
{\rm Sp}(K_1\gamma_\alpha K_2\gamma_{\alpha'})
\varepsilon_n^{*\alpha'} \varepsilon_n^{\nu'} = \hspace*{15mm} \nonumber \\
4(K_1^{\nu}K_2^{\nu'} + K_2^{\nu}K_1^{\nu'} - g^{\nu\nu'}(K_1\cdot K_2)),
\end{eqnarray}
where $(K_1\cdot K_2)=g_{\mu\nu}K_1^{\mu}K_2^{\nu}$, and we used
$\varepsilon_n^{*\nu} \varepsilon_n^{*\alpha} = 
g^{\nu\alpha} - \frac{Q_2^{\nu}Q_2^{\alpha}}{Q_2^2}$. Thus, we arrive at
\begin{equation}
  |{\cal{M}}_{\Un\pi\pi}|^2 = 
\delta_\perp^{\mu\mu'}O_{\mu\nu}R^{\nu\nu'}O_{\nu'\mu'}^*,
\label{eq:m-tot}
\end{equation}
where $\delta_\perp^{\mu\nu}=1$ if $\mu=\nu= 1,2$ and $\delta_\perp^{\mu\nu}=0$ 
otherwise.
The factor $O_{\mu\nu}$ depends on the dynamics of the $\ee\to\Un\pi_1\pi_2$ 
process (see below). In what follows, we consider only the following 
possible contributions to the three-body amplitude: $\ee\to Z_b\pi_1$, 
$Z_b\to \Un\pi_2$ and $\ee\to\Un(\pi_1\pi_2)_{S,D}$, where 
$(\pi_1\pi_2)_{S,D}$ denotes the system of two pions in an $S$- and 
$D$-wave configuration, respectively. We consider the following 
combinations of spin and parity of the intermediate $Z_b$ state: 
$J_{Z_b}^P=1^+$, $1^-$, $2^+$ and $2^-$. Factors $O_{\mu\nu}$ corresponding 
to these six amplitudes are given below.

%%%%%%%%%%%%%%%%%%%%%%%%%%%%%%%%%%%%%
%%%%%%%%%%%%%%%%%%%%%%%%%%%%%%%%%%%%%

%\item[1)] $J_{Z_b}^P=1^-$.
1) $J_{Z_b}^P=1^-$. Although both $P$- and $F$-waves are allowed for 
the $\pi_2$ here (and in the case of $J_{Z_b}^P=2^-$), the $F$-wave is 
substantially suppressed by the phase space factor, so we keep only 
the $P$-wave component of the amplitude
\begin{eqnarray}
O^{\mu\nu}_{\Upsilon\pi_2} = 
\varepsilon^*_{\alpha}\varepsilon^{\mu\alpha\gamma\rho}
P_{0\gamma}Q_{1\rho}
\varepsilon_{\sigma}\varepsilon^{\nu\sigma\delta\kappa}
Q_{1\delta}Q_{2\kappa} = \nonumber \\
g^{\mu\nu}\Bigl((P_0\cdot Q_1)(Q_1\cdot Q_2)-(P_0\cdot Q_2)Q^2_1\Bigr) +
\nonumber \\
Q^\mu_2 P^\nu_0Q^2_1 - Q^\mu_2 Q^\nu_1(P_0\cdot Q_1) + \nonumber \\
Q^\mu_1 Q^\nu_1(P_0\cdot Q_2) - Q^\mu_1 P^\nu_0(Q_1\cdot Q_2).~~~
\end{eqnarray}

%%%%%%%%%%%%%%%%%%%%%%%%%%%%%%%%%%%%%
%%%%%%%%%%%%%%%%%%%%%%%%%%%%%%%%%%%%%

%\item[2)]
2) $J_{Z_b}^P=1^+$. 
In this case (as well as in the case of $J_{Z_b}^P=2^+$) $S$- and $D$-waves
are allowed for the $\pi_2$. We keep only the $S$-wave since the $D$-wave 
is suppressed by the phase space factor. Thus
\begin{eqnarray}
O^{\mu\nu}_{\Upsilon\pi_2} = 
(g^{\mu\alpha}+a_1P_1^\mu P_1^\alpha)
\varepsilon^*_{\alpha}\varepsilon_{\beta}
(g^{\mu\beta}+a_2P_2^\beta P_2^\nu) =   \nonumber \\
g^{\mu\nu} + a_1P^\mu_1P^\nu_1 + a_2P^\mu_2P^\nu_2 - \nonumber \\
\frac{Q_1^\mu Q_1^\nu}{Q_1^2}(1-a_1(Q_1\cdot P_1)+a_2(Q_1\cdot P_2)) +
\nonumber \\
a_0a_1a_2P^\mu_1P^\nu_2,~~~
%\left((P_1\cdot P_2) - 
%\frac{(Q_1\cdot P_1)(Q_1\cdot P_2)}{Q_1^2}\right), \nonumber \\
\label{eq:jp1p}
\end{eqnarray}
where 
\begin{eqnarray}
a_0 & = & (P_1\cdot P_2) - 
      \frac{(Q_1\cdot P_1)(Q_1\cdot P_2)}{Q^2_1}; \nonumber\\
a_1 & = & \frac{(P_0\cdot Q_1) - \sqrt{P^2_0Q^2_1}}
           {(Q_1\cdot P_1)^2-m^2_{\pi}Q^2_1}; \nonumber \\
a_2 & = & \frac{(Q_1\cdot Q_2) - \sqrt{Q^2_1Q^2_2}}
           {(Q_2\cdot P_2)^2-m^2_{\pi}Q^2_2},
\end{eqnarray}
and $\varepsilon^*_{\alpha}\varepsilon_{\beta}=(g_{\alpha\beta}-\frac{Q_{1\alpha}Q_{1\beta}}{Q_1^2})$.
%%%%%%%%%%%%%%%%%%%%%%%%%%%%%%%%%%%%%
%%%%%%%%%%%%%%%%%%%%%%%%%%%%%%%%%%%%%

%\item[3)] $J_{Z_b}^P=2^+$.
3) $J_{Z_b}^P=2^-$.
\begin{eqnarray}
O^{\mu\nu}_{\Upsilon\pi_2} = 
\varepsilon^*_{\alpha\beta}\varepsilon^{\mu\alpha\gamma\rho}
P_{0\gamma}Q_{1\rho}P_0^{\beta} 
\varepsilon_{\sigma\tau}\varepsilon^{\nu\sigma\delta\kappa}
Q_{1\delta}Q_{2\kappa}Q_2^{\tau}.
\end{eqnarray}
Taking into account that
\begin{equation}
\varepsilon^*_{\alpha\beta}\varepsilon_{\sigma\delta} = 
\frac{1}{2}(G_{\alpha\sigma}G_{\beta\delta}+G_{\alpha\delta}G_{\beta\sigma}) - 
\frac{1}{3} G_{\alpha\beta}G_{\sigma\delta},
\end{equation}
where $G_{\alpha\beta} = g_{\alpha\beta} - \frac{Q_{1\alpha}Q_{1\beta}}{Q_1^2}$, 
we obtain
\begin{eqnarray}
O^{\mu\nu}_{\Upsilon\pi_2} = 
\frac{1}{2}\biggl[\Bigl(g^{\mu\nu}[(P_0\cdot Q_1)(Q_1\cdot Q_2) - 
(P_0\cdot Q_2)Q^2_1 ] + \nonumber \\
Q^\mu_2P^\nu_0Q^2_1 - Q^\mu_2Q^\nu_1(P_0\cdot Q_1) +  \nonumber \\
Q^\mu_1Q^\nu_1(P_0\cdot Q_2) - Q^\mu_1P^\nu_0(Q_1\cdot Q_2)\Bigl) \nonumber \\
\Bigl((P_0\cdot Q_2) - 
\frac{(P_0\cdot Q_1)(Q_1\cdot Q_2)}{Q^2_1} \Bigl)-d^\mu d^\nu \biggr],~~~
\end{eqnarray}
where $d^\mu = \varepsilon^\mu_{\nu\alpha\beta}P^\nu_0Q^\alpha_1Q^\beta_2$ and 
$\varepsilon^\mu_{\nu\alpha\beta} = g^{\mu\sigma}\varepsilon_{\sigma\nu\alpha\beta}$ 
and $\varepsilon_{\sigma\nu\alpha\beta}$ is an antisymmetric tensor.

%%%%%%%%%%%%%%%%%%%%%%%%%%%%%%%%%%%%%
%%%%%%%%%%%%%%%%%%%%%%%%%%%%%%%%%%%%%

%\item[4)] 
4) $J_{Z_b}^P=2^+$. 
\begin{eqnarray}
O^{\mu\nu}_{\Upsilon\pi_2} = 
\varepsilon^*_{\kappa\sigma}
(g^{\mu\kappa} + a_1P_1^\mu P_1^\kappa)P_1^\sigma
\varepsilon_{\alpha\beta}
(g^{\alpha\nu} + a_2P_2^\alpha P_2^\nu)P_2^\beta
\nonumber
\end{eqnarray}
and
\begin{eqnarray}
O^{\mu\nu}_{\Upsilon\pi_2} = \hspace*{68mm} \nonumber \\
g^{\mu\nu}\frac{a_0}{2} + \frac{P^\mu_2P^\nu_1}{2} +
P^\mu_2P^\nu_2\left(a_0a_2 - \frac{(Q_1\cdot P_1)}{2Q^2_1}\right) + \nonumber \\
P^\mu_1P^\nu_1\left(a_0a_1 - \frac{(Q_1\cdot P_2)}{2Q^2_1}\right) +
\nonumber \\
\frac{1}{3}\frac{Q^\mu_1Q^\nu_1}{Q^4_1}
\biggl[(P_0\cdot Q_1)(Q_1\cdot Q_2) +  \nonumber \\
3(Q_1\cdot P_1)(Q_1\cdot P_2)-\frac{3}{2}(P_1\cdot P_2)Q^2_1 +
\nonumber \\
a_2\Bigl((P_0\cdot Q_1)(m^2_{\pi}Q^2_1-(Q_1\cdot P_2)^2)\Bigr) + \nonumber \\
a_1\Bigl((Q_1\cdot Q_2)(m^2_{\pi}Q^2_1-(Q_1\cdot P_1)^2)\Bigr) + \nonumber \\
3a_0Q^2_1\Bigl(a_1(Q_1\cdot P_1) - a_2(Q_1\cdot P_2)\Bigr) - \nonumber \\
a_1a_2\Bigl(3a^2_0Q^4_1 - \hspace*{45mm} \nonumber \\
(m_\pi^2Q_1^2 - (Q_1\cdot P_1)^2)(m_\pi^2Q_1^2-(Q_1\cdot P_2)^2)\Bigr)\biggr],~~~
% \nonumber \\
\end{eqnarray}
where factors $a_0$, $a_1$, and $a_2$ are the same as in 
Eq.~\ref{eq:jp1p}.

%%%%%%%%%%%%%%%%%%%%%%%%%%%%%%%%%%%%%
%%%%%%%%%%%%%%%%%%%%%%%%%%%%%%%%%%%%%

In the case of production of the $\pp$ system with defined spin and parity, 
we assume that spin structure of the $b\bar{b}$ pair is not modified and 
the $\pp$ system is produced in an $S$-wave with respect to the $\Un$ 
state and decays depending on its spin.
We consider two cases: the relative angular momentum of the two pions being
equal to zero (decay in an $S$-wave) and equal to two (decay in a $D$-wave). 
The $O^{\mu\nu}$ factor for these parts of the three-body $\ee\to\Un\pp$ 
amplitude can be written as:

%\item[5)] 
5) $S$-wave.
\begin{equation}
O_S^{\mu\nu} = g^{\mu\nu} + 
Q^\mu_0Q^\nu_0
\frac{(P_0\cdot Q_2)-\sqrt{P^2_0Q^2_2}}
     {(Q_0\cdot Q_2)^2-Q^2_0Q^2_2}.
\end{equation}

%%%%%%%%%%%%%%%%%%%%%%%%%%%%%%%%%%%%%
%%%%%%%%%%%%%%%%%%%%%%%%%%%%%%%%%%%%%

%\item[6)] 
6) $D$-wave.

\begin{eqnarray}
O_D^{\mu\nu} = 
O_S^{\mu\nu}\biggl[(P_0\cdot P_1)^2 -
\frac{2(P_0\cdot P_1)(Q_0\cdot P_1)(P_0\cdot Q_0)}{Q^2_0}+ \hspace*{-10mm}
\nonumber \\
\frac{(P_0\cdot Q_0)^2(Q_0\cdot P_1)^2}{Q^4_0} - \nonumber \\
\frac{1}{3}\left(P^2_0 - \frac{(P_0\cdot Q_0)}{Q^2_0}\right)
\left(m^2_\pi - \frac{(Q_0\cdot P_1)^2}{Q^2_0}\right)\biggr].~~~
\end{eqnarray}

%\end{itemize}
%
%\vspace*{3mm}
%

%%%%%%%%%%%%%%%%%%%%%%%%%%%%%%%%%%%%%
%%%%%%%%%%%%%%%%%%%%%%%%%%%%%%%%%%%%%

The combined $O^{\mu\nu}$ in Eq.~\ref{eq:m-tot} is then calculated as
\begin{eqnarray}
O^{\mu\nu} = 
                 a_S(s_{23})O_S^{\mu\nu} + a_D(s_{23})O_D^{\mu\nu} ~+ \nonumber \\
         c_{Z_1}e^{i\delta_{Z_1}}\left(a_{Z_1}(s_{12})O^{\mu\nu}_{\Upsilon\pi_1} +
                 a_{Z_1}(s_{13})O^{\mu\nu}_{\Upsilon\pi_2}\right) + \nonumber \\
         c_{Z_2}e^{i\delta_{Z_2}}\left(a_{Z_2}(s_{12})O^{\mu\nu}_{\Upsilon\pi_1} +
                 a_{Z_2}(s_{13})O^{\mu\nu}_{\Upsilon\pi_2}\right),~~
%\nonumber
\label{eq:m_tot}
\end{eqnarray}
where $s_{12} = M^2(\Un\pi_1)$, $s_{13} = M^2(\Un\pi_2)$, and 
$s_{23} = M^2(\pi^+\pi^-)$ ($s_{23}$ can be expressed via $s_{12}$ and 
$s_{13}$ but we prefer to keep it here for clarity); $c_{Z_k}$ and 
$\delta_{z_k}$ are free parameters of the fit. Note that the $Z_k$ 
amplitudes in Eq.~\ref{eq:m_tot} are symmetrized with respect to $\pi_1$ 
and $\pi_2$ interchange to respect isospin symmetry.

In this analysis, the $S$-wave part of the amplitude is comprised of
the following possible modes: $\Un\sigma(500)$, $\Un f_0(980)$ and a 
non-resonant one, that is,
\begin{eqnarray}
a_S(s_{23}) = c_\sigma e^{i\delta_\sigma}a_\sigma(s_{23}) 
           + c_{f_0} e^{i\delta_{f_0}}a_{f_0}(s_{23}) 
           + {\cal A}^{\rm NR}(s_{23}), \hspace*{-10mm} \nonumber \\
\label{eq:s-wave}
\end{eqnarray}
where $a_\sigma(s_{23})$ is a Breit-Wigner function with mass and width 
fixed at $600$~MeV/$c^2$ and $400$~MeV, respectively; $a_{f_0}(s_{23})$ is 
parametrized by a Flatt\'{e} function with the mass and coupling constants 
fixed at values defined from the analysis of $B^+\to K^+\pi^+\pi^-$:
$M(f_0(980))=950$~MeV/$c^2$, $g_{\pi\pi}=0.23$, $g_{KK}=0.73$~\cite{Belle_kpp}.
Following the suggestion given in Refs.~\cite{Vol:2006ce,Vol:2007dx}, 
the non-resonant amplitude ${\cal A}^{\rm NR}(s_{23})$ is parametrized as 
\begin{equation}
{\cal A}^{\rm NR}(s_{23}) = c^{\rm NR}_1 e^{i\delta^{\rm NR}_1} +
                            c^{\rm NR}_2 e^{i\delta^{\rm NR}_2} s_{23}.
\label{eq:anr}
\end{equation}

The $D$-wave part of the three-body amplitude consists of only the
$\Un f_2(1270)$ mode
\begin{equation}
a_D(s_{23}) = c_{f_2} e^{i\delta_{f_2}}a_{f_2}(s_{23}),
\label{eq:d-wave}
\end{equation}
where $a_{f_2}(s_{23})$ is a Breit-Wigner function with the mass and width 
fixed at world average values~\cite{PDG}. In the study of a model related 
uncertainty, we also fit the data with $a_{f_2}(s_{23})$ replaced by just 
an $s_{23}$ term to represent a possible $D$-wave component of the 
non-resonant amplitude. Parameters $c_X$, $c^{\rm NR}_k$, and phases 
$\delta_X$ and $\delta^{\rm NR}_k$ in Eqs.~\ref{eq:s-wave}-\ref{eq:d-wave} 
are free parameters of the fit. Finally, terms 
$a_{Z_k}(s)$ in Eq.~\ref{eq:m_tot} are parametrized by Breit-Wigner 
functions with masses and widths to be determined from the fit.

Since we are sensitive to the relative phases and amplitudes only, we 
are free to fix one phase and one amplitude in Eq.~(\ref{eq:m_tot}). 
In the analysis of the $\Uu\pp$ mode, we fix $c^{\rm NR}_1=1$ and 
$\delta^{NR}_1=0$; in the analysis of the $\Ud\pp$ and $\Ut\pp$ modes, 
we fix the amplitude and the phase of the $Z_b(10610)$ component to
$c_{Z_1}=1$ and $\delta_{Z_1}=0$.

%======================================================================
%======================================================================
%======================================================================

\end{document}